\documentclass[Letter]{aa}

\usepackage{graphicx}
\usepackage{txfonts}
\usepackage{natbib}
\usepackage[breaklinks=true, colorlinks=true, allcolors=blue]{hyperref}
\usepackage{amsmath}    
\usepackage{amssymb}    
\usepackage{multirow}
\usepackage{longtable}
\usepackage{threeparttable}

\usepackage{multicol}

\graphicspath{{./}{figures/}}
\usepackage{graphicx}
\usepackage{epstopdf}
\usepackage{subfigure}
\usepackage{amssymb}
\usepackage{float}
\usepackage{needspace}
\usepackage{etoolbox}
\usepackage{placeins}

\begin{document}

  \title{$^{14}$N$/^{15}$N abundance ratio toward  massive star-forming regions with different Galactic distances}

\author{Chang Ruan\inst{1}, Junzhi Wang\inst{1}, Chao Ou\inst{1}, Juan Li\inst{2, 3}, Bo Zhang\inst{2, 3}}

 \institute{Guangxi Key Laboratory for Relativistic Astrophysics, School of Physical Science and Technology, Guangxi University, Nanning 530004,  China 
        \\ \email{junzhiwang@gxu.edu.cn, changruan@st.gxu.edu.cn}
        \and
        Shanghai Astronomical Observatory, Chinese Academy of Sciences,80 Nandan Road, Shanghai, 200030, China
         \and
            Key Laboratory of Radio Astronomy, Chinese Academy of Sciences, Nanjing 210033, China}

 \date{Received xx; accepted xxx}

 \authorrunning{Ruan et al.}


  \abstract
   {The abundance ratio of $^{14}$N$/^{15}$N is, in principle, a powerful tool for tracing stellar nucleosynthesis.}
   {This work aims to measure and analyze  ($^{14}$N/$^{15}$N)$\times$($^{13}$C/$^{12}$C) and $^{14}$N$/^{15}$N abundance ratios in massive star-forming regions across a range of galactocentric distances to provide constraints on galactic chemical evolution (GCE)  models.
}
   {We present  H$^{13}$CN and HC$^{15}$N J=2-1 results  toward 51massive star-forming regions obtained with the Institut de Radioastronomie Millimétrique (IRAM) 30 meter telescope.\ We used these results to derive  ($^{14}$N/$^{15}$N)$\times$($^{13}$C/$^{12}$C) abundance ratios as well as  $^{14}$N$/^{15}$N ratios  using the double isotope method.}
   {We find an overall decreasing trend in the  ($^{14}$N/$^{15}$N)$\times$($^{13}$C/$^{12}$C) abundance ratio  and an increasing trend in the $^{14}$N$/^{15}$N ratio with increasing galactocentric distance ($D_{\rm GC}$), which provides a good constraint for the GCE model based on high signal to noise ratio measurements. While the predicted  ($^{14}$N/$^{15}$N)$\times$($^{13}$C/$^{12}$C) ratios between  6 and 12 kpc determined using current GCE models are consistent with our observational results, the ratios from models for $D_{\rm GC}$ less than 6 kpc  are significantly higher than the observational results, which indicates GCE models for $^{14}$N/$^{15}$N and/or $^{13}$C/$^{12}$C ratios need to be updated for at least this range.}
   {}


\maketitle
%
\section{Introduction} \label{sec1}

        Galactic chemical evolution (GCE) models, which track the distribution of chemical elements of the interstellar medium in galaxies and their formation in stars  \citep{matteucci_modelling_2021}, help us understand the evolution of stars and galaxies  \citep{1980FCPh....5..287T, matteucci_modelling_2021}. 
        Carbon, nitrogen, and oxygen in the  interstellar medium originate from the evolution of stars via nucleosynthesis processes \citep{burbidge_synthesis_1957, wallerstein_synthesis_1997}. 
        The triple-$\rm \alpha$ capture process leads to the formation of $^{12}$C  \citep{1952ApJ...115..326S},
        while isotopes $^{13}$C, $^{15}$N, and $^{17}$O are produced in hydrogen-burning zones through either the cold or hot CNO cycles (\citealt{wiescher_cold_2010}; \citealt{romano_evolution_2022} provides a comprehensive review of CNO nucleosynthesis in stars). 
        Throughout their evolutionary phases, stars release these nucleosynthetic products into the surrounding medium, notably during the asymptotic giant branch and supernova phases \citep[see, e.g.,][]{nomoto_nucleosynthesis_2013, di_criscienzo_studying_2016, woosley_evolution_2019}.

        The isotopic abundance ratios of carbon, nitrogen, and oxygen depend heavily on the growth history of the Galactic disk and thus exhibit strong evolutionary features  \citep[e.g.,][]{romano_nova_2003, romano_evolution_2017}.
        On various timescales, stars with variable initial masses and chemical compositions create isotopes of different species in varying amounts \citep{1979ApJ...229.1046T, 1980ARA&A..18..399W}.
        Hence, measuring the isotopic abundance ratio is an efficient method for tracing the chemical history and constraining GCE models \citep{1994ARA&A..32..191W, romano_evolution_2022}. 
        For example, the $^{12}$C/$^{13}$C ratio is the consequence of contributions from stars of varying masses, which collectively contribute to the current radial gradient in the Milky Way over different timescales  \citep{milam_12_2005, yan_systematic_2019}.
        The $^{18}$O/$^{17}$O ratio can be used to trace stellar nucleosynthesis and metal enrichment processes  \citep{1994LNP...439...72H, 1994ARA&A..32..191W, jimenez-donaire_13_2017, ou_18o17o_2023}. 
        The $^{14}$N/$^{15}$N ratio can also be used to study stellar nucleosynthesis and constrain the latest GCE models \citep{ritchey_c_2015, colzi_chemout_2022}.
        $^{14}$N is produced through processes like hot bottom burning in asymptotic giant branch stars  \citep{1981A&A....94..175R, romano_evolution_2022} and the cold CNO cycle during hydrogen-burning stages  \citep{colzi_nitrogen_2018-1}, while $^{15}$N is mainly formed in nova outbursts via the hot CNO cycle  \citep{2003hic..book.....C, colzi_nitrogen_2018-1}.
        
        The abundance ratio of  isotopes can be obtained using three alternative approaches: the singly substituted molecules, doubly substituted molecules, and double isotope methods \citep{2014A&A...572A..24W}.
        For the singly substituted molecules method, which involves molecules such as CN/C$^{15}$N  \citep[see, e.g.,][]{2012ApJ...744..194A, ritchey_c_2015, sun_improved_2024}, the main isotopologs are usually abundant, resulting in strong lines. However, a high abundance frequently means a high optical depth, which must be corrected via radiative transfer modeling or the intensity ratios of the hyperfine transitions \citep{2014A&A...572A..24W}.
        The lines of doubly substituted molecules such as C$^{18}$O/$^{13}$C$^{18}$O \citep[see, e.g.,][]{paron_mapping_2018, jimenez-donaire_13_2017} are usually very weak as a result of their low abundances.
        The double isotope method, which involves, for example, HN$^{13}$C/H$^{15}$NC or H$^{13}$CN/HC$^{15}$N  \citep[see, e.g.,][]{2012ApJ...744..194A, 2014A&A...572A..24W, colzi_nitrogen_2018, colzi_nitrogen_2018-1, colzi_chemout_2022}, avoids the high opacities of the main isotopolog and the weak lines of doubly substituted molecules. 
        However, the abundance ratio of $^{12}$C/$^{13}$C must be known to derive the $^{14}$N/$^{15}$N ratio.
        
        We used the double isotope method to derive $^{14}$N/$^{15}$N ratios from the intensity ratio of H$^{13}$CN/HC$^{15}$N J=2-1 lines.  
        Assuming H$^{13}$CN and HC$^{15}$N 2-1  to be  optically thin with similar excitation properties,  the abundance ratios of $^{14}$N/$^{15}$N can be derived from velocity-integrated intensities of H$^{13}$CN/HC$^{15}$N 2-1 by multiplying by $^{12}$C/$^{13}$C  \citep[see, e.g.,][]{goldsmith_determination_1981, 2014A&A...572A..24W} as a function of $D_{\rm GC}$,
        \begin{equation}
        \frac{^{14}\rm N}{^{15}\rm N} = \frac{\int T_{\rm mb}\mathrm{d}v \rm (H^{13}CN)}{\int T_{\rm mb}\mathrm{d}v \rm(HC^{15}N)} \times \frac{^{12}\rm C}{^{13}\rm C}
        .\end{equation}
        The $^{12}$C/$^{13}$C  value in each source was obtained  from \cite{sun_improved_2024} as 
        \begin{equation}
        ^{12}\mathrm{C}/^{13}\mathrm{C} = (4.08^{+0.86}_{-0.43})\, \mathrm{kpc}^{-1} \times D_{\mathrm{GC}} + (18.8^{+2.6}_{-6.6})
        .\end{equation}
        
        We report the abundance ratios of $^{14}$N/$^{15}$N derived using the double isotope method from data obtained with the Institut de Radioastronomie Millimétrique (IRAM) 30 m telescope toward 51 massive star-forming regions.
        The IRAM 30 m observations are described in Sect. \ref{sec2}.
        Section \ref{sec3} provides a description of the main results.
        A discussion and our conclusions are presented in Sects. \ref{sec4} and \ref{sec5}.
        
\section{Observation}\label{sec2}
\subsection{IRAM 30 m observations}       
        The sample of 51 late-stage massive star-forming regions with parallactic distances from \citet{reid_trigonometric_2014, reid_trigonometric_2019} was selected to observe the spectral lines of H$^{13}$CN J=2-1 at 172.6778512 GHz and HC$^{15}$N J=2-1 at 172.1079571 GHz using the IRAM 30 meter telescope on Pico Veleta, Spain.\ The observations were carried out in June 2016, October 2016, and August 2017, except for G211.59+01.05, which  was observed in August 2020. 
        These sources, except for G211.59+01.05, are the same as those presented in \citet{li_deuterated_2022} for NH$_2$D molecules.
        The 2 mm (E1) band of the Eight Mixer Receiver (EMIR) and the Fourier Transform Spectrometers (FTS) backend were used to cover  from about 166.0 GHz to 173.8 GHz  with 195 kHz  channel spacing and dual polarization.
        The standard position-switching mode with an azimuth OFF of -600 arcsec was used.

        The IRAM 30 m telescope, with a beam size of  $\sim$14.3 arcsec at 172 GHz and a typical system temperature of  400 K in the 2 mm band, utilizes nearby quasi-stellar objects for pointing correction every two hours. 
        Focus was checked and corrected at the start of each run, as well as during sunsets and sunrises.
        The main beam brightness temperature ($T_{\rm mb}$) was calculated using $T_{\rm mb} = T_{\rm A}^{\ast} \cdot F_{\rm eff} / B_{\rm eff}$, where $T_{\rm A}^{\ast}$ is the antenna temperature, the forward efficiency ($F_{\rm eff}$) is 0.93, and the beam efficiency ($B_{\rm eff}$) is 0.73 for the 2 mm band.
        Each scan lasted 2 minutes, and the total on-source time ranged from 12 to 234 minutes per source.

\subsection{Data reduction}  
        Data reduction was based on the CLASS package in the GILDAS\footnote{\url{http://www.iram.fr/IRAMFR/GILDAS}} software,
        a comprehensive tool designed for (sub)millimeter radio-astronomical applications, which facilitates the efficient manipulation and visualization of spectra. 
        GILDAS is widely used to process data from the IRAM 30 m telescope and the NOrthern Extended Millimeter Array (NOEMA), excluding very long-baseline interferometry observations. 
        The spectra from each source were averaged into one spectrum using the CLASS package and subtracted from a first-order baseline. The spectra in each source were plotted on the same velocity scale to facilitate comparisons between H$^{13}$CN and HC$^{15}$N J=2-1.
        The line fluxes were determined through a direct integration of the emission lines, with errors calculated using the equation $\sigma = \rm{rms}
\sqrt{\delta \rm{v} \cdot \Delta \rm{v}}$, where $\delta \rm{v}$ represents the channel separation in velocity, $\Delta \rm{v}$  the velocity 
range for integration, and $\rm{rms}$  the root mean square noise obtained with baseline fitting for each spectrum with $\delta \rm{v}$ channel spacing.

\section{Results}\label{sec3} 
        The spectra with $T_{\rm peak} > 3 \sigma$ at 195 kHz channel spacing, which corresponds to  0.34 km s$^{-1}$ at 172 GHz,  were considered detections, where $T_{\rm peak}$ is peak $T_{\mathrm{mb}}$ and $\sigma$ is the rms noise. 
        Neither  H$^{13}$CN J=2-1 nor HC$^{15}$N J=2-1 were  detected in IRAS 05137+3919 or G012.90-00.24.
        HC$^{15}$N J=2-1 was not detected, while a clear double-peak structure of H$^{13}$CN J=2-1 was found in Sgr B2, which was likely due to the self-absorption of hot core regions \citep[see, e.g.,][]{2012ApJ...744..194A}.
        Both H$^{13}$CN J=2-1 and HC$^{15}$N J=2-1 lines were contaminated by nearby lines in G010.47+00.02, which hindered the use of these two lines for scientific analysis.
       H$^{13}$CN J=2-1 and HC$^{15}$N J=2-1 were detected and included in the ratio calculations for the remaining 47 sources.
        
         The velocity-integrated intensities of H$^{13}$CN J=2-1 range from 1.6±0.03 K km\,s$^{-1}$ in G211.59+01.05 to 115.7±0.7 K km\,s$^{-1}$ in G005.88-00.39, with a mid-value of 10.9±0.1 K km\,s$^{-1}$, as shown in Table \ref{table1}. 
        The velocity-integrated intensities of HC$^{15}$N J=2-1 range from 0.3±0.03 K km\,s$^{-1}$ in G211.59+01.05 to 38.2±0.2 K km\,s$^{-1}$ in W51M, with a mid-value of 2.4±0.1 K km\,s$^{-1}$.

\begin{figure*}
\includegraphics[width=0.48\linewidth]{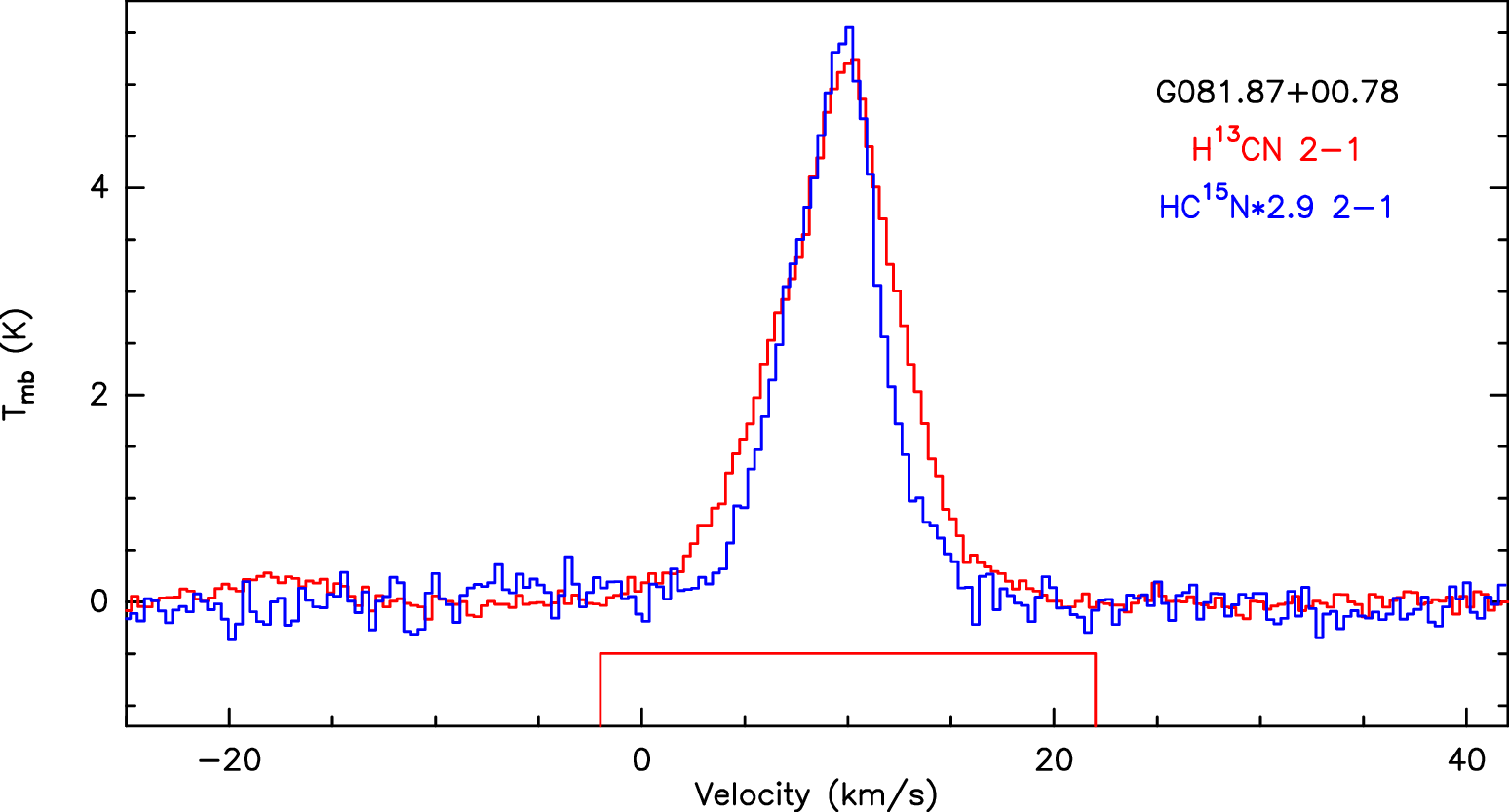}{}
\includegraphics[width=0.48\linewidth]{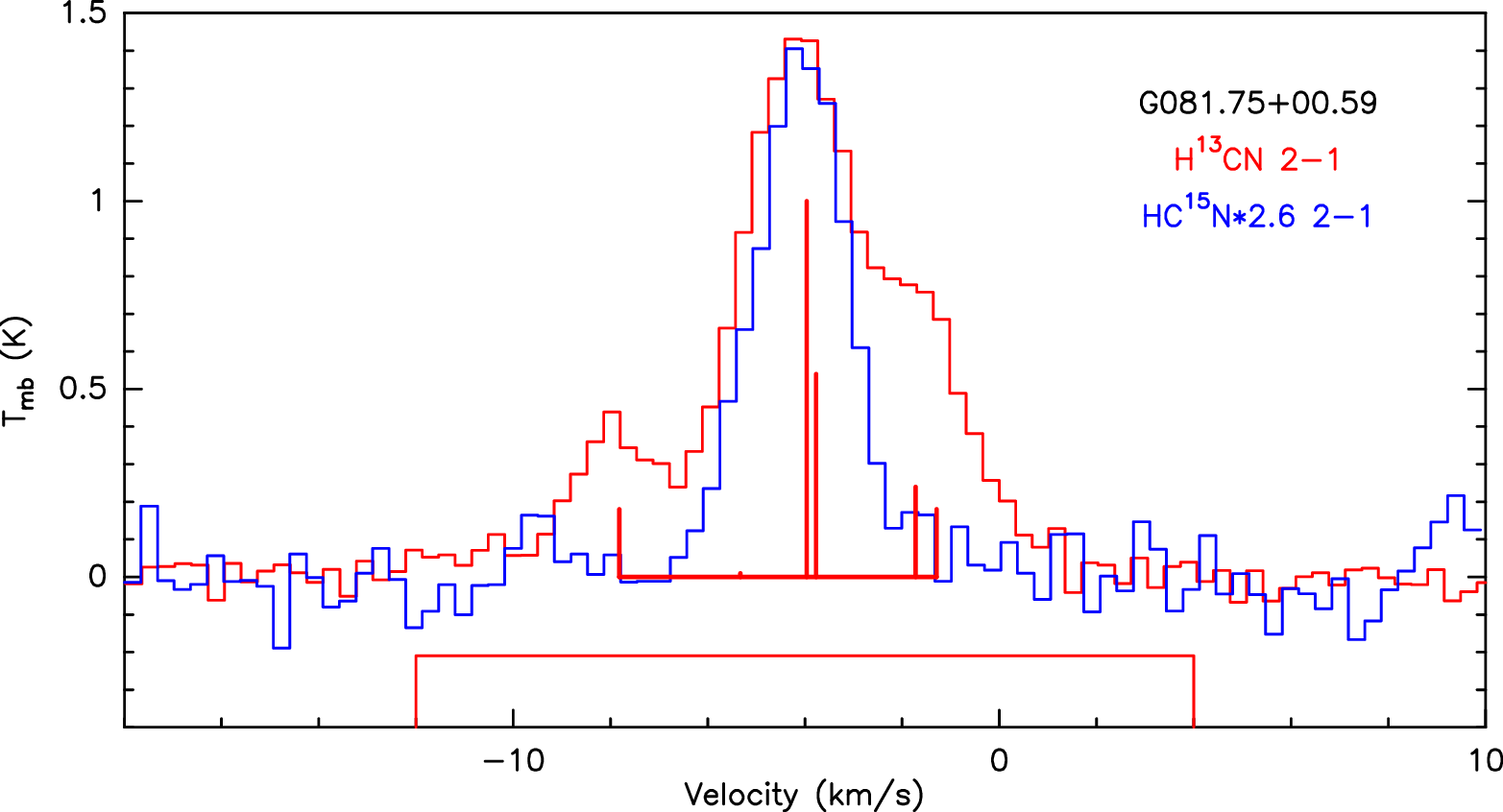}{}     
\caption{
Spectra of H$^{13}$CN J=2-1 and HC$^{15}$N J=2-1 for two sources as examples.
Left: H$^{13}$CN J=2-1 hfs in G081.87+00.78 cannot be resolved due to line broadening.
Right:  H$^{13}$CN J=2-1 hfs in G081.75+00.59 can be resolved with a narrow line width.
The vertical bold red lines indicate the hyperfine transitions for  H$^{13}$CN J=2-1: F=1-1, 1-2, 3-2, 2-1, 1-0, and 2-2 (from left to right).
}
\label{fig:hfs}
\end{figure*}

The hyperfine structure (hfs)  of H$^{13}$CN  2-1  \citep{fuchs_high_2004, cazzoli_lamb-dip_2005} can cause an overestimation of line widths if single-component Gaussians are used.
        Thus, only full widths at half maximum (FWHMs)  of HC$^{15}$N J=2-1 in each source were derived through single-component Gaussian fitting using the CLASS package. 
        The derived FWHMs  of HC$^{15}$N J=2-1 range from 2.0±0.2 km s$^{-1}$ in G183.72-03.66 to 15.8±0.2 km s$^{-1}$ in W49N, with a mid-value of 4.6±0.5 km s$^{-1}$.
        Obvious hfs features of H$^{13}$CN J=2-1 were found in 16 sources with HC$^{15}$N J=2-1 FWHMs less than 5.3 km s$^{-1}$, but they were blurred into component due to line broadening. 
        Figure \ref{fig:hfs}a shows an example of this, with the  H$^{13}$CN J=2-1 spectrum in  G081.87+00.78. Another example (G081.75+00.59), in which the hfs feature of  H$^{13}$CN J=2-1 can be distinguished, is shown in  Fig. \ref{fig:hfs}b.
        
\begin{figure*}
        \includegraphics[width=0.48\linewidth]{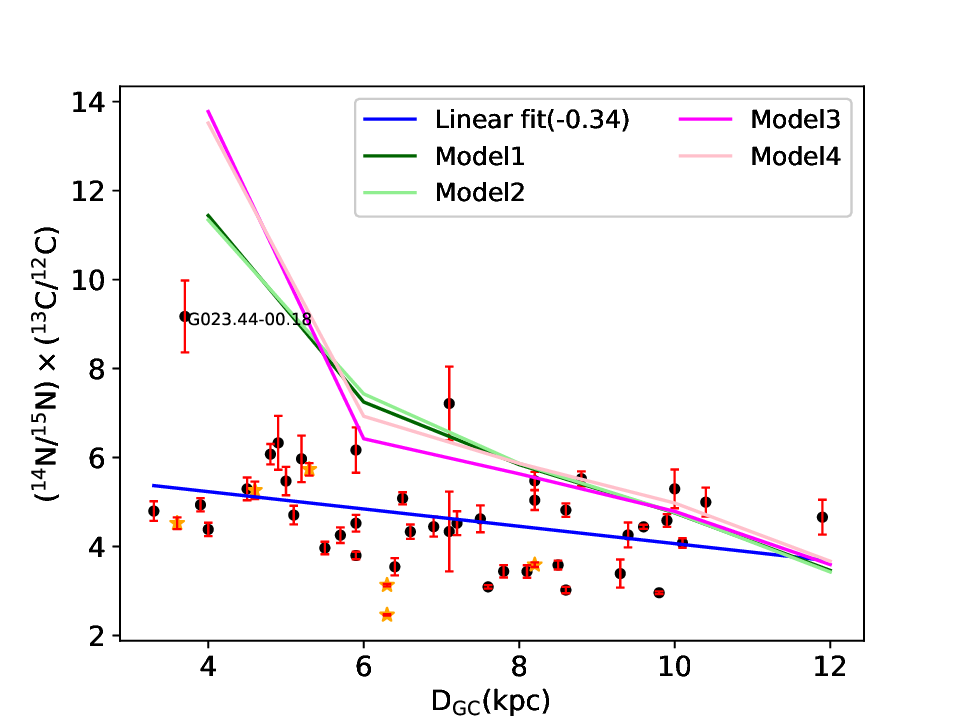}{}
        \includegraphics[width=0.48\linewidth]{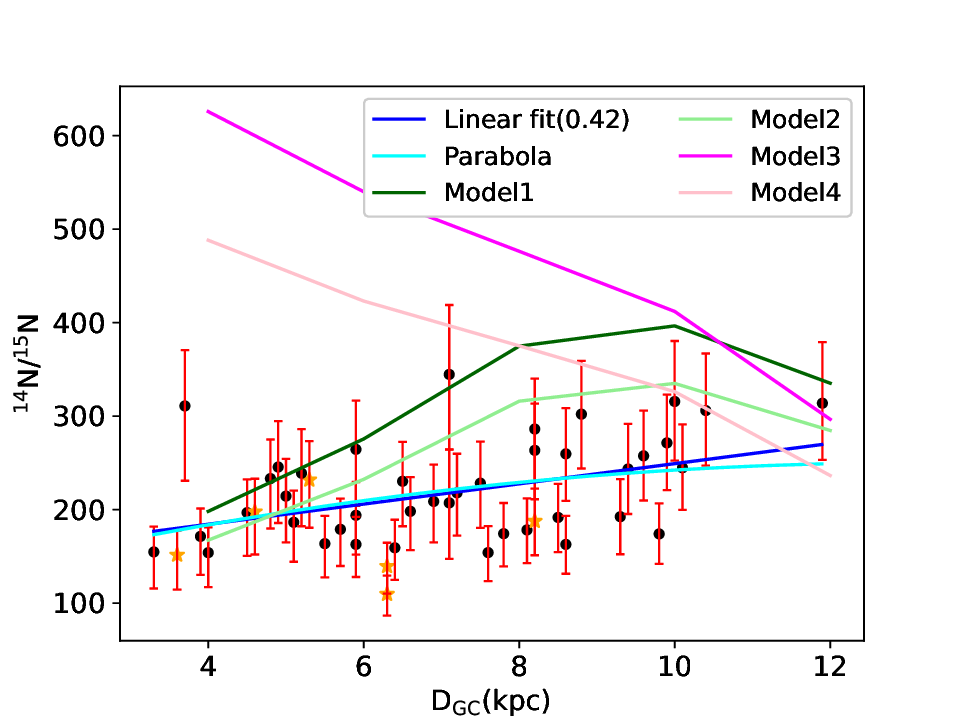}{}     
        \caption{       
        Abundance ratios of ($^{14}$N/$^{15}$N)$\times$($^{13}$C/$^{12}$C) (left) and $^{14}$N/$^{15}$N with $D_{\rm GC}$ (right). 
        The filled black circles are sources with peak $T_{\rm mb}$  of H$^{13}$CN 2-1 lower than 4K, and the orange stars sources with peak $T_{\rm mb}$ higher than 4 K.
        The blue lines are linear fitting results. The cyan parabola was obtained by multiplying the linear regression fit of ($^{14}$N/$^{15}$N)$\times$($^{13}$C/$^{12}$C) given in Eq. (3) by the $^{12}$C/$^{13}$C ratio (Eq. 2).
        The green, dark green, pink, and magenta lines are the models from \citet{colzi_chemout_2022}.
        }
        \label{fig:plot}        
        \end{figure*}

        The line ratio of H$^{13}$CN/HC$^{15}$N 2-1 can be used as the abundance ratio of ($^{14}$N/$^{15}$N)$\times$($^{13}$C/$^{12}$C)  \citep{goldsmith_determination_1981}, assuming the molecular isotopic abundances ratio represents the isotopic ratio of elements,      the same excitation condition as H$^{13}$CN and HC$^{15}$N molecules, and both lines are optically thin. The relationship between the ($^{14}$N/$^{15}$N)$\times$($^{13}$C/$^{12}$C) abundance ratio  derived from the H$^{13}$CN/HC$^{15}$N 2-1 line ratio and $D_{\rm GC}$, as well as the $^{14}$N/$^{15}$N ratios and $D_{\rm GC}$, is presented in Fig. \ref{fig:plot}. 
        The $D_{\rm GC}$ values range from 3.3 kpc in G009.62+00.19 to 11.9 kpc in G211.59+01.05 \citep{reid_trigonometric_2014, reid_trigonometric_2019}. 
        The ($^{14}$N/$^{15}$N)$\times$($^{13}$C/$^{12}$C)  ratio varies from 2.5±0.02 in W51M to 9.2±0.8 in G023.44-00.18.
        The unweighted linear regression fits to these data were obtained as follows:
        \begin{equation}
        \mathrm{(^{14}N/^{15}N)\times(^{13}C/^{12}C)} = (-0.2 \pm 0.1)\, \mathrm{kpc}^{-1} \times D_{\mathrm{GC}} + (6.0 \pm 0.9)
        ,\end{equation}
        with a correlation coefficient of -0.34.
        
\begin{figure}
        \centering
        \includegraphics[width=1\linewidth]{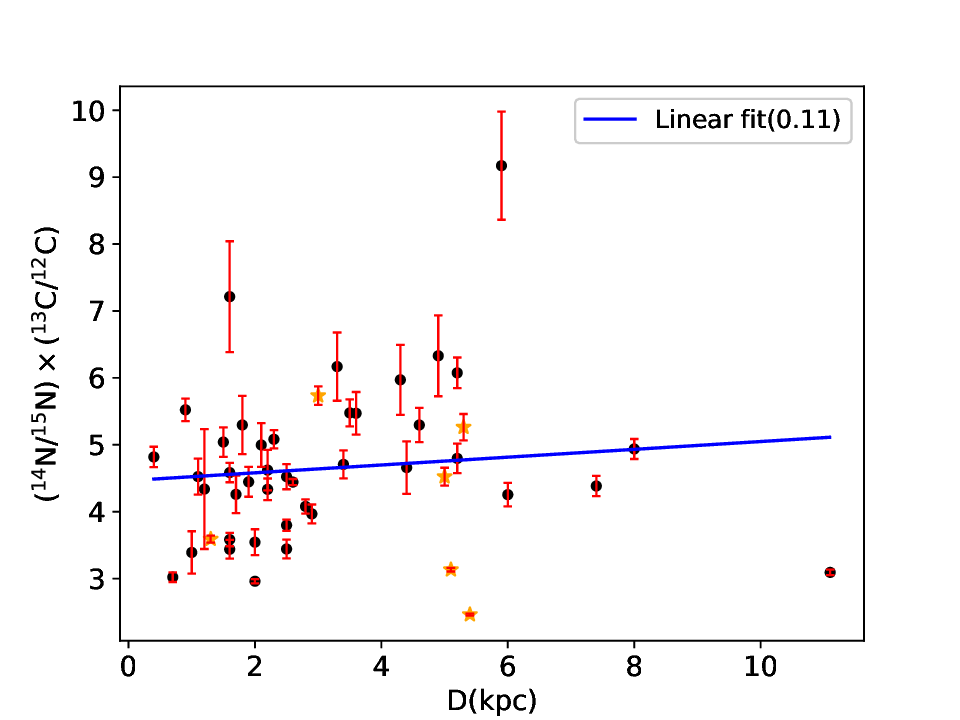}
        \caption{
        Abundance ratio of ($^{14}$N/$^{15}$N)$\times$($^{13}$C/$^{12}$C) versus heliocentric distance.
        The symbols are the same as in Fig. \ref{fig:plot}, and the blue line is the linear fitting result.
        }
        \label{fig:plot2} 
        \end{figure}

        The relation between  $D_{\rm GC}$ and the   $^{14}$N/$^{15}$N ratio, derived from the ($^{14}$N/$^{15}$N)$\times$($^{13}$C/$^{12}$C) ratio by multiplying by Eq. (2), spans from 109.6$^{+19.8}_{-22.9}$ in W51M to 344.6$^{+74.2}_{-80.1}$ in G034.39+00.22. It is presented in Fig. \ref{fig:plot}b.
        The unweighted linear regression fits for these data are        \begin{equation}
        ^{14}\mathrm{N}/^{15}\mathrm{N} = (10.8 \pm 6.0)\, \mathrm{kpc}^{-1} \times D_{\mathrm{GC}} + (141.1 \pm 43.8)
        .\end{equation}
        As shown in Figs. \ref{fig:plot}a and \ref{fig:plot}b, there is a significant decreasing trend in ($^{14}$N/$^{15}$N)$\times$($^{13}$C/$^{12}$C) and an increasing trend in $^{14}$N/$^{15}$N ratios with increasing $D_{\rm GC}$. The
        ($^{14}$N/$^{15}$N)$\times$($^{13}$C/$^{12}$C) ratio versus   the heliocentric distance instead of $D_{\rm GC}$ is presented in Fig. \ref{fig:plot2}, which does  not show any significant trend.
        
\section{Discussion}\label{sec4}

The  ($^{14}$N/$^{15}$N)$\times$($^{13}$C/$^{12}$C) ratios derived from the observed H$^{13}$CN/HC$^{15}$N 2-1  line ratio versus $D_{\rm GC}$ in these 47 sources are compared with four GCE models, shown in Fig. \ref{fig:plot}, with $D_{\rm GC}$  ranging from 4 to 12 kpc with data points for each 2 kpc \citep{colzi_chemout_2022}, in which different initial mass ranges for nova system white dwarf progenitors and various masses ejected in the form of $^{13}$C and $^{15}$N in each outburst were considered.
         The decreasing trend in the  ($^{14}$N/$^{15}$N)$\times$($^{13}$C/$^{12}$C) ratio with increments of  $D_{\rm GC}$  in the four models is consistent with our observational results (see Fig. \ref{fig:plot}a). On the other hand, the poor correlation coefficient between the ($^{14}$N/$^{15}$N)$\times$($^{13}$C/$^{12}$C) ratio and the source distance (see Fig. \ref{fig:plot2}) indicates that the decreasing trend in the  ($^{14}$N/$^{15}$N)$\times$($^{13}$C/$^{12}$C)  ratio with increments of  $D_{\rm GC}$ is not caused by observational effects of difference source distances to the Sun. 
         
The overall model values of the ($^{14}$N/$^{15}$N)$\times$($^{13}$C/$^{12}$C) ratio are higher than the measured ones (see Fig. \ref{fig:plot}a), which may be caused by the underestimation of the H$^{13}$CN/HC$^{15}$N ratio since H$^{13}$CN J=2-1 lines are not really optically thin in many sources, especially for sources with strong H$^{13}$CN 2-1 emission. 
There are six sources with peak H$^{13}$CN 2-1  $T_{\mathrm{mb}}$   values higher than 4 K. Because of the possible underestimation of the H$^{13}$CN/HC$^{15}$N ratio caused by the moderate opacity of H$^{13}$CN J=2-1,  our observational results in the range 6-12 kpc should be considered to be generally consistent with the models since they follow a similar trend.  However,  the  ($^{14}$N/$^{15}$N)$\times$($^{13}$C/$^{12}$C) ratios predicted by all four models are significantly higher than the observational results, except for that in G023.44-00.18, which has a large error bar. We suggest  that the model results of the $^{14}$N/$^{15}$N and/or $^{13}$C/$^{12}$C ratios at 4 kpc need to be updated. Similar results of low ($^{14}$N/$^{15}$N)$\times$($^{13}$C/$^{12}$C) ratios in three sources (IRAS 15567-5236, IRAS 17160-3707, and IRAS 17220-3609) with small  $D_{\rm GC}$   had been reported,  with H$^{13}$CN and HC$^{15}$N 1-0 observations  \citep{1995A&A...295..194D}. However, no scientific comparisons of observational and model results are available for these sources  \citep{1995A&A...295..194D}, mainly because the models had not been developed by that time.   
        
        The $^{14}$N/$^{15}$N ratio in each source can be derived using the already obtained  ($^{14}$N/$^{15}$N)$\times$($^{13}$C/$^{12}$C) ratios and the $^{13}$C/$^{12}$C ratio, which is unknown. We obtained the  $^{14}$N/$^{15}$N ratio of each source by assuming the  $^{13}$C/$^{12}$C ratio is related to $D_{\rm GC}$ and using the fitting result from  \cite{sun_improved_2024},  we which plot in  Fig. \ref{fig:plot}b.  The slightly tighter correlation between the $^{14}$N/$^{15}$N and the $D_{\rm GC}$,  with a correlation coefficient of  0.42, than that of ($^{14}$N/$^{15}$N)$\times$($^{13}$C/$^{12}$C) and $D_{\rm GC}$, with a correlation coefficient of -0.34, is due to the use of the $^{13}$C/$^{12}$C ratio from the fitting result in  \cite{sun_improved_2024}. An accurate $^{14}$N/$^{15}$N  ratio for each source should be obtained with the measured $^{13}$C/$^{12}$C value in individual sources.  Even though the comparison between measured and model ratios of  ($^{14}$N/$^{15}$N)$\times$($^{13}$C/$^{12}$C) versus  $D_{\rm GC}$ should be more accurate than that of the  $^{14}$N/$^{15}$N ratio versus $D_{\rm GC}$, due to the uncertainties of the assumed $^{13}$C/$^{12}$C ratio, the relation between  $^{14}$N/$^{15}$N  and $D_{\rm GC}$ (see Fig. \ref{fig:plot}b) can help us find the differences between the four models. When the assumed $^{13}$C/$^{12}$C ratio from \cite{sun_improved_2024} is similar to that of the real value in each source, the  $^{14}$N/$^{15}$N ratios from observational results are generally consistent with models 1 and 2.\ However,  they are significantly different to that at 4 and 6 kpc with models 3 and 4, which should therefore be updated. 
        
        The range in  $D_{\rm GC}$ in our sample is similar to that in  \citet{colzi_nitrogen_2018-1}, while sources with $D_{\rm GC}$ from 12 kpc to 19 kpc were included in  \cite{colzi_chemout_2022}.  However, the ($^{14}$N/$^{15}$N)$\times$($^{13}$C/$^{12}$C) ratio has much lower uncertainties thanks to the longer typical observing time on each source than that in \citet{colzi_nitrogen_2018-1}, and as such our results can better constrain predicted ($^{14}$N/$^{15}$N)$\times$($^{13}$C/$^{12}$C)  and $^{14}$N/$^{15}$N ratios from GCE models, especially when $D_{\rm GC}$ is less than 6 kpc,  using the values measured in each source.  Future observations with other tracers, which can avoid optically thick effects, will be useful for deriving more accurate  $^{14}$N/$^{15}$N ratios than observations with H$^{13}$CN lines. 
        GCE models of the relation between chemical abundances and $D_{\rm GC}$  were also tested with various tracers in ionized gas, such as radio recombination lines \citep[RRLs;][]{balser_h_2011}, RRLs and optical lines \citep{mendez-delgado_gradients_2022},   and RRLs and infrared hfs lines  \citep{pineda_nitrogen_2024}.   The relationship between N/H and $D_{\rm GC}$  shows a clear gradient in regions of $D_{\rm GC}$ greater than 6 kpc,  consistent with GCE models, while large scatter is found for $D_{\rm GC}$  less than $\sim$ 6 kpc \citep{pineda_nitrogen_2024}.  The He, C, N, O, Ne, S, Cl, and Ar abundances varying with $D_{\rm GC}$ \citep{mendez-delgado_gradients_2022}  agrees well with GCE models for    $D_{\rm GC}$ greater than 6 kpc; however, they lack data points with $D_{\rm GC}$ less than 6 kpc, mainly because of dust obscuration in the Galactic plane. Thus, isotopic ratios in the regions with $D_{\rm GC}$ less than 6 kpc in molecular gas derived with millimeter wavelengths are ideal for constraining GCE models.

\section{Summary and conclusion}\label{sec5}
         H$^{13}$CN 2-1 and HC$^{15}$N 2-1 lines are detected in 47 of the 51 massive star-forming regions observed with the IRAM 30m telescope, and they
         were used to derive the abundance ratios of ($^{14}$N/$^{15}$N)$\times$($^{13}$C/$^{12}$C)  as well as $^{14}$N/$^{15}$N via the double isotope method and assumed  $^{13}$C/$^{12}$C ratio.
        The  ($^{14}$N/$^{15}$N)$\times$($^{13}$C/$^{12}$C) ratio shows a decreasing trend with increasing  $D_{\rm GC}$,  while  no trend is found for this ratio and distance from the Sun.
        The $^{14}$N/$^{15}$N ratio  shows an increasing  trend with increasing $D_{\rm GC}$. With a small error bar for the measured ($^{14}$N/$^{15}$N)$\times$($^{13}$C/$^{12}$C) ratio, our results can constrain the  GCE model of ($^{14}$N/$^{15}$N)$\times$($^{13}$C/$^{12}$C) well, especially for $D_{\rm GC}$ less than 6kpc; for that range, updated GCE models are needed. 
        
\begin{acknowledgements}
We thank Dr. Donatella Romano for helpful discussion and provide the data GCE model.  This work is supported by National Key R$\&$D Program of China under grant 2023YFA1608204 and the National Natural Science Foundation of China grant 12173067. This work is based on observations carried out under project numbers 012-16, 023-17, and 005-20, with the IRAM 30m telescope. IRAM is supported by INSU/CNRS (France), MPG (Germany) and IGN (Spain).
\end{acknowledgements}

\bibliography{NitrogenAbundanceRatio}
\bibliographystyle{aa}

\begin{appendix}

\onecolumn
\section{Data parameters}

\begin{table*}[h!]
\caption{
        Source names, equatorial coordinates, distances, molecular species, and ratio data.
        \label{table1}
        }
\setlength{\tabcolsep}{2pt}
\tiny
\begin{tabular}{ccccccccccccc}
        \hline
        \hline
        Source & 
        R.A. & 
        Decl. & 
        $D$ & 
        $D_{\rm GC}$ & 
        Line & 
        $\int T_{\rm mb}\mathrm{d}v$ & 
        Velocity range & 
        Rms & 
        $T_{\rm peak}$ & 
        $V_{\rm LSR}$ & 
        ($^{14}$N/$^{15}$N)$\times$($^{13}$C/$^{12}$C) &
        $^{14}$N/$^{15}$N \\ 
        Alias & (hh:mm:ss) & (dd:mm:ss) & (kpc) & (kpc) & (2-1) & (K·km\,s$^{-1}$) & (km\,s$^{-1}$) & (10$^{-2}$K) & (k) & (km\,s$^{-1}$) &  &  \\
        \hline
G005.88-00.39 & 18:00:30.31 & -24:04:04.5 & 3.0 & 5.3 & H$^{13}$CN & 115.73(0.69) & -10 40 & 16.8 & 10.2 & 10.02(0.03) & 5.73(0.14) & 231.8($^{+41.43}_{-51.22}$)\\
 &  &  &  &  & HC$^{15}$N & 20.18(0.47) & 3 18 & 21.0 & 3.4 & 9.29(0.06) &  & \\
G009.62+00.19 & 18:06:14.66 & -20:31:31.7 & 5.2 & 3.3 & H$^{13}$CN & 32.77(0.39) & -11 18 & 12.4 & 2.51 & 4.62(0.07) & 4.8(0.22) & 154.74($^{+27.04}_{-39.11}$)\\
 &  &  &  &  & HC$^{15}$N & 6.83(0.3) & -7 15 & 11.1 & 0.96 & 3.77(0.12) &  & \\
G010.62-00.38 & 18:10:28.55 & -19:55:48.6 & 5.0 & 3.6 & H$^{13}$CN & 63.45(0.44) & -20 20 & 11.9 & 6.21 & -4.54(0.03) & 4.52(0.13) & 151.48($^{+26.15}_{-37.13}$)\\
W 31 &  &  &  &  & HC$^{15}$N & 14.03(0.4) & -20 20 & 10.9 & 1.86 & -3.49(0.06) &  & \\
G011.49-01.48 & 18:16:22.13 & -19:41:27.2 & 1.2 & 7.1 & H$^{13}$CN & 7.02(0.37) & 5 17 & 18.5 & 1.41 & 10.39(0.11) & 4.34(0.9) & 207.2($^{+57.06}_{-59.86}$)\\
 &  &  &  &  & HC$^{15}$N & 1.62(0.32) & 5 15 & 17.5 & 0.57 & 10.16(0.21) &  & \\
G011.91-00.61 & 18:13:58.12 & -18:54:20.3 & 3.4 & 5.1 & H$^{13}$CN & 12.49(0.15) & 20 50 & 4.65 & 0.95 & 35.91(0.07) & 4.71(0.21) & 186.42($^{+33.91}_{-42.21}$)\\
 &  &  &  &  & HC$^{15}$N & 2.65(0.11) & 27 48 & 4.26 & 0.3 & 35.48(0.17) &  & \\
G012.80-00.20 & 18:14:14.23 & -17:55:40.5 & 2.9 & 5.5 & H$^{13}$CN & 13.68(0.13) & 20 50 & 4.07 & 1.21 & 36.19(0.04) & 3.97(0.14) & 163.59($^{+29.65}_{-36.03}$)\\
 &  &  &  &  & HC$^{15}$N & 3.45(0.12) & 25 50 & 4.04 & 0.61 & 35.35(0.06) &  & \\
G012.88+00.48 & 18:11:51.42 & -17:31:29.0 & 2.5 & 5.9 & H$^{13}$CN & 17.12(0.12) & 23 49 & 3.89 & 1.84 & 33.21(0.04) & 3.8(0.08) & 162.83($^{+29.37}_{-34.89}$)\\
IRAS 18089-1732 &  &  &  &  & HC$^{15}$N & 4.51(0.09) & 25 41 & 4.04 & 0.79 & 33.23(0.08) &  & \\
G012.90-00.26 & 18:14:39.57 & -17:52:00.4 & 2.5 & 5.9 & H$^{13}$CN & 18.87(0.22) & 23 52 & 7.17 & 2.04 & 38.19(0.04) & 4.52(0.19) & 193.92($^{+35.63}_{-42.11}$)\\
 &  &  &  &  & HC$^{15}$N & 4.17(0.17) & 30 45 & 7.35 & 0.77 & 37.83(0.08) &  & \\
G014.33-00.64 & 18:18:54.67 & -16:47:50.3 & 1.1 & 7.2 & H$^{13}$CN & 27.57(0.62) & 6 35 & 19.8 & 3.09 & 22.41(0.07) & 4.52(0.27) & 217.96($^{+41.84}_{-45.74}$)\\
 &  &  &  &  & HC$^{15}$N & 6.09(0.34) & 15 27 & 16.6 & 1.43 & 22.13(0.09) &  & \\
G015.03-00.67 & 18:20:24.81 & -16:11:35.3 & 2.0 & 6.4 & H$^{13}$CN & 9.02(0.17) & 10 27 & 7.25 & 1.7 & 19.52(0.04) & 3.54(0.19) & 159.17($^{+30.01}_{-34.27}$)\\
M 17 &  &  &  &  & HC$^{15}$N & 2.54(0.13) & 13 25 & 6.45 & 0.78 & 19.23(0.06) &  & \\
G016.58-00.05 & 18:21:09.08 & -14:31:48.8 & 3.6 & 5.0 & H$^{13}$CN & 11.54(0.17) & 47 73 & 5.56 & 1.28 & 59.77(0.05) & 5.47(0.32) & 214.45($^{+39.75}_{-49.46}$)\\
 &  &  &  &  & HC$^{15}$N & 2.11(0.12) & 52 68 & 5.09 & 0.38 & 59.94(0.12) &  & \\
G023.00-00.41 & 18:34:40.20 & -09:00:37.0 & 4.6 & 4.5 & H$^{13}$CN & 11.04(0.14) & 65 90 & 4.64 & 1.03 & 78.22(0.06) & 5.3(0.26) & 196.78($^{+35.55}_{-46.18}$)\\
 &  &  &  &  & HC$^{15}$N & 2.08(0.1) & 70 85 & 4.3 & 0.28 & 77.62(0.17) &  & \\
G023.44-00.18 & 18:34:39.19 & -08:31:25.4 & 5.9 & 3.7 & H$^{13}$CN & 8.39(0.13) & 88 115 & 4.14 & 0.74 & 103.01(0.08) & 9.17(0.81) & 310.88($^{+59.69}_{-79.96}$)\\
 &  &  &  &  & HC$^{15}$N & 0.91(0.08) & 97 107 & 4.31 & 0.22 & 102.16(0.15) &  & \\
 G027.36-00.16 & 18:41:51.06 & -05:01:43.4 & 8.0 & 3.9 & H$^{13}$CN & 18.71(0.15) & 79 110 & 4.59 & 1.6 & 92.45(0.06) & 4.94(0.15) & 171.39($^{+29.85}_{-41.2}$)\\
 &  &  &  &  & HC$^{15}$N & 3.79(0.11) & 80 110 & 3.46 & 0.39 & 92.85(0.12) &  & \\
G028.86+00.06 & 18:43:46.22 & -03:35:29.6 & 7.4 & 4.0 & H$^{13}$CN & 10.63(0.11) & 87 113 & 3.71 & 1.09 & 103.92(0.04) & 4.38(0.15) & 153.97($^{+27.01}_{-36.86}$)\\
 &  &  &  &  & HC$^{15}$N & 2.43(0.08) & 95 112 & 3.31 & 0.33 & 103.64(0.1) &  & \\
G029.95-00.01 & 18:46:03.74 & -02:39:22.3 & 5.3 & 4.6 & H$^{13}$CN & 37.16(0.4) & 85 110 & 13.9 & 4.08 & 97.88(0.04) & 5.26(0.2) & 197.7($^{+35.29}_{-45.74}$)\\
W 43S &  &  &  &  & HC$^{15}$N & 7.06(0.25) & 90 105 & 11.2 & 1.07 & 97.79(0.1) &  & \\
G031.28+00.06 & 18:48:12.39 & -01:26:30.7 & 4.3 & 5.2 & H$^{13}$CN & 10.92(0.24) & 100 116 & 10.5 & 1.22 & 109.1(0.1) & 5.97(0.52) & 238.94($^{+47.13}_{-56.76}$)\\
 &  &  &  &  & HC$^{15}$N & 1.83(0.15) & 105 113 & 9.39 & 0.39 & 109.16(0.21) &  & \\
G031.58+00.07 & 18:48:41.68 & -01:09:59.0 & 4.9 & 4.9 & H$^{13}$CN & 6.99(0.14) & 88 103 & 6.04 & 0.89 & 96.39(0.08) & 6.33(0.61) & 245.55($^{+49.11}_{-59.91}$)\\
W 43Main &  &  &  &  & HC$^{15}$N & 1.1(0.1) & 90 103 & 4.91 & 0.29 & 96.48(0.14) &  & \\
G032.04+00.05 & 18:49:36.58 & -00:45:46.9 & 5.2 & 4.8 & H$^{13}$CN & 15.38(0.15) & 85 112 & 5.05 & 1.36 & 95.51(0.05) & 6.08(0.23) & 233.18($^{+41.81}_{-53.36}$)\\
 &  &  &  &  & HC$^{15}$N & 2.53(0.09) & 90 102 & 4.56 & 0.4 & 96.01(0.12) &  & \\
G034.39+00.22 & 18:53:18.77 & 01:24:08.8 & 1.6 & 7.1 & H$^{13}$CN & 3.6(0.09) & 48 67 & 3.48 & 0.51 & 57.42(0.07) & 7.21(0.83) & 344.6($^{+74.22}_{-80.09}$)\\
 &  &  &  &  & HC$^{15}$N & 0.5(0.06) & 52 62 & 3.04 & 0.13 & 57.13(0.18) &  & \\
G035.02+00.34 & 18:54:00.67 & 02:01:19.2 & 2.3 & 6.5 & H$^{13}$CN & 12.11(0.09) & 40 62 & 3.32 & 1.46 & 52.69(0.03) & 5.08(0.14) & 230.33($^{+42.08}_{-48.15}$)\\
 &  &  &  &  & HC$^{15}$N & 2.38(0.06) & 45 59 & 2.82 & 0.39 & 52.46(0.07) &  & \\
 G035.19-00.74 & 18:58:13.05 & 01:40:35.7 & 2.2 & 6.6 & H$^{13}$CN & 17.55(0.2) & 22 44 & 7.48 & 1.82 & 33.99(0.05) & 4.33(0.16) & 198.18($^{+36.62}_{-41.57}$)\\
 &  &  &  &  & HC$^{15}$N & 4.05(0.14) & 23 42 & 5.65 & 0.64 & 33.23(0.09) &  & \\
 G035.20-01.73 & 19:01:45.54 & 01:13:32.5 & 3.3 & 5.9 & H$^{13}$CN & 5.92(0.08) & 36 50 & 3.81 & 1.08 & 43.58(0.03) & 6.17(0.51) & 264.46($^{+52.15}_{-60.46}$)\\
 &  &  &  &  & HC$^{15}$N & 0.96(0.08) & 37 50 & 3.72 & 0.29 & 43.27(0.09) &  & \\
 G037.43+01.51 & 18:54:14.35 & 04:41:41.7 & 1.9 & 6.9 & H$^{13}$CN & 7.67(0.11) & 35 52 & 4.78 & 1.3 & 44.15(0.03) & 4.45(0.22) & 208.73($^{+39.36}_{-43.8}$)\\
 &  &  &  &  & HC$^{15}$N & 1.73(0.08) & 38 50 & 4.09 & 0.49 & 43.97(0.06) &  & \\
G043.16+00.01 & 19:10:13.41 & 09:06:12.8 & 11.1 & 7.6 & H$^{13}$CN & 44.25(0.24) & -20 35 & 5.59 & 2.54 & 5.19(0.04) & 3.09(0.04) & 154.04($^{+28.33}_{-30.58}$)\\
W 49N &  &  &  &  & HC$^{15}$N & 14.31(0.17) & -13 28 & 4.54 & 0.86 & 6.31(0.09) &  & \\
G043.79-00.12 & 19:11:53.99 & 09:35:50.3 & 6.0 & 5.7 & H$^{13}$CN & 10.89(0.13) & 33 56 & 4.6 & 1.14 & 44.41(0.05) & 4.25(0.18) & 178.93($^{+32.77}_{-39.21}$)\\
OH 43.8-0.1 &  &  &  &  & HC$^{15}$N & 2.56(0.1) & 35 52 & 4.22 & 0.39 & 43.44(0.11) &  & \\
\hline
\end{tabular}
\tablefoot{
        The table lists the source names, aliases, and equatorial coordinates (Cols. 1-3), the heliocentric distance determined via the parallax method from \citet[Col. 4]{reid_trigonometric_2014, reid_trigonometric_2019}, the galactocentric distance ($D_{\rm GC}$) calculated from the heliocentric distances (Col. 5), the molecular species data, including their velocity-integrated intensities and velocity ranges (Cols. 6-8), the rms noise level obtained through baseline fitting (Col. 9), the peak $T_{\mathrm{mb}}$ and central velocity derived from single Gaussian fitting of the spectra (Cols. 10-11), and   the ($^{14}$N/$^{15}$N)$\times$($^{13}$C/$^{12}$C) and $^{14}$N/$^{15}$N ratios (Cols. 12-13).
}
\end{table*}

\begin{table*}[h!]
\addtocounter{table}{-1}
\caption{Continued.}
\setlength{\tabcolsep}{2pt}
\tiny
\begin{tabular}{ccccccccccccc}
        \hline
        \hline
        Source & 
        R.A. & 
        Decl. & 
        $D$ & 
        $D_{\rm GC}$ & 
        Line & 
        $\int T_{\rm mb}\mathrm{d}v$ & 
        Velocity range & 
        Rms & 
        $T_{\rm peak}$ & 
        $V_{\rm LSR}$ & 
        ($^{14}$N/$^{15}$N)$\times$($^{13}$C/$^{12}$C) &
        $^{14}$N/$^{15}$N \\ 
        Alias & (hh:mm:ss) & (dd:mm:ss) & (kpc) & (kpc) & (2-1) & (K·km\,s$^{-1}$) & (km\,s$^{-1}$) & (10$^{-2}$K) & (k) & (km\,s$^{-1}$) &  &  \\
        \hline
G049.48-00.36 & 19:23:39.82 & 14:31:05.0 & 5.1 & 6.3 & H$^{13}$CN & 59.91(0.23) & 40 80 & 6.35 & 5.5 & 60.47(0.03) & 3.13(0.03) & 139.42($^{+25.15}_{-29.19}$)\\
W 51 IRS2 &  &  &  &  & HC$^{15}$N & 19.12(0.15) & 50 75 & 5.04 & 2.18 & 60.38(0.07) &  & \\
G049.48-00.38 & 19:23:43.87 & 14:30:29.5 & 5.4 & 6.3 & H$^{13}$CN & 94.12(0.32) & 40 80 & 8.8 & 6.84 & 55.63(0.09) & 2.46(0.02) & 109.58($^{+19.75}_{-22.93}$)\\
W 51M &  &  &  &  & HC$^{15}$N & 38.23(0.2) & 35 77 & 5.26 & 3.2 & 56.48(0.2) &  & \\
G059.78+00.06 & 19:43:11.25 & 23:44:03.3 & 2.2 & 7.5 & H$^{13}$CN & 5.48(0.1) & 15 30 & 4.29 & 0.98 & 22.38(0.04) & 4.62(0.3) & 228.3($^{+44.4}_{-47.79}$)\\
 &  &  &  &  & HC$^{15}$N & 1.19(0.07) & 16 28 & 3.69 & 0.36 & 22.16(0.07) &  & \\
G069.54-00.97 & 20:10:09.07 & 31:31:36.0 & 2.5 & 7.8 & H$^{13}$CN & 8.9(0.11) & 0 23 & 3.87 & 1.23 & 12.05(0.03) & 3.44(0.14) & 174.27($^{+32.81}_{-34.99}$)\\
ON 1 &  &  &  &  & HC$^{15}$N & 2.58(0.1) & 0 23 & 3.58 & 0.46 & 11.92(0.07) &  & \\
G075.76+00.33 & 20:21:41.09 & 37:25:29.3 & 3.5 & 8.2 & H$^{13}$CN & 9.61(0.07) & -11 7 & 2.88 & 1.53 & -1.7(0.02) & 5.48(0.2) & 286.19($^{+53.91}_{-56.46}$)\\
 &  &  &  &  & HC$^{15}$N & 1.75(0.06) & -9 5 & 2.92 & 0.43 & -1.81(0.05) &  & \\
G078.12+03.63 & 20:14:26.07 & 41:13:32.7 & 1.6 & 8.1 & H$^{13}$CN & 10.73(0.15) & -13 7 & 5.69 & 1.49 & -3.47(0.04) & 3.44(0.14) & 178.26($^{+33.69}_{-35.43}$)\\
IRAS 20126+4104 &  &  &  &  & HC$^{15}$N & 3.12(0.12) & -13 5 & 4.87 & 0.45 & -3.71(0.11) &  & \\
G081.75+00.59 & 20:39:01.99 & 42:24:59.3 & 1.5 & 8.2 & H$^{13}$CN & 6.67(0.08) & -12 4 & 3.55 & 1.23 & -3.77(0.03) & 5.04(0.22) & 263.4($^{+50.0}_{-52.32}$)\\
DR 21 &  &  &  &  & HC$^{15}$N & 1.32(0.06) & -8 0 & 3.37 & 0.54 & -4.06(0.04) &  & \\
G081.87+00.78 & 20:38:36.43 & 42:37:34.8 & 1.3 & 8.2 & H$^{13}$CN & 36.71(0.22) & -2 22 & 7.81 & 4.78 & 9.61(0.02) & 3.59(0.05) & 187.62($^{+34.76}_{-36.46}$)\\
W 75N &  &  &  &  & HC$^{15}$N & 10.22(0.13) &  0 18 & 5.39 & 1.76 & 9.4(0.03) &  & \\
G092.67+03.07 & 21:09:21.73 & 52:22:37.1 & 1.6 & 8.5 & H$^{13}$CN & 8.59(0.08) & -15 2 & 3.5 & 1.31 & -6.25(0.03) & 3.58(0.1) & 191.64($^{+35.91}_{-37.13}$)\\
 &  &  &  &  & HC$^{15}$N & 2.4(0.06) & -11 0 & 3.22 & 0.61 & -6.23(0.04) &  & \\
G109.87+02.11 & 22:56:18.10 & 62:01:49.5 & 0.7 & 8.6 & H$^{13}$CN & 15.44(0.15) & -21 3 & 5.21 & 1.85 & -10.06(0.03) & 3.02(0.07) & 162.8($^{+30.45}_{-31.35}$)\\
Cep A &  &  &  &  & HC$^{15}$N & 5.11(0.11) & -20 -1 & 4.36 & 0.81 & -10.09(0.06) &  & \\
G111.54+00.77 & 23:13:45.36 & 61:28:10.6 & 2.6 & 9.6 & H$^{13}$CN & 15.84(0.05) & -70 -48 & 1.72 & 2.49 & -57.5(0.02) & 4.44(0.05) & 257.51($^{+48.3}_{-47.73}$)\\
NGC 7538 &  &  &  &  & HC$^{15}$N & 3.57(0.04) & -65 -51 & 1.63 & 0.77 & -57.8(0.04) &  & \\
G121.29+00.65 & 00:36:47.35 & 63:29:02.2 & 0.9 & 8.8 & H$^{13}$CN & 5.76(0.04) & -27 -10 & 1.52 & 1.0 & -17.33(0.01) & 5.52(0.17) & 302.1($^{+56.9}_{-58.08}$)\\
L 1287 &  &  &  &  & HC$^{15}$N & 1.04(0.03) & -25 -12 & 1.48 & 0.31 & -17.6(0.03) &  & \\
G123.06-06.30 & 00:52:24.70 & 56:33:50.5 & 2.8 & 10.1 & H$^{13}$CN & 7.21(0.05) & -40 -20 & 2.04 & 1.16 & -30.51(0.02) & 4.08(0.11) & 244.74($^{+46.47}_{-45.08}$)\\
NGC 281 &  &  &  &  & HC$^{15}$N & 1.77(0.04) & -40 -20 & 1.7 & 0.39 & -30.57(0.04) &  & \\
G133.94+01.06 & 02:27:03.82 & 61:52:25.2 & 2.0 & 9.8 & H$^{13}$CN & 24.06(0.11) & -60 -37 & 4.01 & 2.78 & -47.0(0.03) & 2.96(0.03) & 174.03($^{+32.7}_{-32.07}$)\\
W 3OH &  &  &  &  & HC$^{15}$N & 8.13(0.08) & -55 -40 & 3.52 & 1.31 & -47.02(0.03) &  & \\
G176.51+00.20 & 05:37:52.14 & 32:00:03.9 & 1.0 & 9.3 & H$^{13}$CN & 1.88(0.06) & -23 -12 & 3.08 & 0.4 & -18.01(0.07) & 3.39(0.32) & 192.41($^{+40.17}_{-40.17}$)\\
 &  &  &  &  & HC$^{15}$N & 0.55(0.05) & -23 -14 & 2.78 & 0.17 & -18.25(0.11) &  & \\
G183.72-03.66 & 05:40:24.23 & 23:50:54.7 & 1.8 & 10.0 & H$^{13}$CN & 2.34(0.04) & -5 8 & 1.81 & 0.5 & 2.41(0.03) & 5.3(0.44) & 315.63($^{+64.74}_{-63.29}$)\\
 &  &  &  &  & HC$^{15}$N & 0.44(0.04) & -3 7 & 1.93 & 0.19 & 1.99(0.06) &  & \\
G188.94+00.88 & 06:08:53.35 & 21:38:28.7 & 2.1 & 10.4 & H$^{13}$CN & 3.99(0.07) & -3 10 & 3.36 & 0.83 & 3.58(0.03) & 5.0(0.33) & 305.95($^{+61.04}_{-58.82}$)\\
S 252 &  &  &  &  & HC$^{15}$N & 0.8(0.05) & -1 8 & 2.87 & 0.24 & 3.34(0.09) &  & \\
G192.60-00.04 & 06:12:54.02 & 17:59:23.3 & 1.6 & 9.9 & H$^{13}$CN & 8.8(0.09) & 0 15 & 3.78 & 1.63 & 7.59(0.02) & 4.58(0.14) & 271.37($^{+51.67}_{-50.51}$)\\
S 255 &  &  &  &  & HC$^{15}$N & 1.92(0.06) & 1 13 & 2.86 & 0.47 & 7.4(0.05) &  & \\
G209.00-19.38 & 05:35:15.80 & -05:23:14.1 & 0.4 & 8.6 & H$^{13}$CN & 5.89(0.05) & 2 15 & 2.38 & 1.36 & 8.97(0.02) & 4.82(0.15) & 259.64($^{+48.85}_{-50.29}$)\\
Orion Nebula &  &  &  &  & HC$^{15}$N & 1.22(0.04) & 3 14 & 1.92 & 0.47 & 8.81(0.03) &  & \\
G232.62+00.99 & 07:32:09.78 & -16:58:12.8 & 1.7 & 9.4 & H$^{13}$CN & 3.5(0.1) & 10 24 & 4.66 & 0.72 & 17.09(0.06) & 4.26(0.28) & 243.43($^{+48.24}_{-48.07}$)\\
 &  &  &  &  & HC$^{15}$N & 0.82(0.05) & 14 21 & 3.15 & 0.34 & 17.0(0.06) &  & \\
G211.59+01.05 & 06:52:45.30 & 01:40:23.1 & 4.4 & 11.9 & H$^{13}$CN & 1.55(0.03) & 39 51 & 1.62 & 0.31 & 45.62(0.05) & 4.66(0.39) & 313.81($^{+65.38}_{-60.66}$)\\
 &  &  &  &  & HC$^{15}$N & 0.33(0.03) & 41 50 & 1.55 & 0.09 & 45.76(0.12) &  & \\
\hline
\end{tabular}
\end{table*}

\begin{table*}[h!]
\caption{
Same as Table \ref{table1} but for the four sources excluded from the ratio calculations.
\label{table2}
}
\setlength{\tabcolsep}{2pt}
\tiny
\begin{tabular}{ccccccccccccc}
        \hline
        \hline
        Source & 
        R.A. & 
        Decl. & 
        $D$ & 
        $D_{\rm GC}$ & 
        Line & 
        $\int T_{\rm mb}\mathrm{d}v$ & 
        Velocity range & 
        Rms & 
        $T_{\rm peak}$ & 
        $V_{\rm LSR}$ & 
        ($^{14}$N/$^{15}$N)$\times$($^{13}$C/$^{12}$C) &
        $^{14}$N/$^{15}$N \\ 
        Alias & (hh:mm:ss) & (dd:mm:ss) & (kpc) & (kpc) & (2-1) & (K·km\,s$^{-1}$) & (km\,s$^{-1}$) & (10$^{-2}$K) & (k) & (km\,s$^{-1}$) &  &  \\
        \hline
G000.67-00.03 & 17:47:20.00 & -28:22:40.0 & 7.8 & 0.2 & H$^{13}$CN & ... & ... & ... & ... & ... & ... & ... \\
Sgr B2 &  &  &  &  & HC$^{15}$N & 4.46(0.61) & 45  75 & 19.1 & 0.38 & 57.11(0.64) &  & \\
G010.47+00.02 & 18:08:38.23 & -19:51:50.3 & 8.5 & 1.6 & H$^{13}$CN & ... & ... & ... &  ... & ... & ... & ... \\
 &  &  &  &  & HC$^{15}$N & ... & ... & ... & ... & ... & & \\
G012.90-00.24 & 18:14:34.42 & -17:51:51.9 & 2.5 & 5.9 & H$^{13}$CN & 1.74(0.43) & 30 45 & 19.1 & 0.23 & 37.37(1.0) & ... & ... \\
&  &  &  &  & HC$^{15}$N & ... & ... & ... & ...  & ... &  & \\
G168.06+00.82 & 05:17:13.74 & 39:22:19.9 & 7.7 & 15.9 & H$^{13}$CN & 0.78(0.17) & -28 -22 & 12.2 & 0.26 & -25.76(0.32) & 1.14(0.4) & 95.24($^{+38.39}_{-36.94}$)\\
IRAS 05137+3919 &  &  &  &  & HC$^{15}$N & 0.68(0.19) & -32 -22 & 10.1 & 0.19 & -24.76(0.34) &  & \\
\hline
\end{tabular}
\tablefoot{
IRAS 05137+3919, G012.90-00.24, Sgr B2, and G010.47+00.02 were excluded from the ratio calculations due to non-detections or contamination in the H$^{13}$CN J=2-1 and/or HC$^{15}$N J=2-1 lines.
}
\end{table*}

\onecolumn
\section{Spectral lines}

\begin{figure}[H]
\centering
\includegraphics[width=0.24\textwidth]{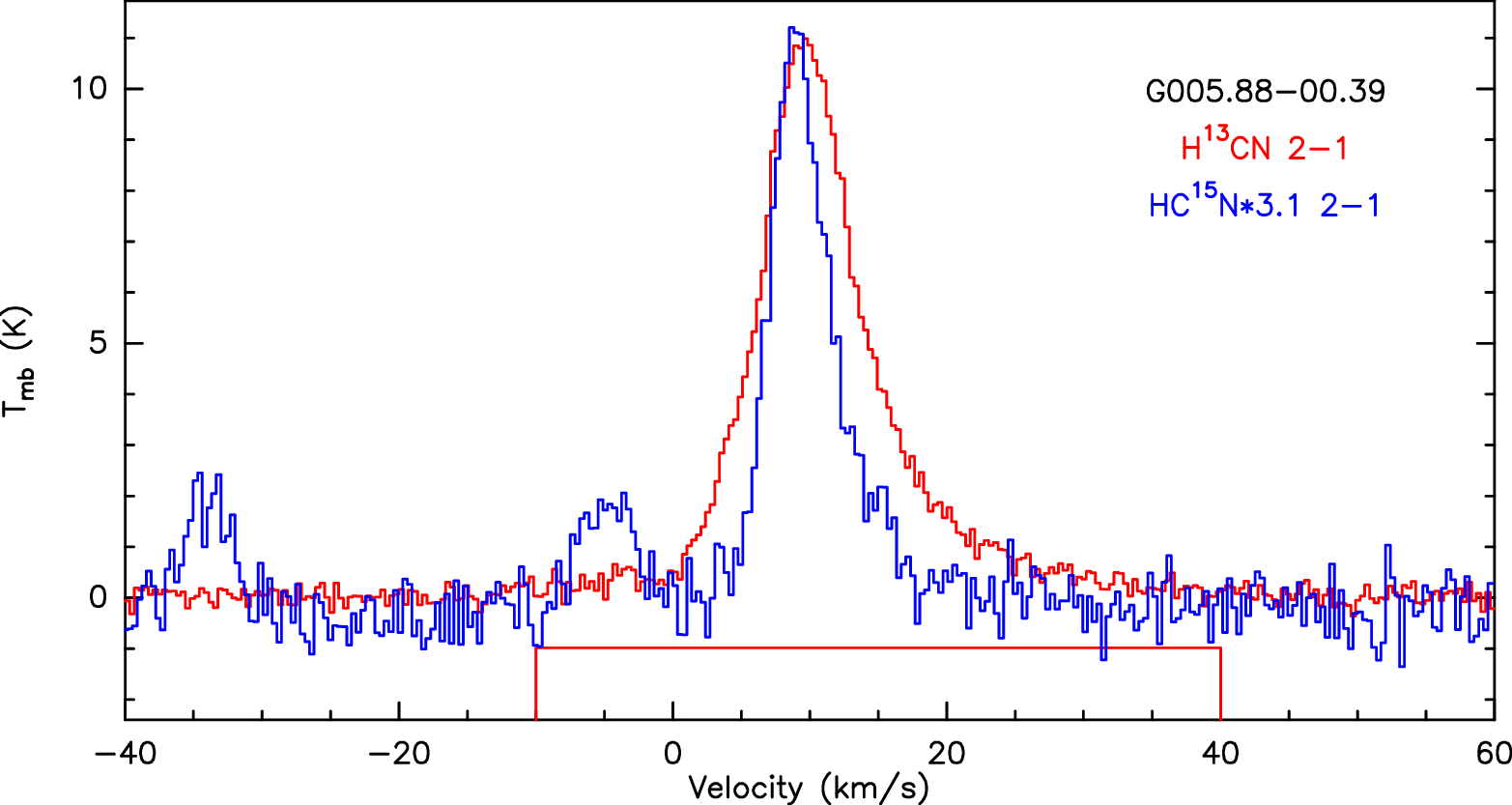}\includegraphics[width=0.24\textwidth]{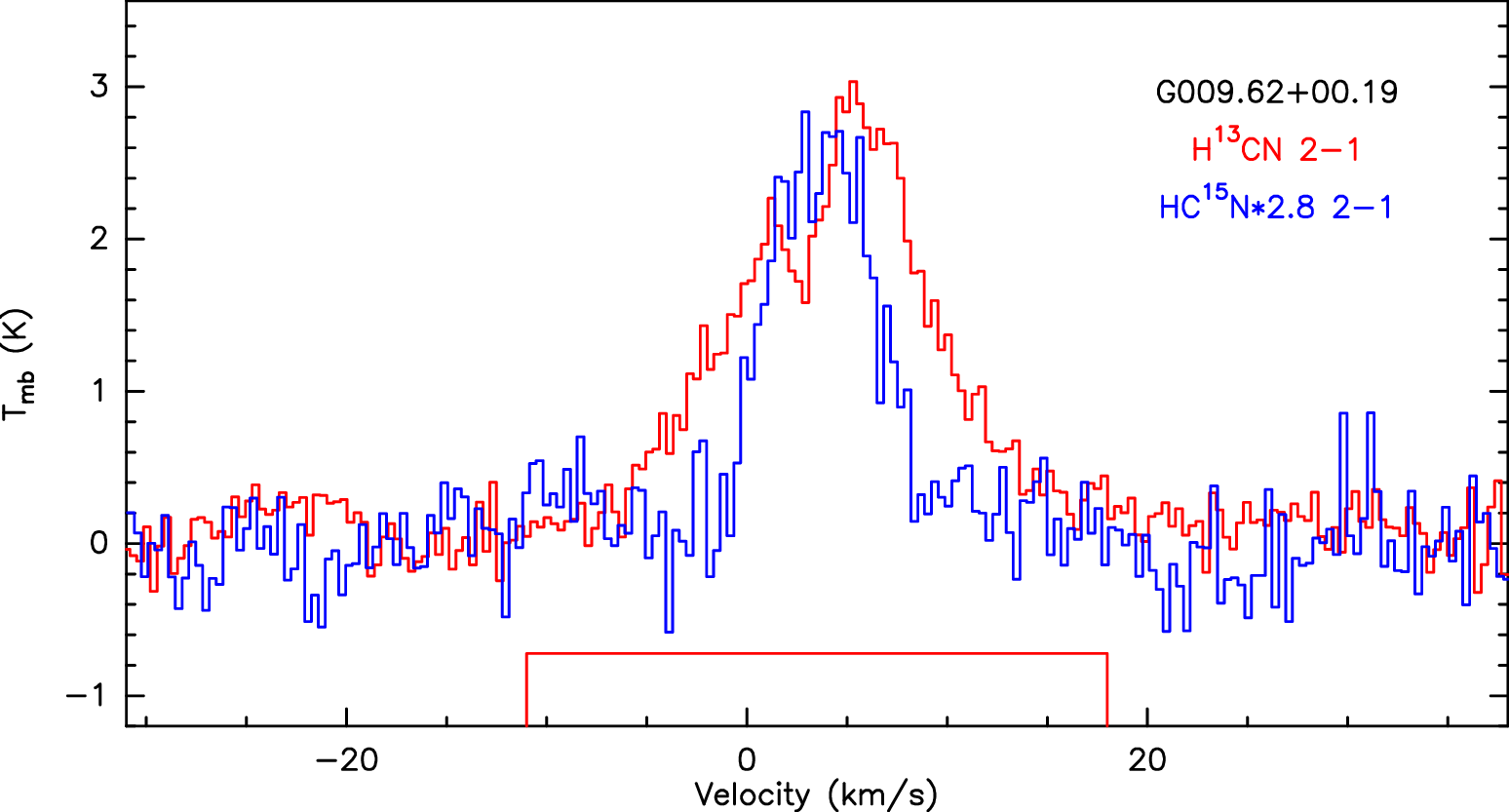}
\includegraphics[width=0.24\textwidth]{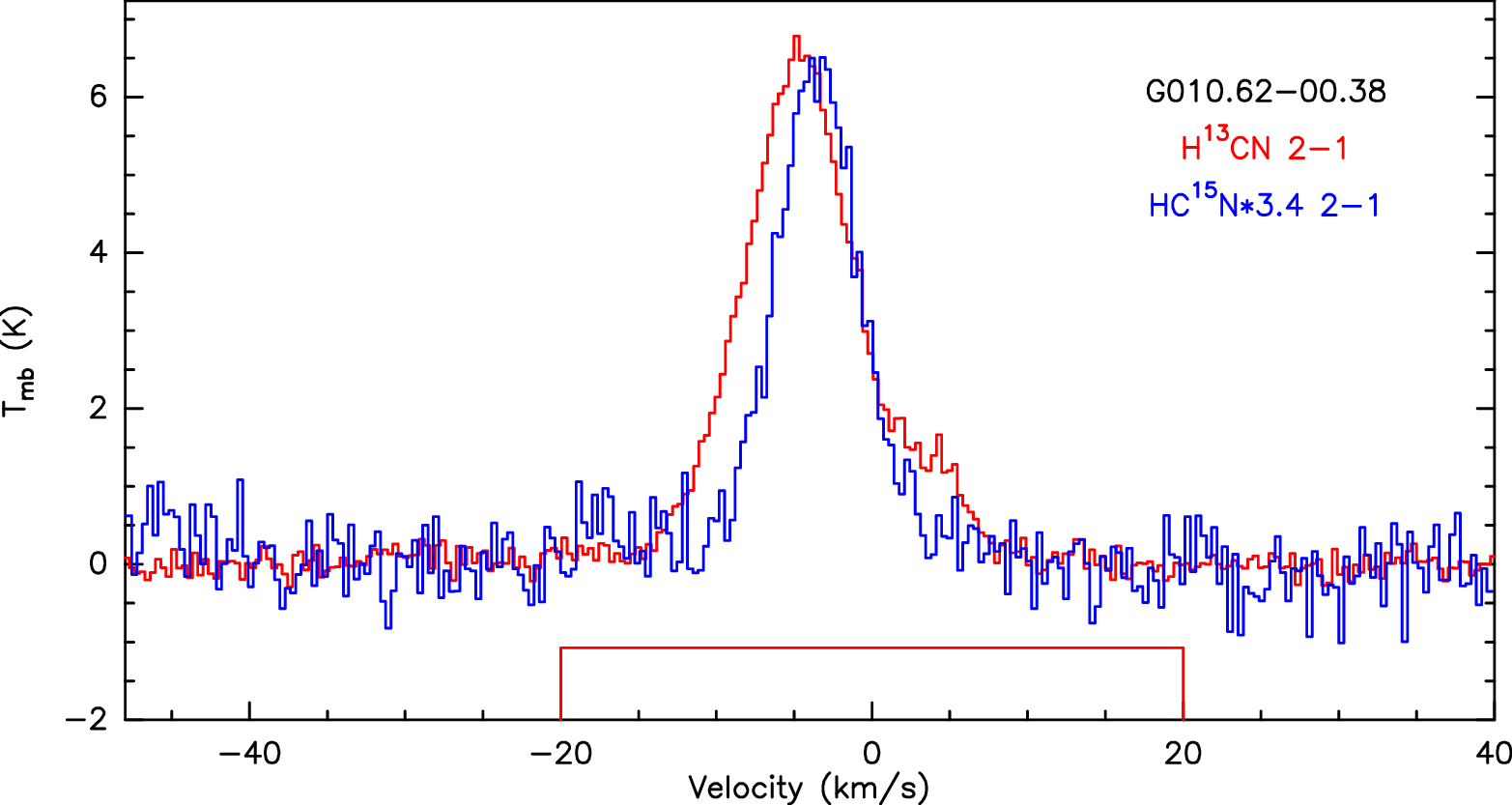}\includegraphics[width=0.24\textwidth]{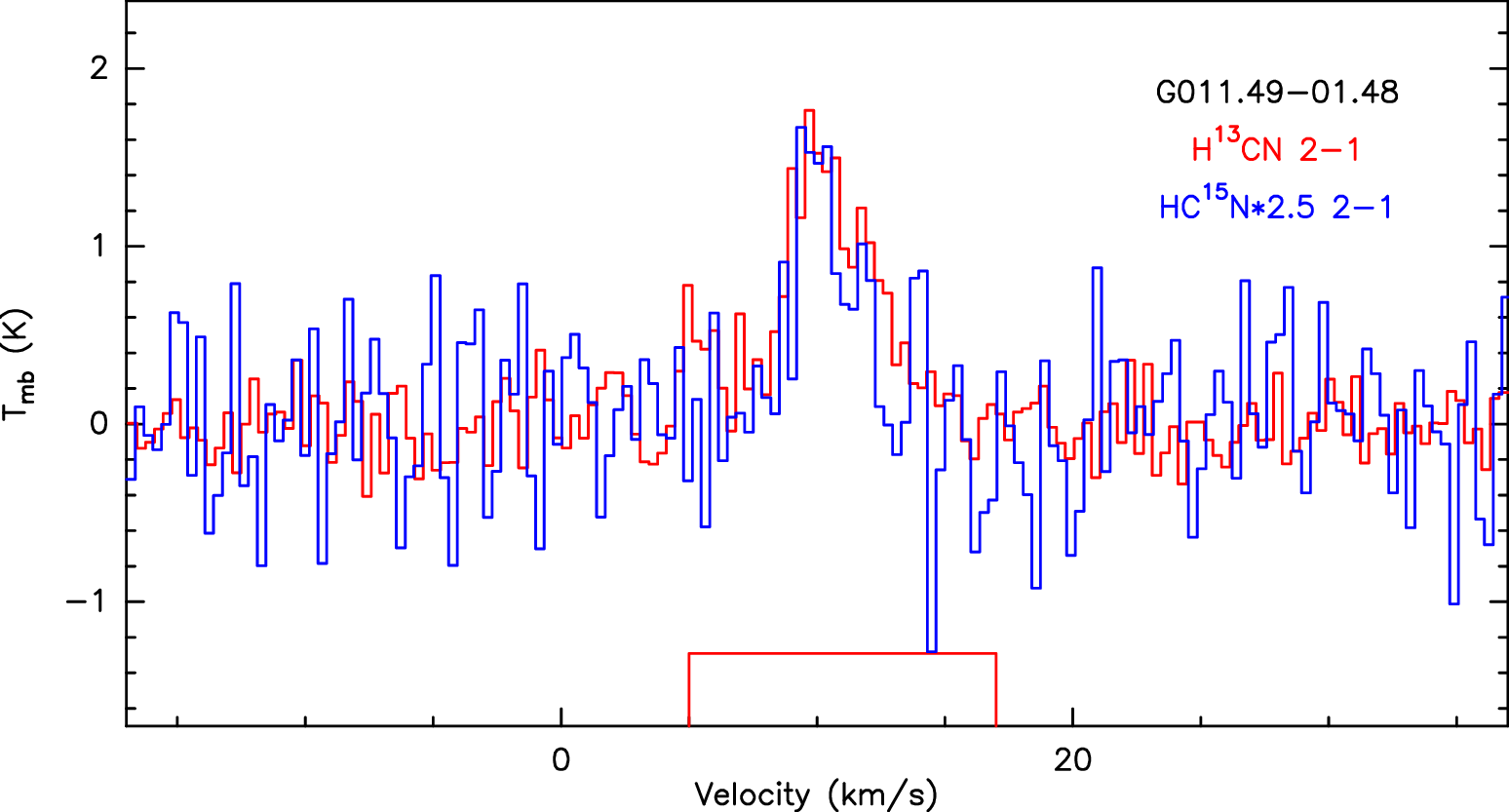}\\
\includegraphics[width=0.24\textwidth]{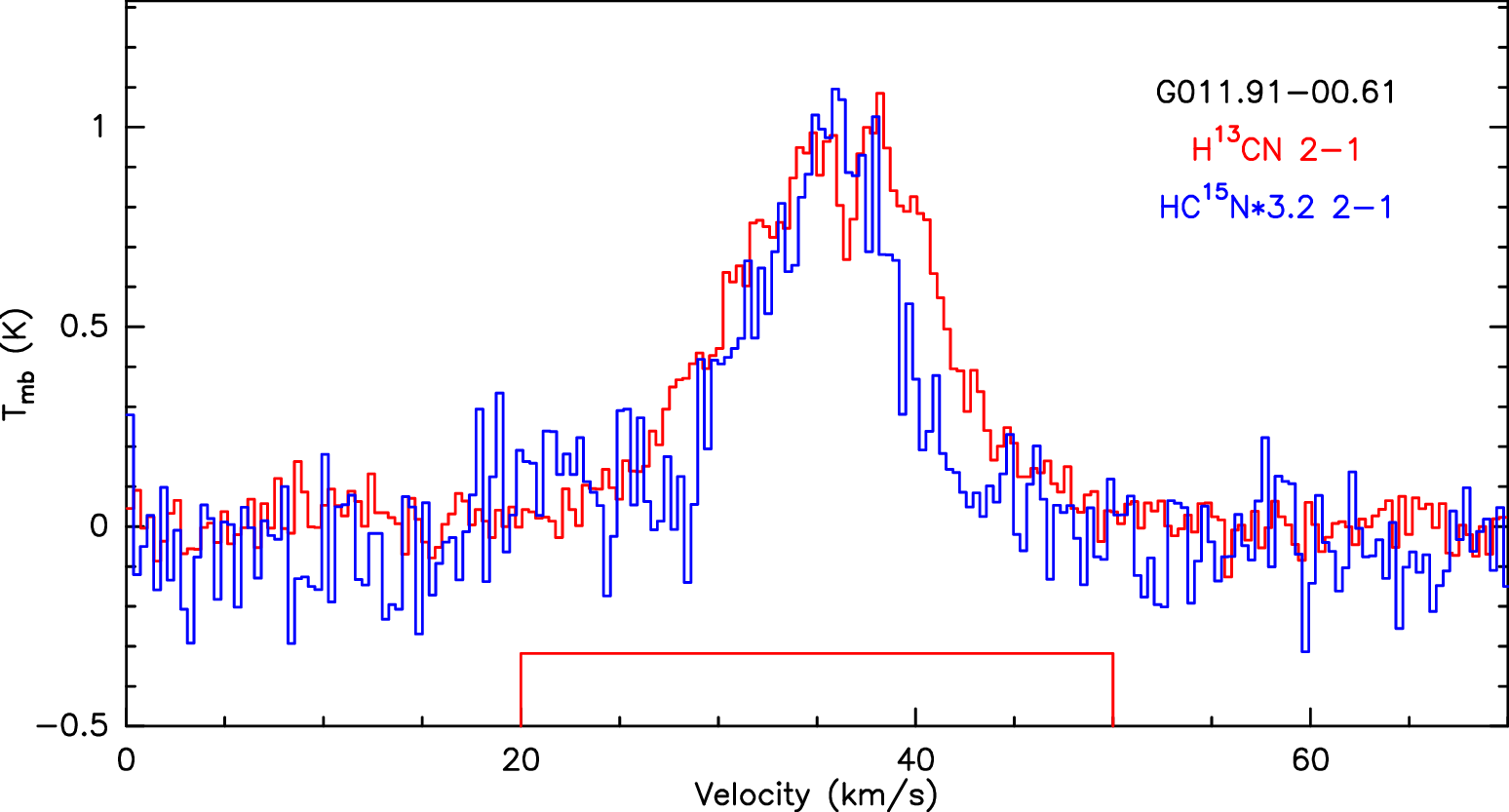}\includegraphics[width=0.24\textwidth]{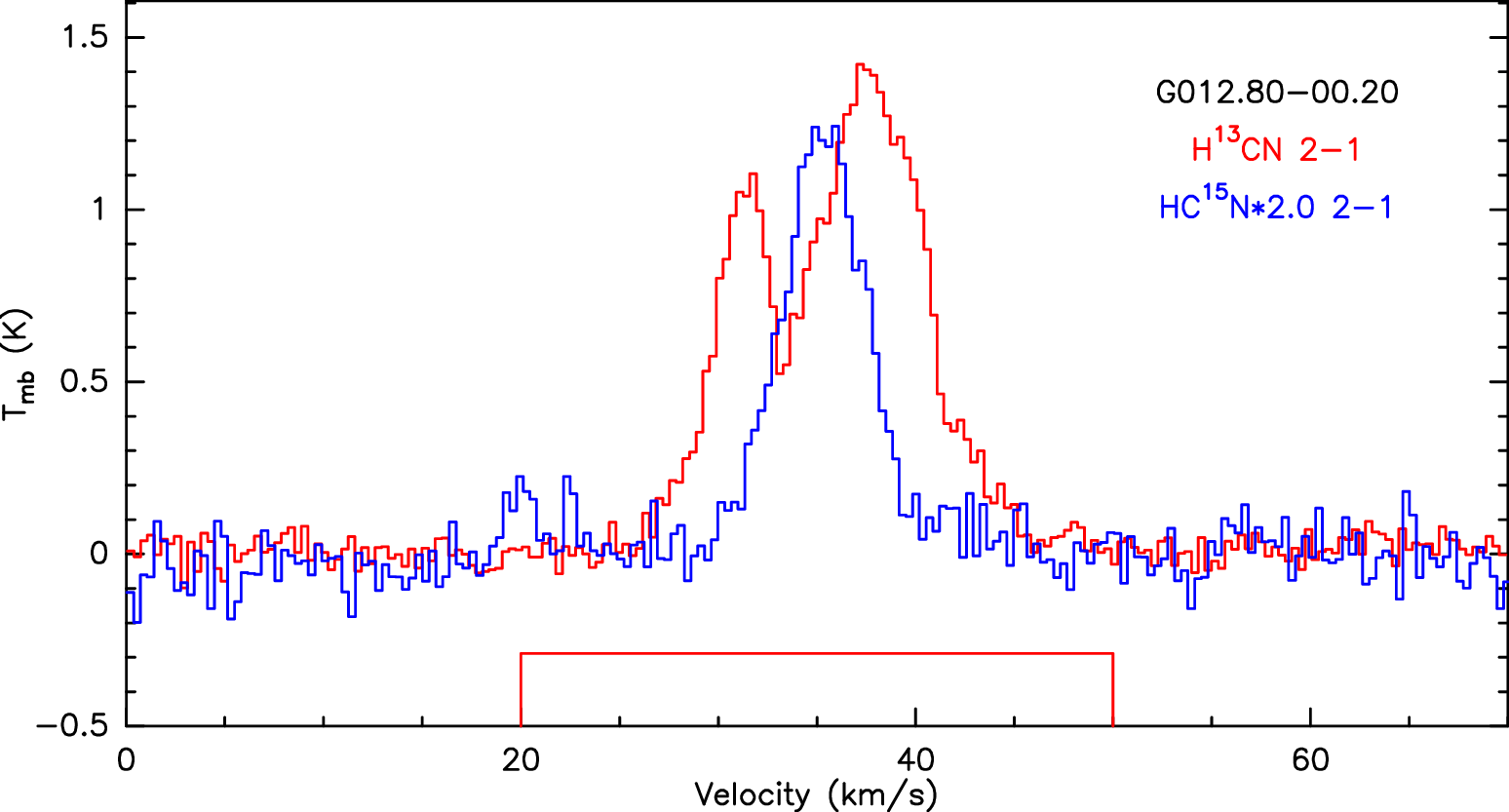}
\includegraphics[width=0.24\textwidth]{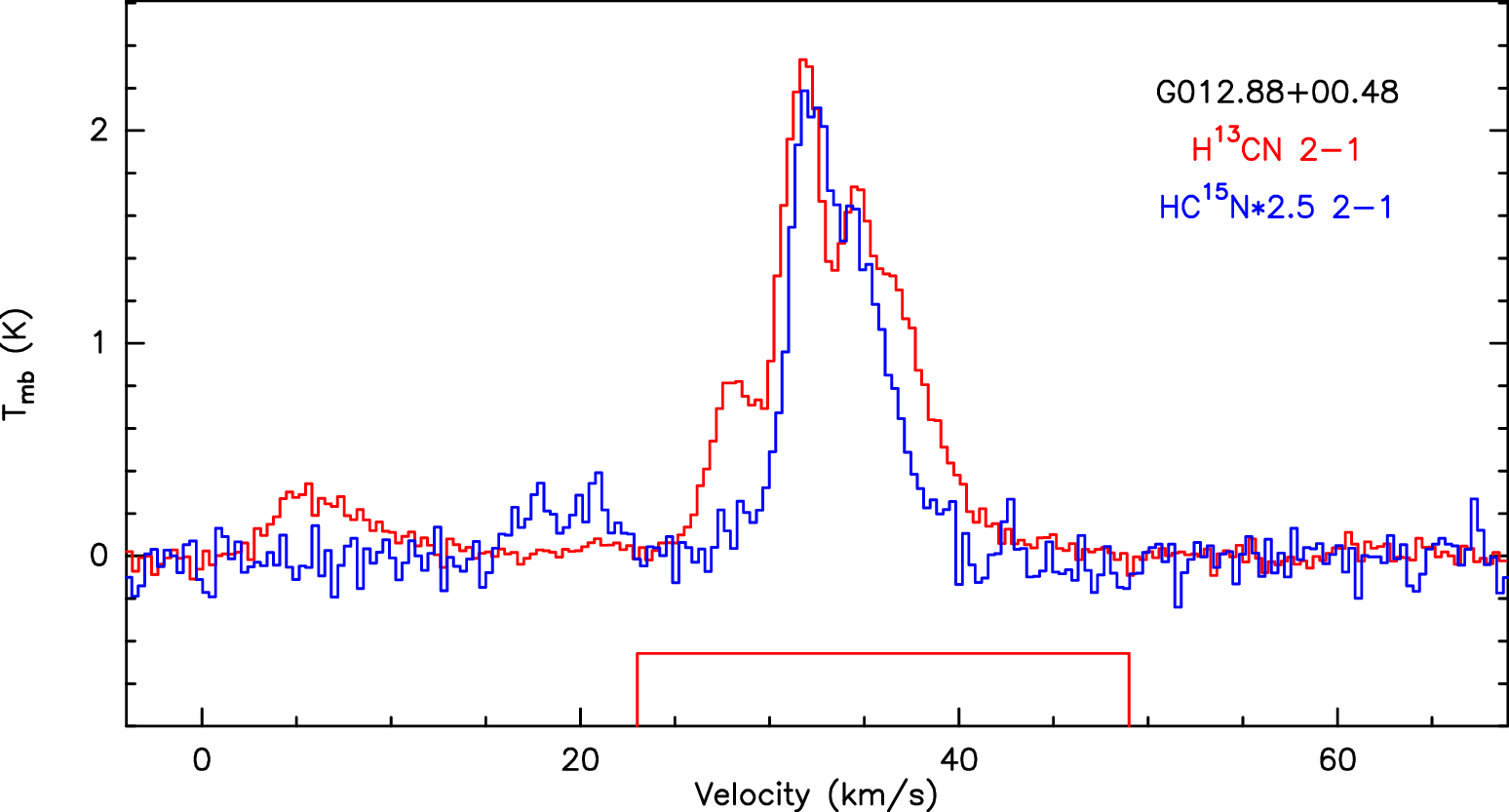}\includegraphics[width=0.24\textwidth]{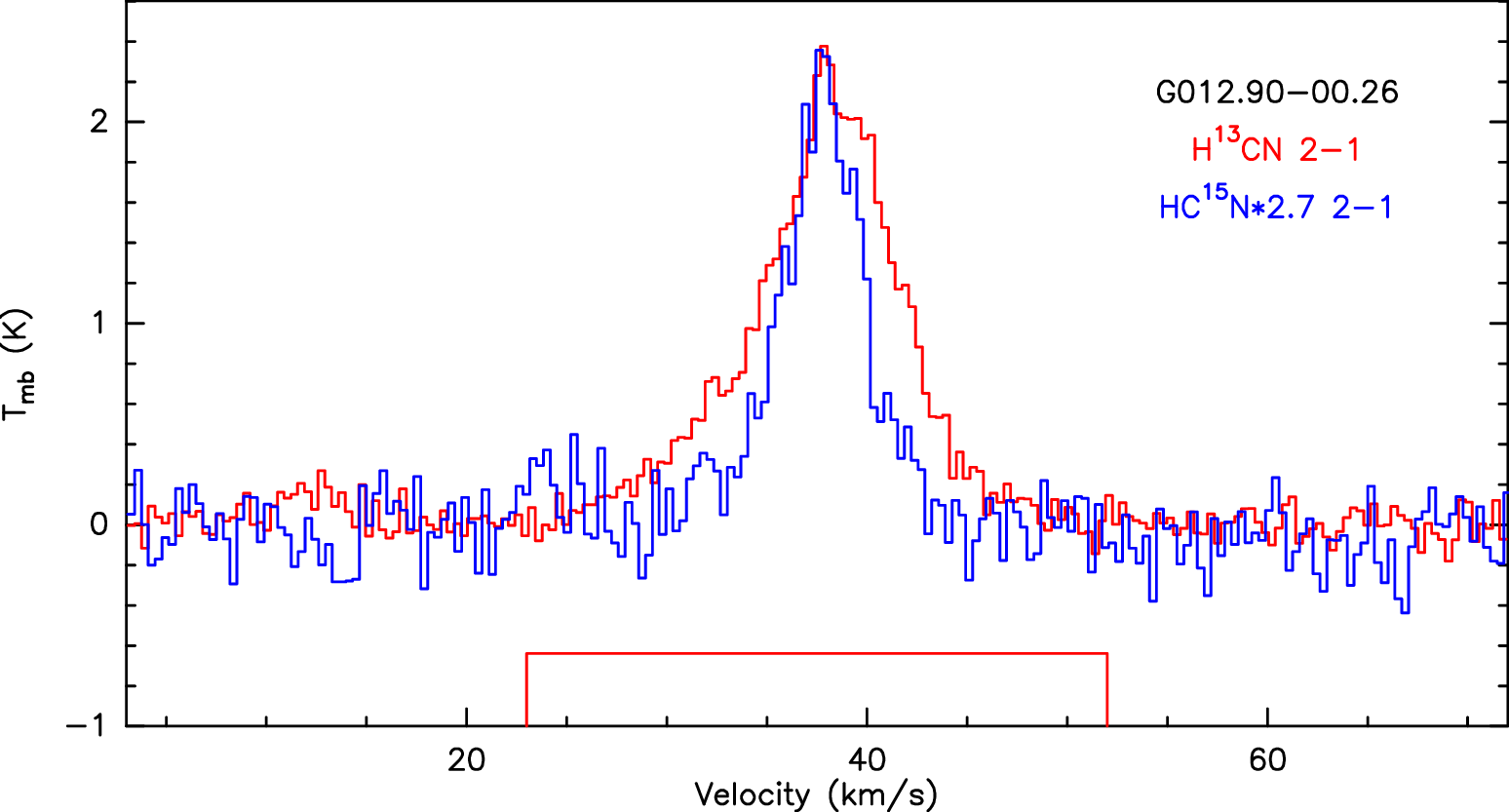}\\
\includegraphics[width=0.24\textwidth]{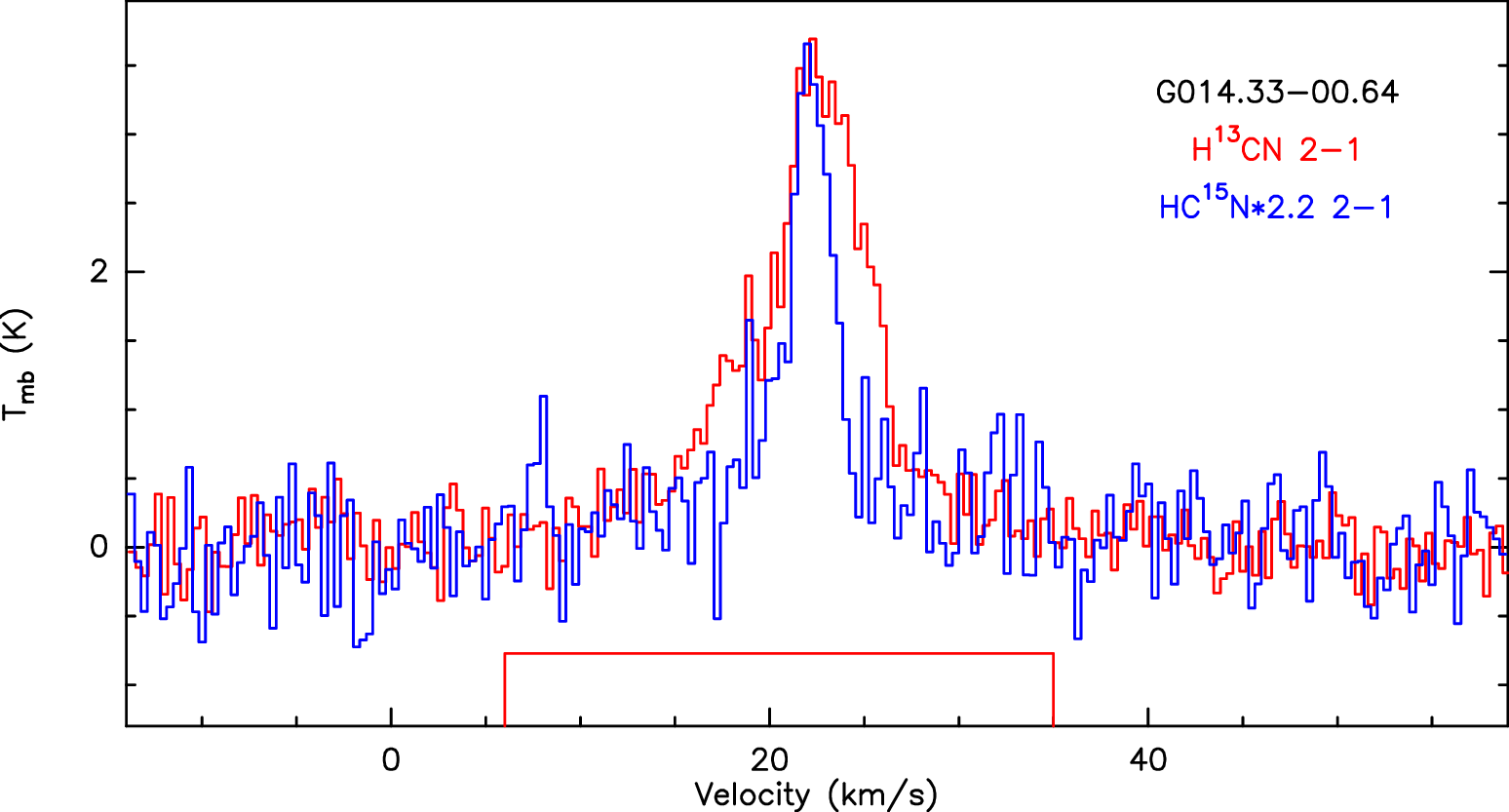}\includegraphics[width=0.24\textwidth]{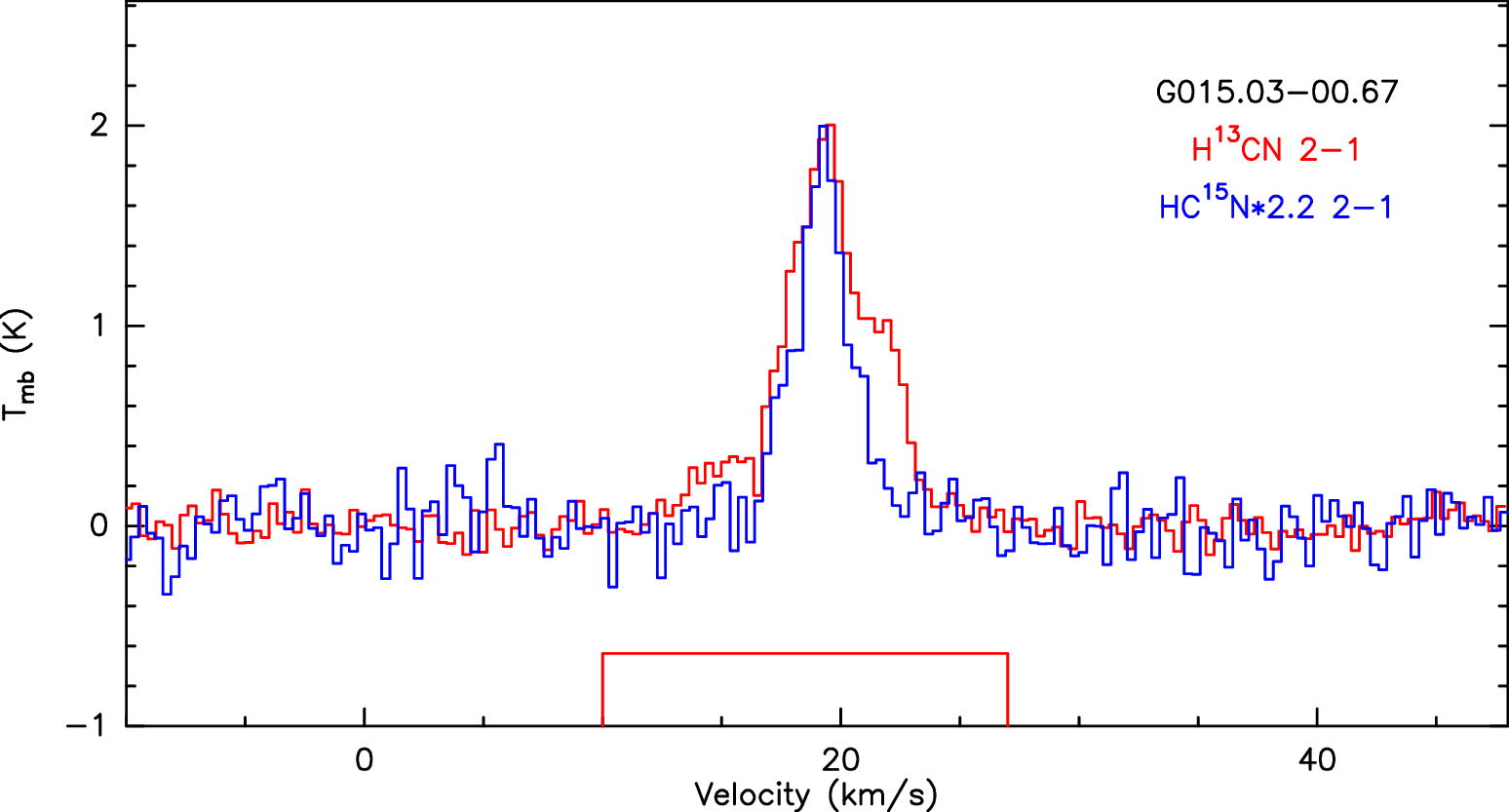}
\includegraphics[width=0.24\textwidth]{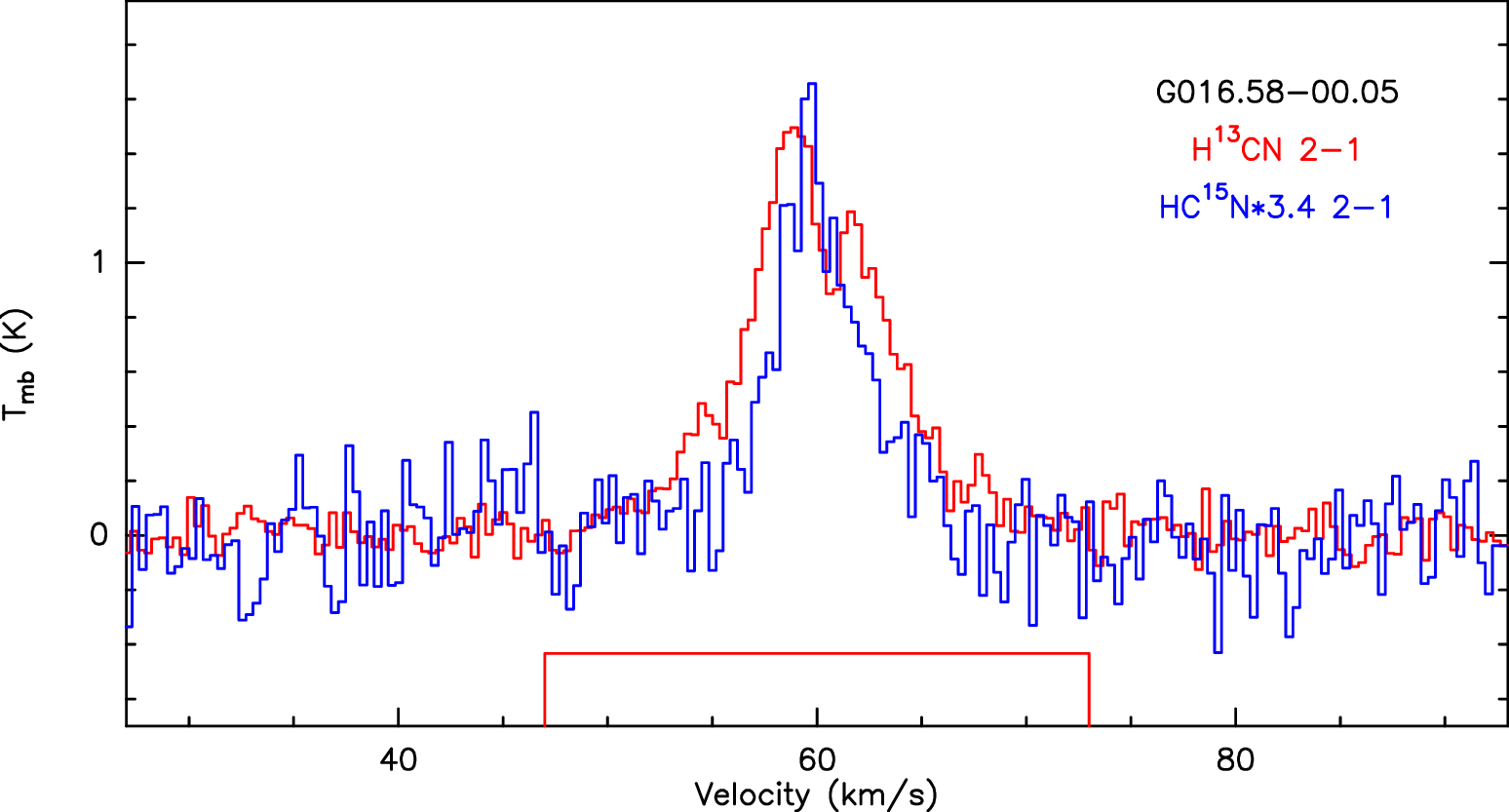}\includegraphics[width=0.24\textwidth]{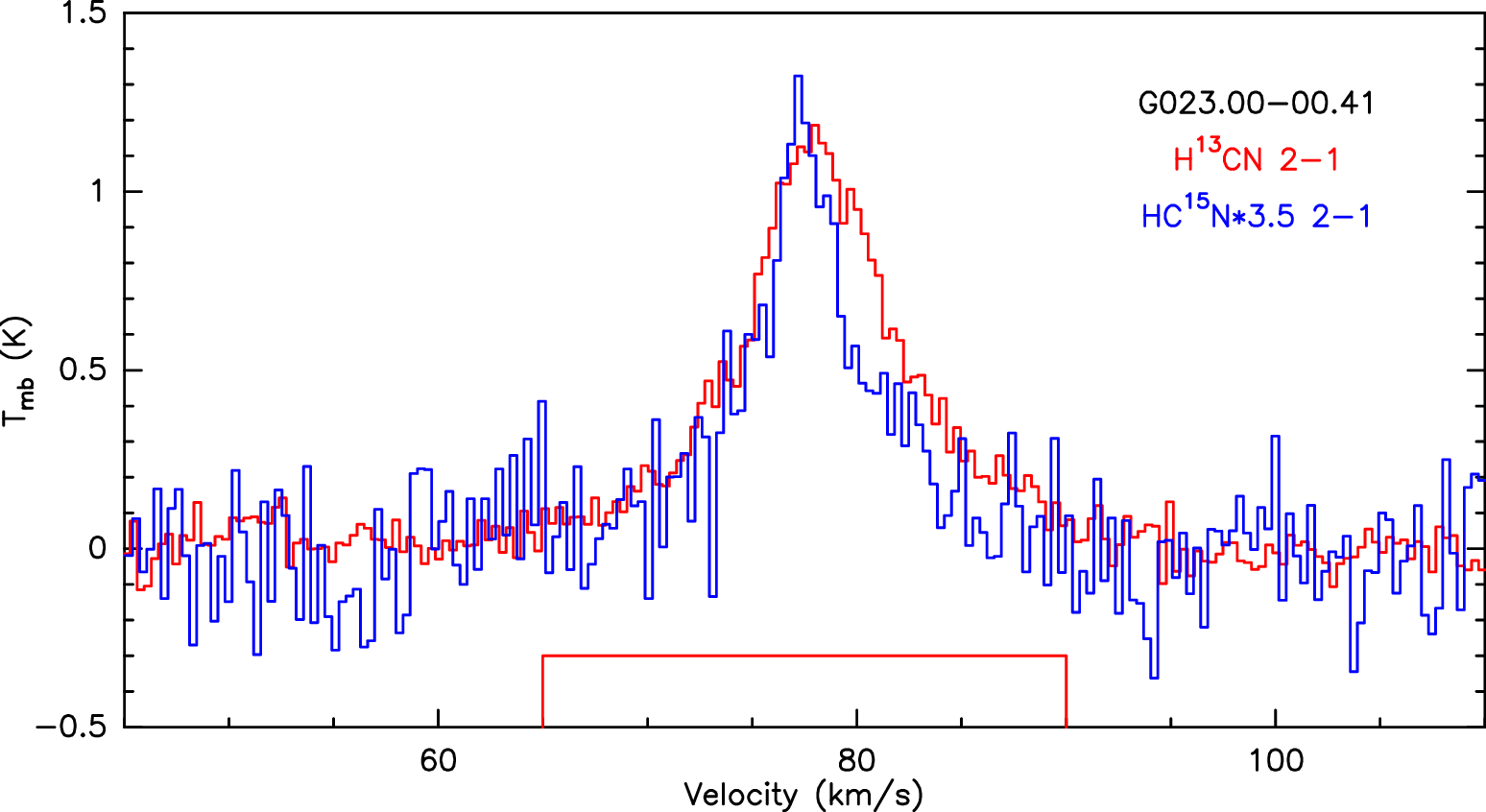}\\
\includegraphics[width=0.24\textwidth]{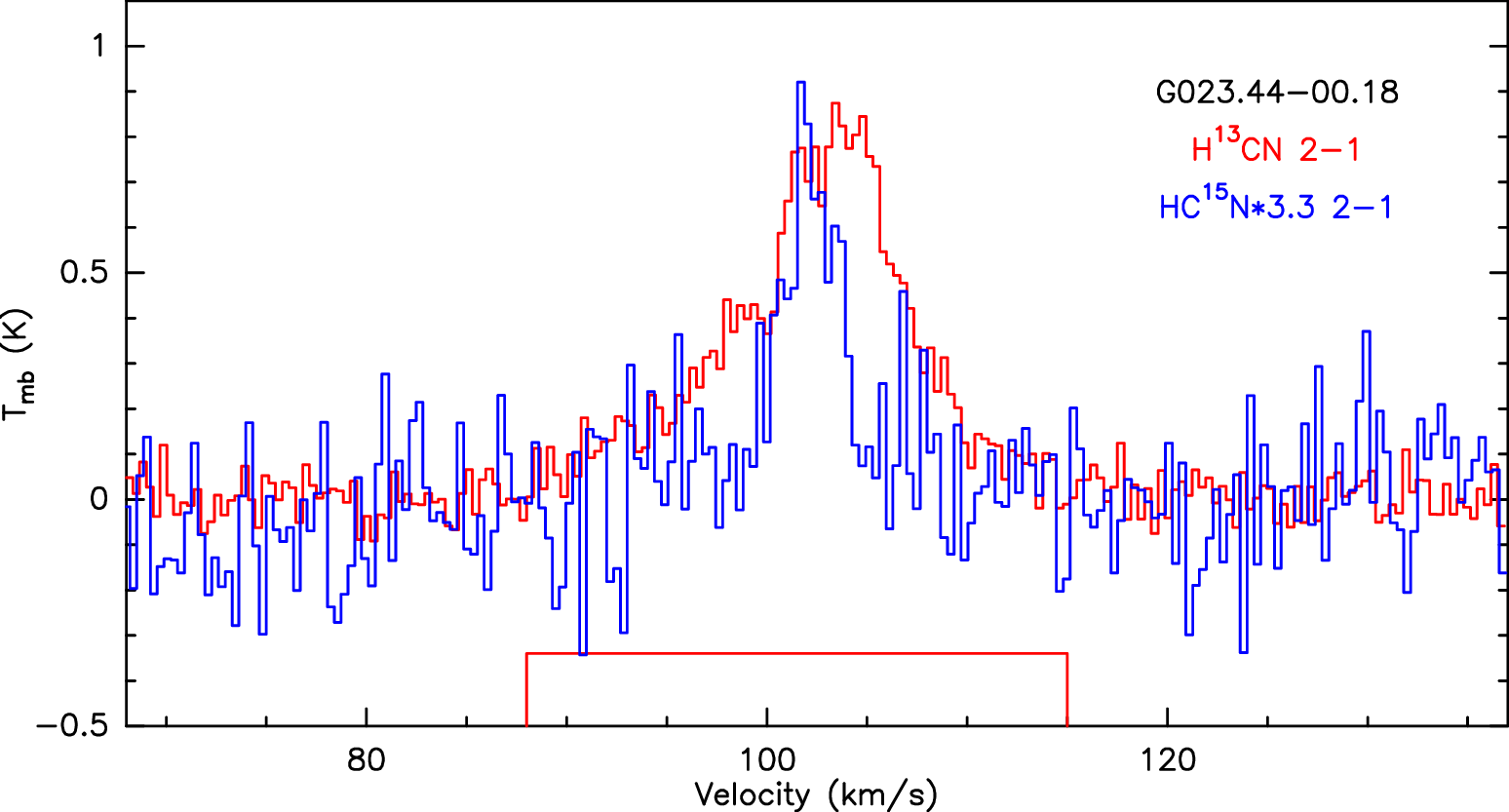}\includegraphics[width=0.24\textwidth]{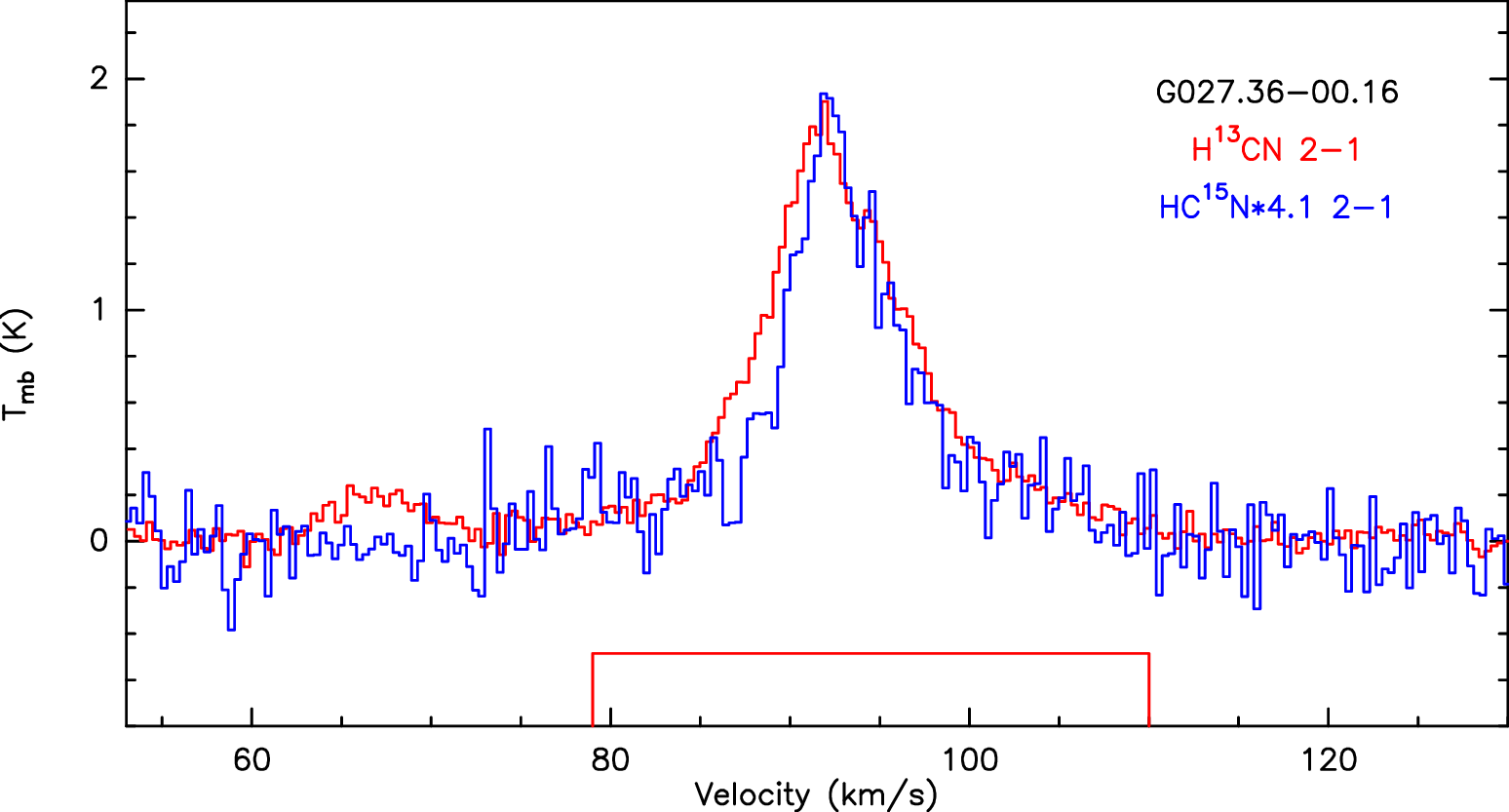}
\includegraphics[width=0.24\textwidth]{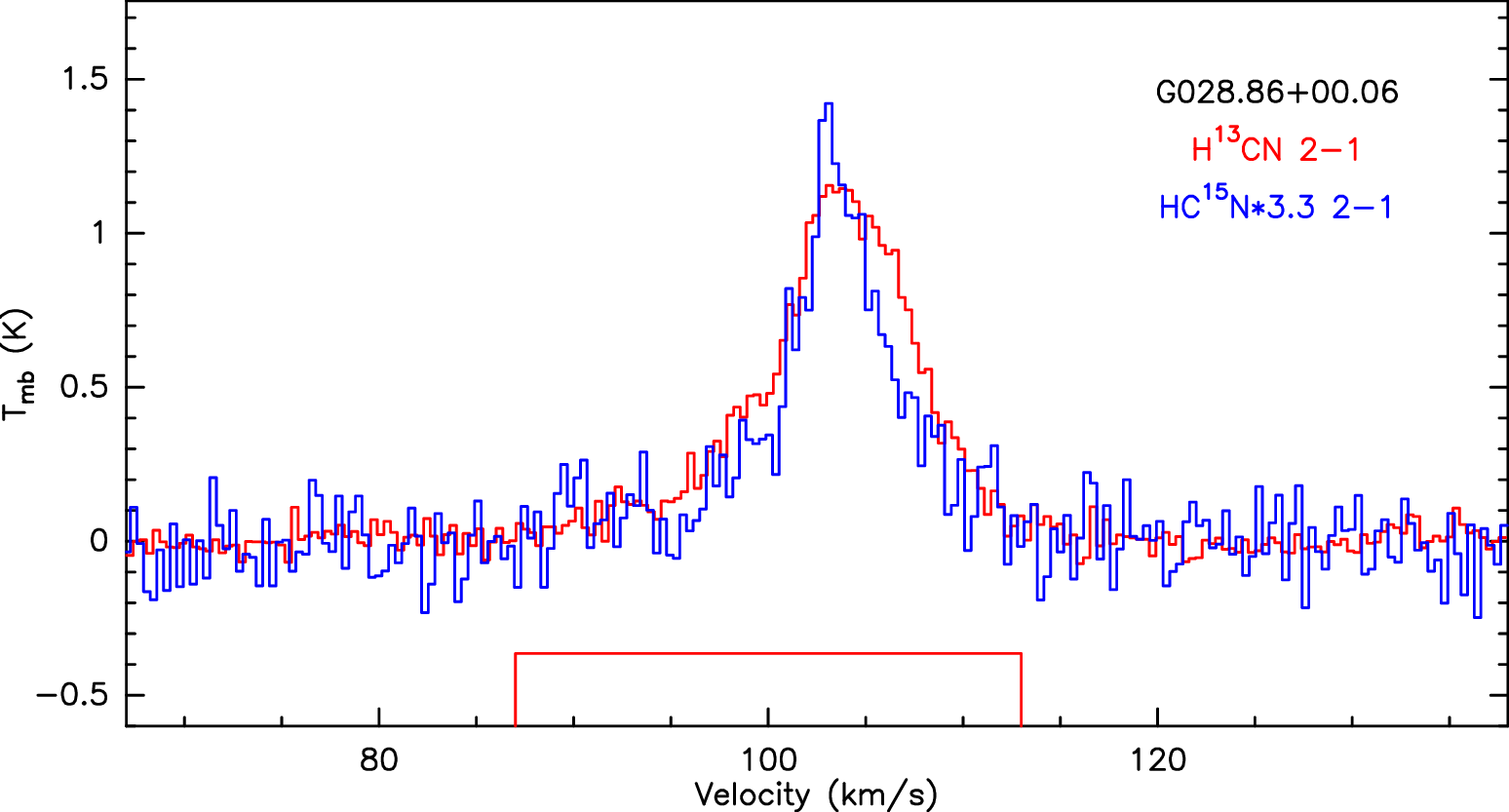}\includegraphics[width=0.24\textwidth]{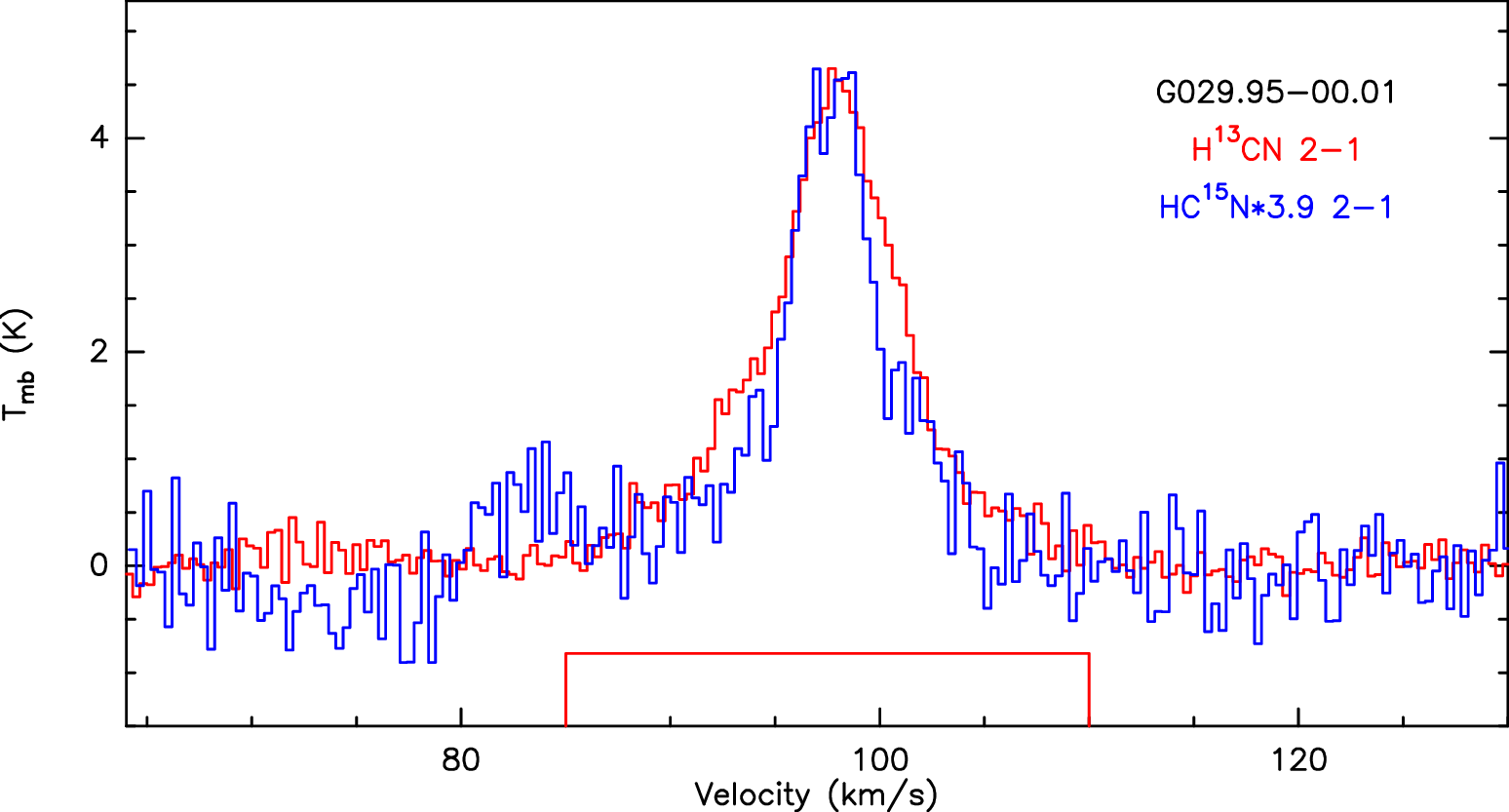}\\
\includegraphics[width=0.24\textwidth]{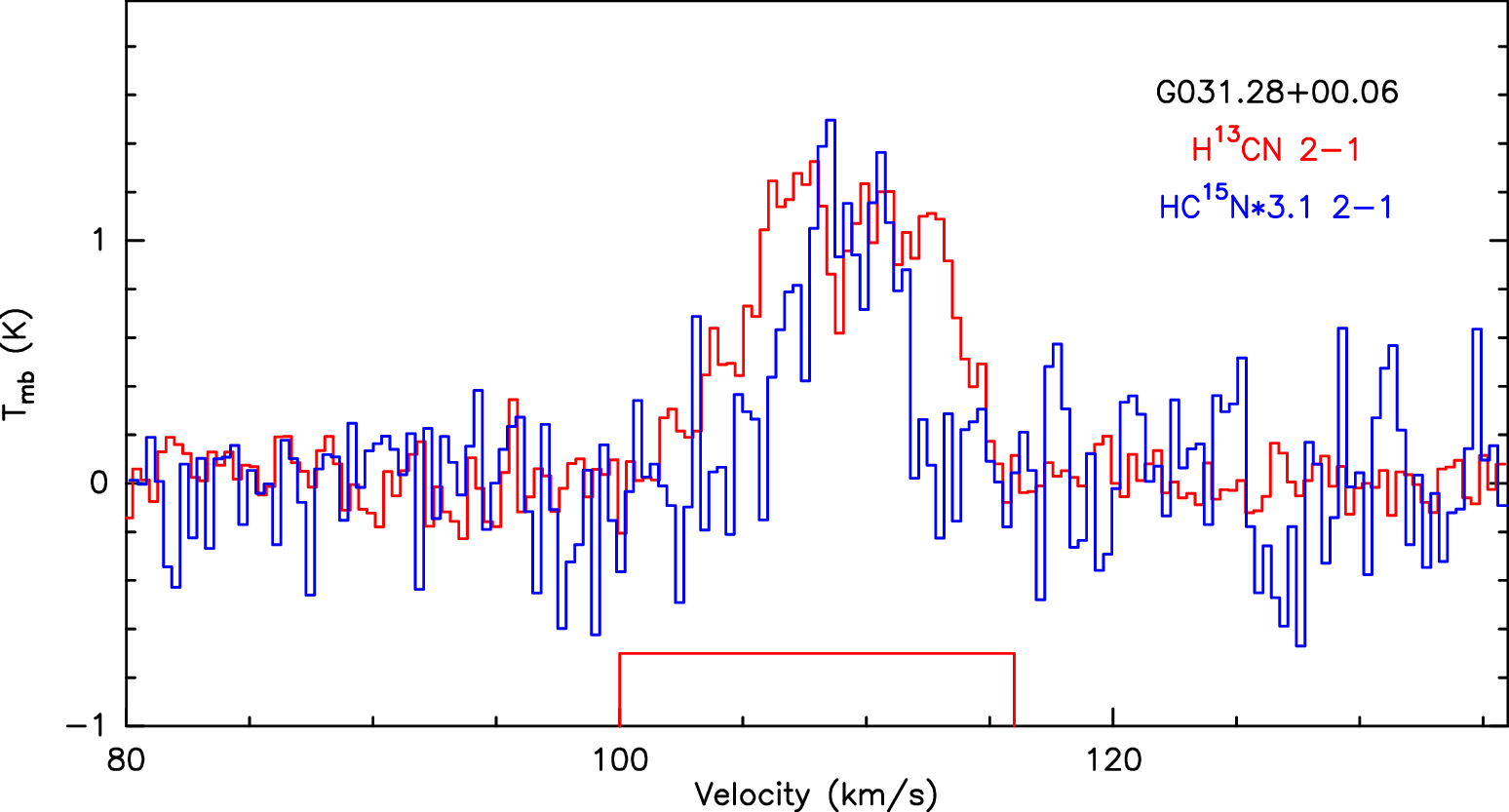}\includegraphics[width=0.24\textwidth]{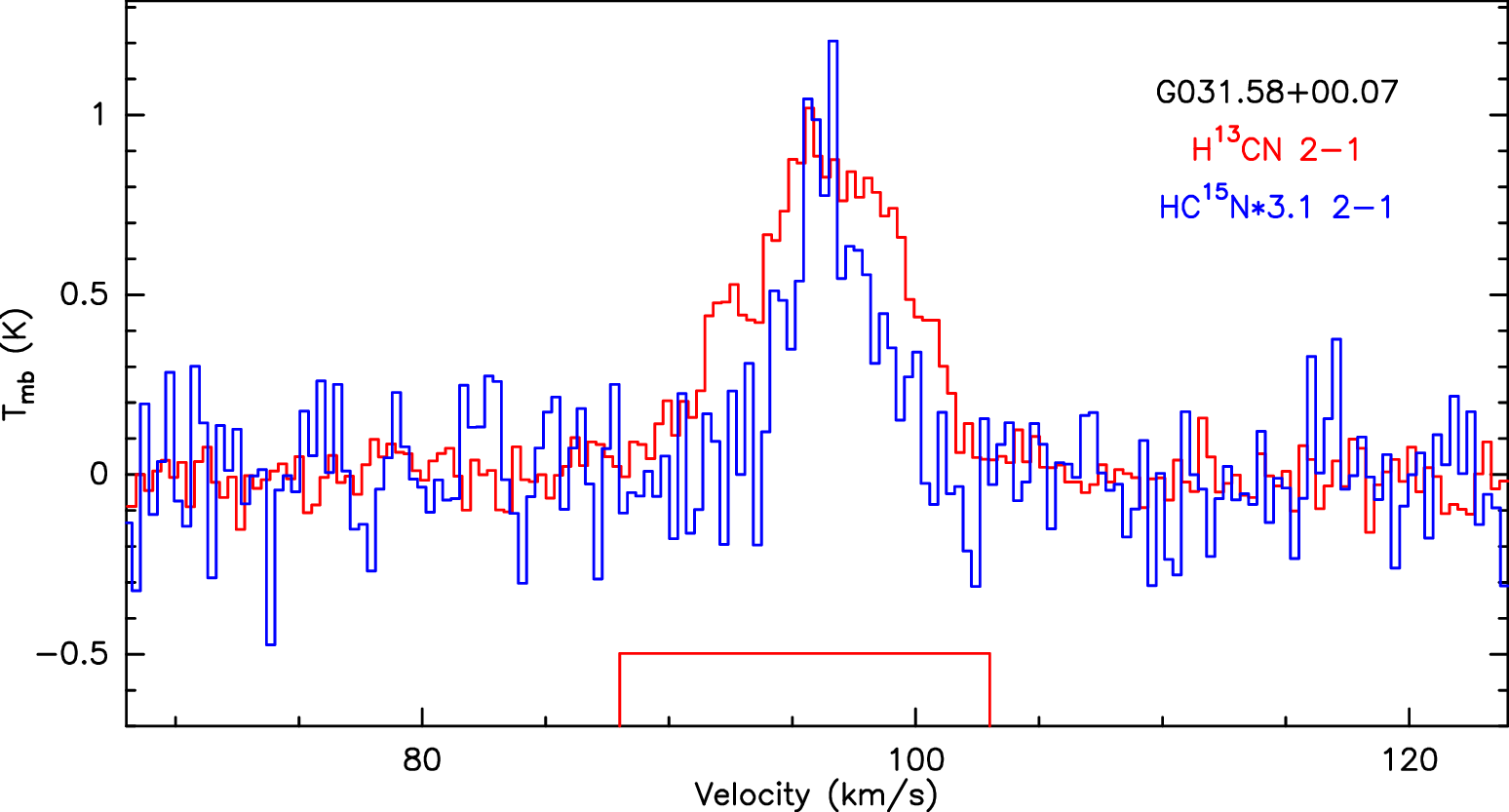}
\includegraphics[width=0.24\textwidth]{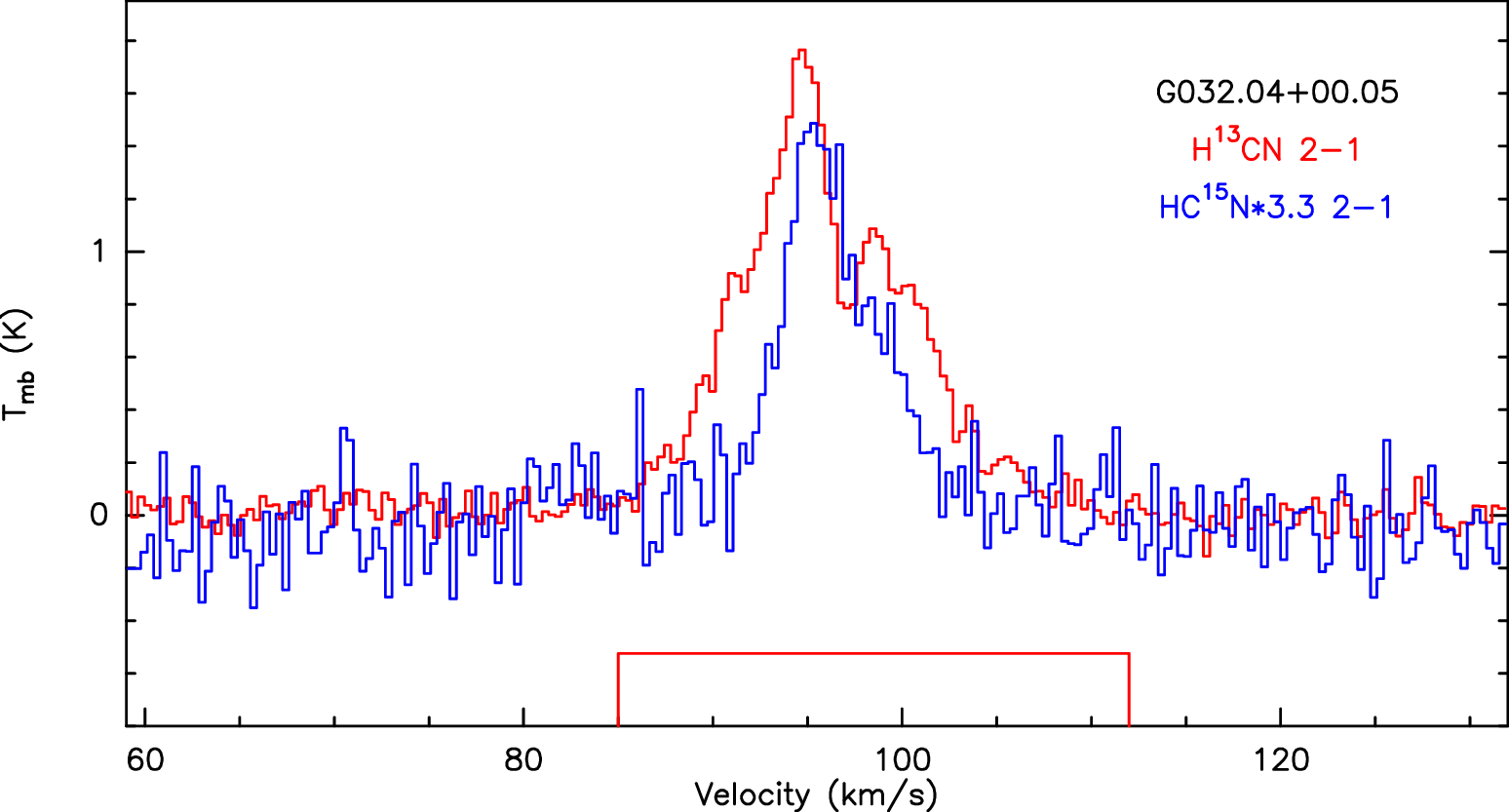}\includegraphics[width=0.24\textwidth]{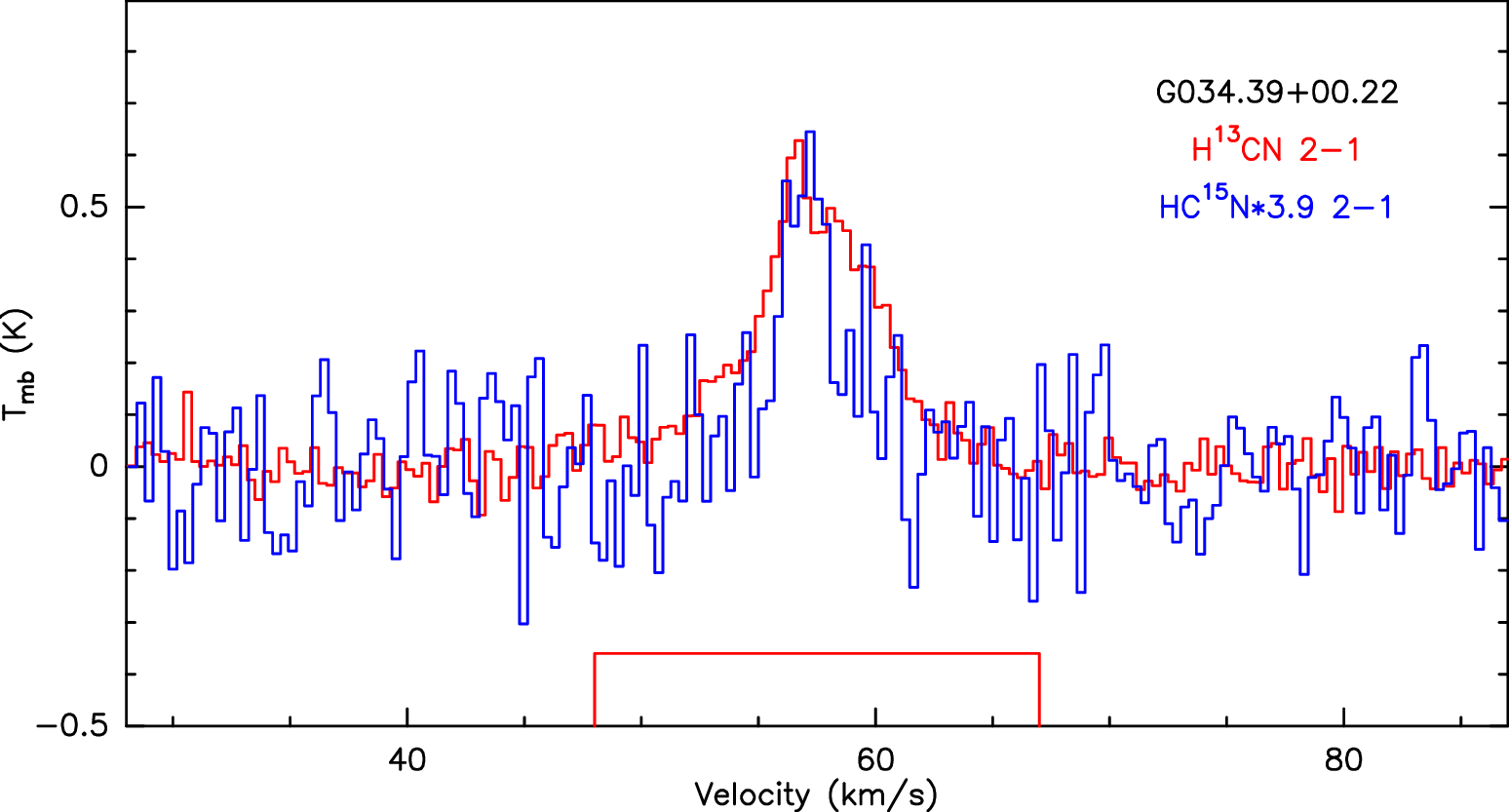}\\
\includegraphics[width=0.24\textwidth]{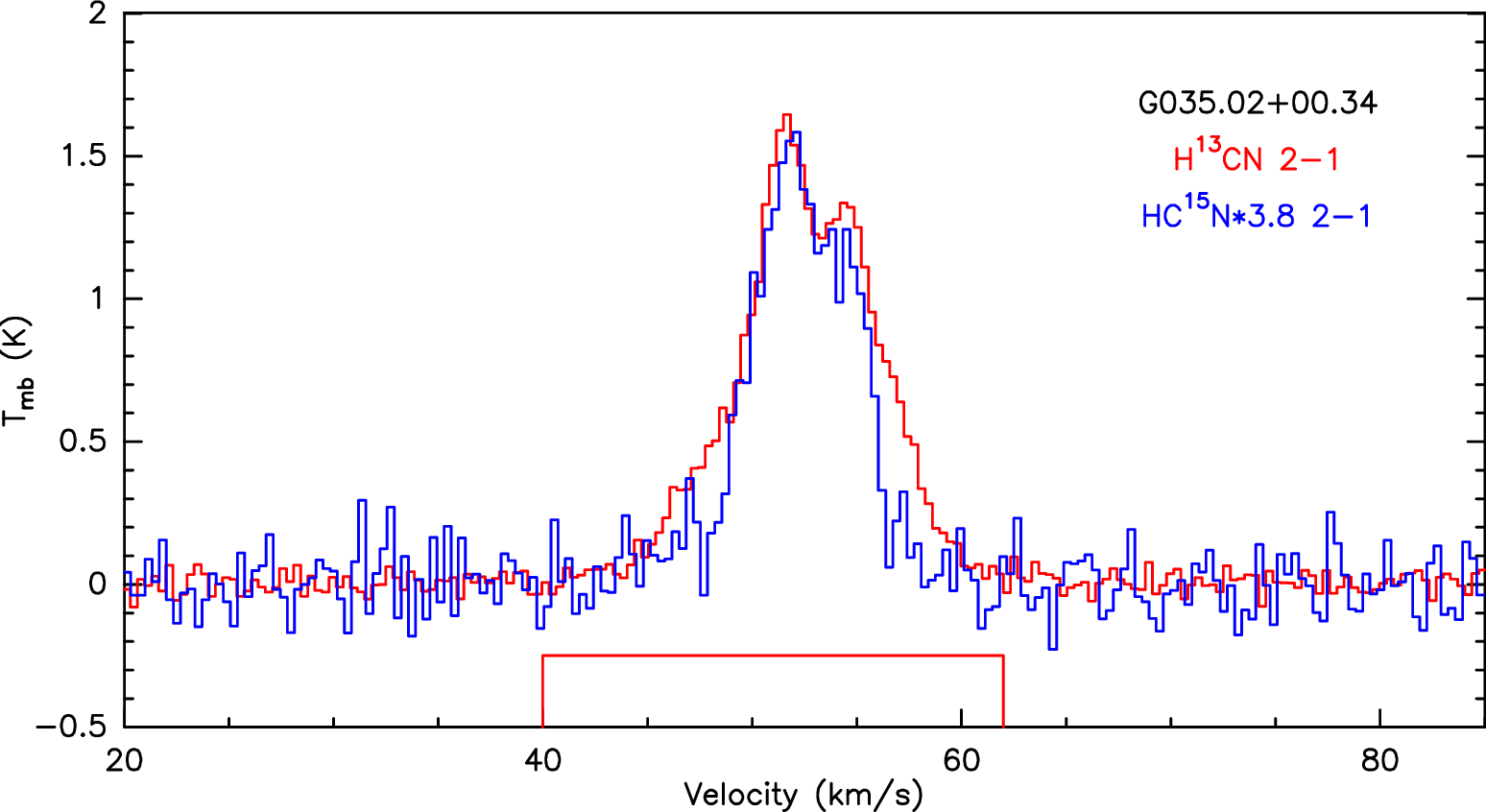}\includegraphics[width=0.24\textwidth]{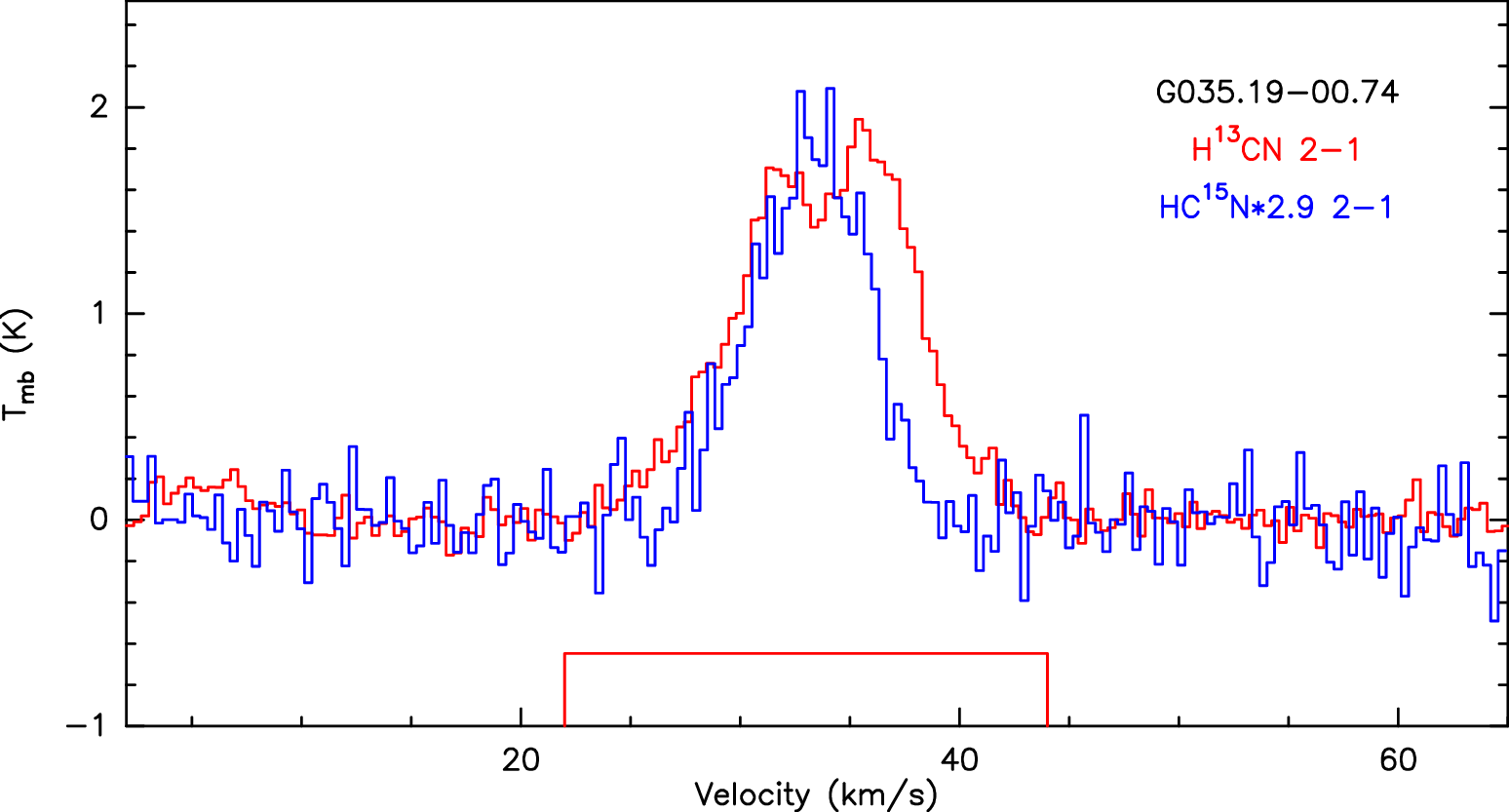}
\includegraphics[width=0.24\textwidth]{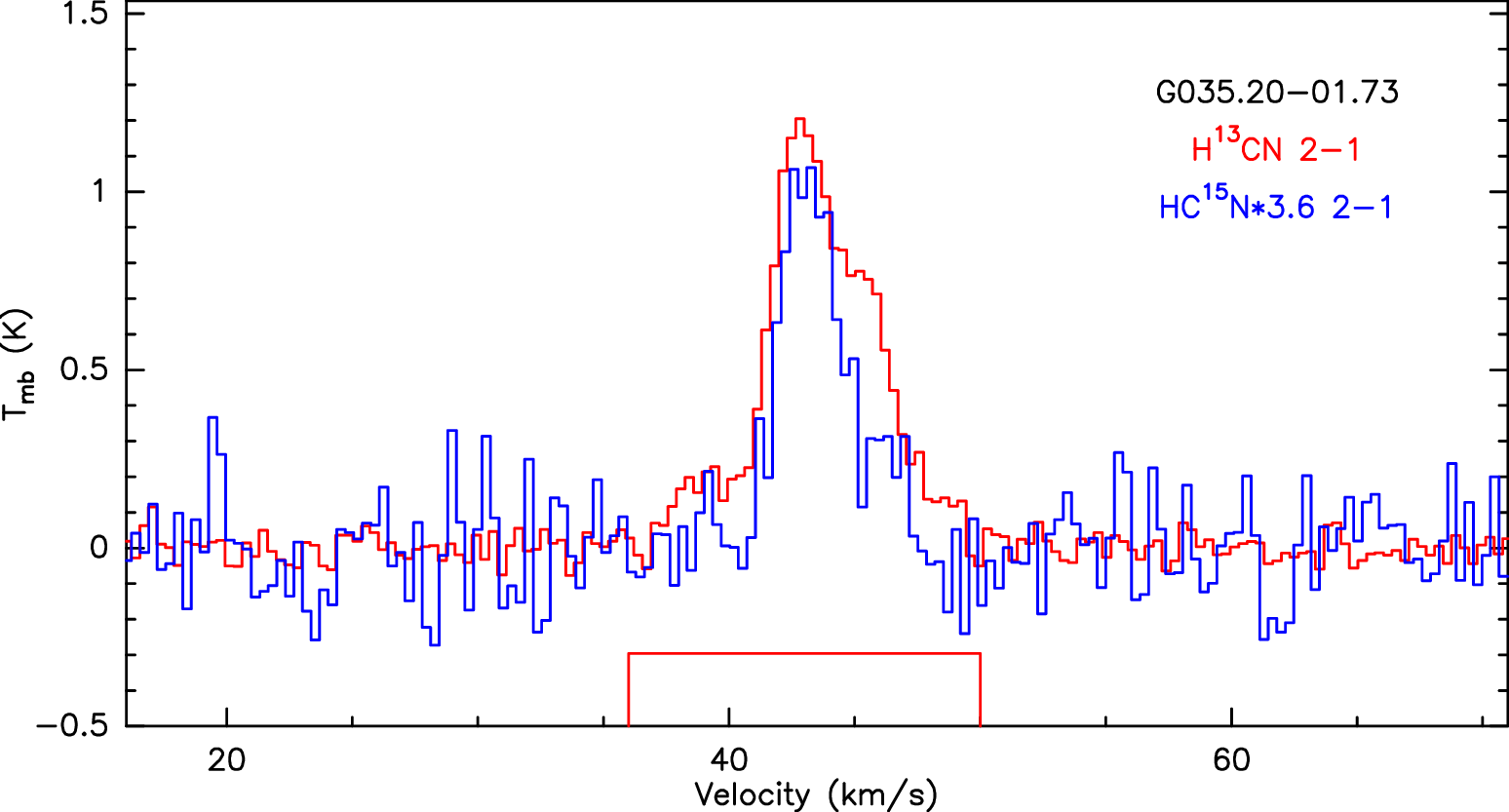}\includegraphics[width=0.24\textwidth]{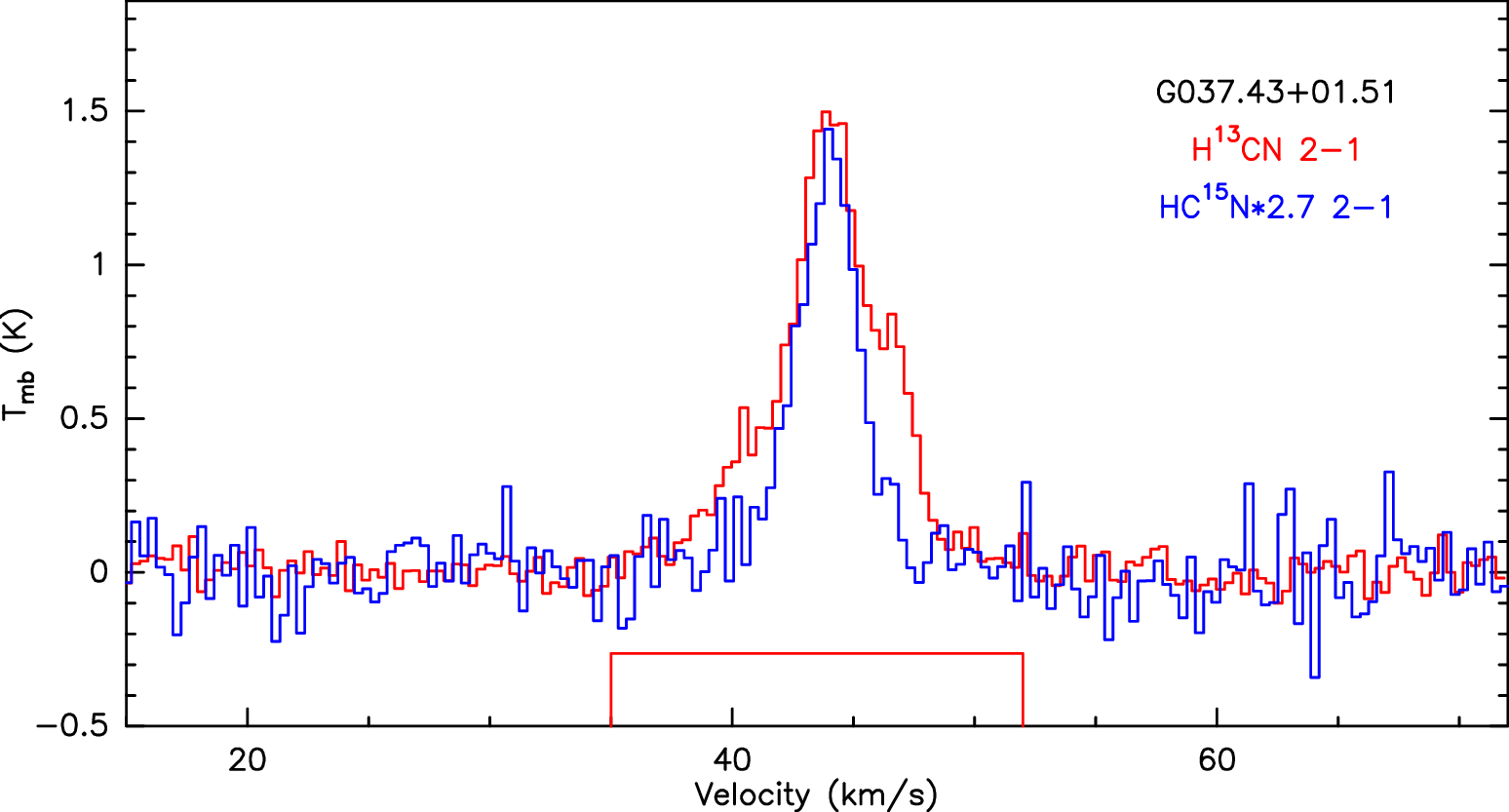}\\
\includegraphics[width=0.24\textwidth]{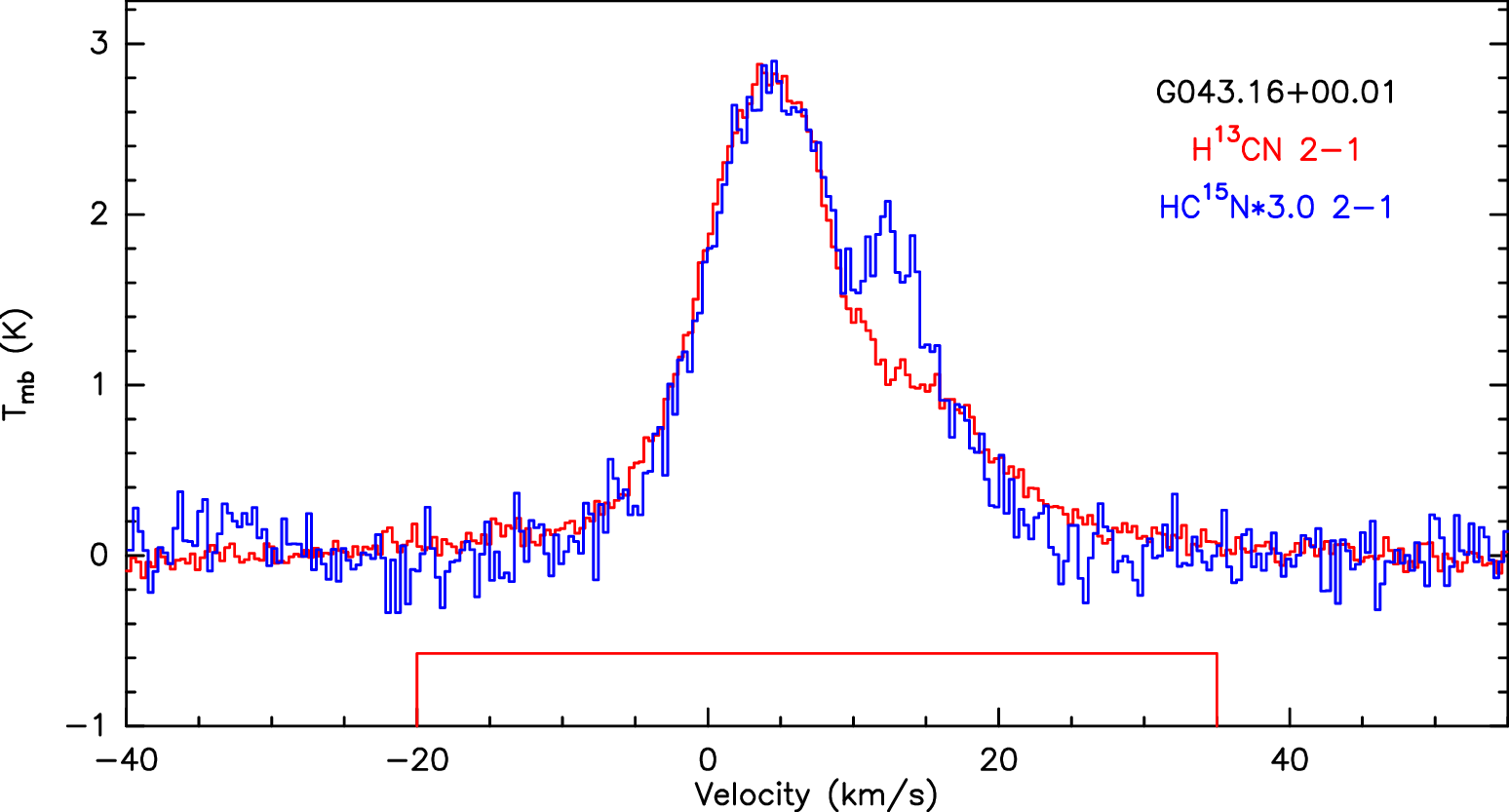}\includegraphics[width=0.24\textwidth]{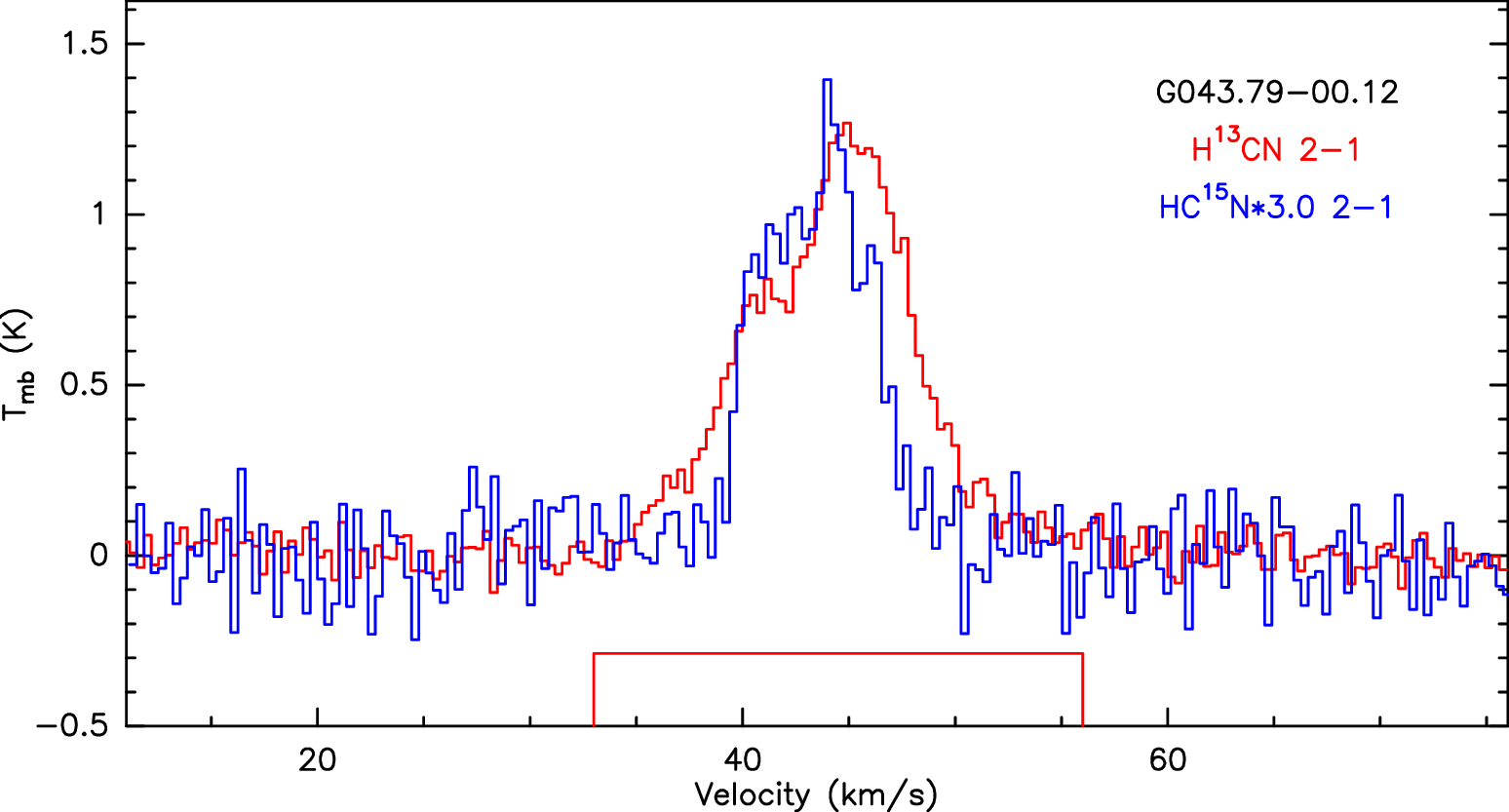}
\includegraphics[width=0.24\textwidth]{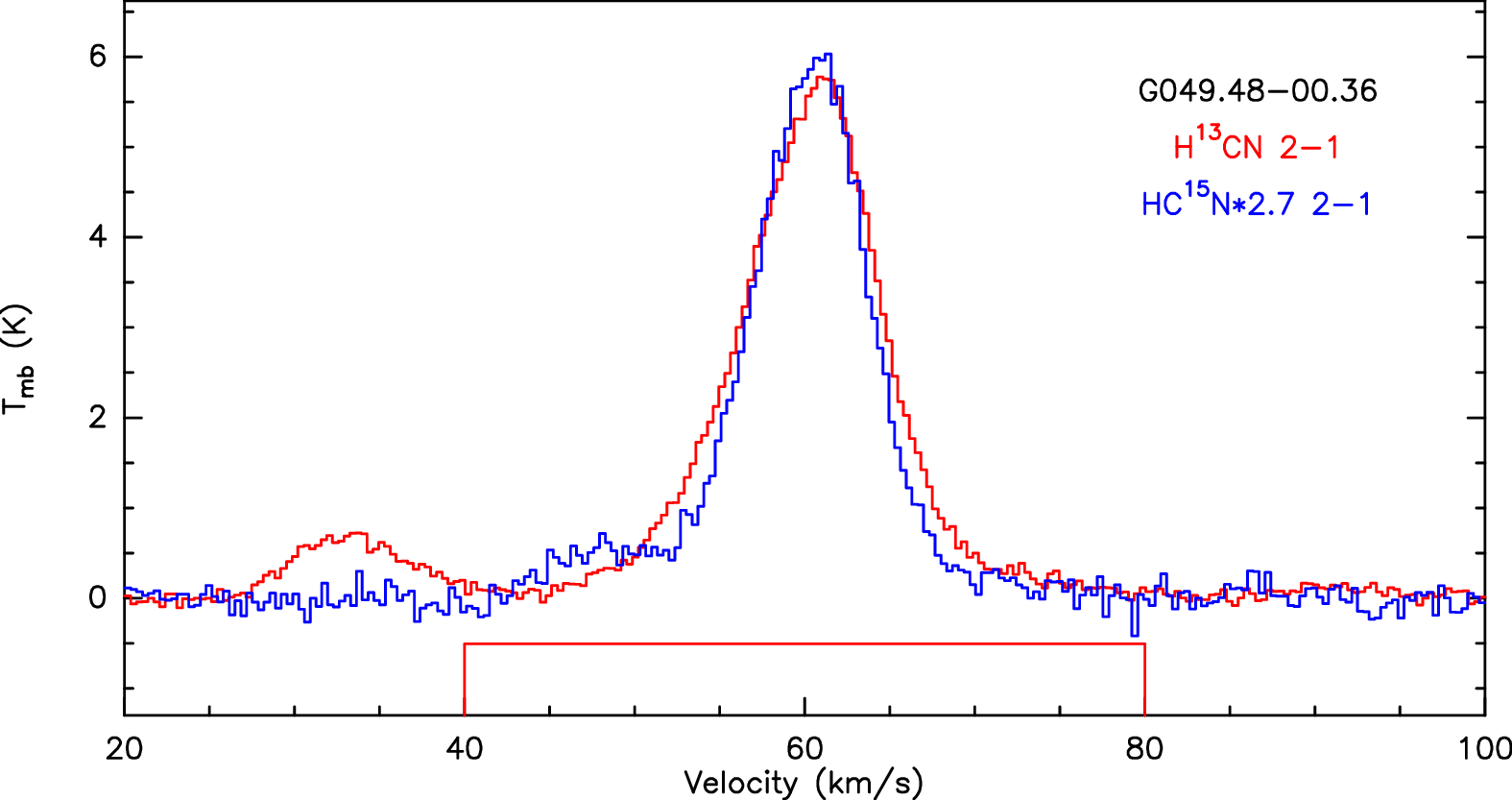}\includegraphics[width=0.24\textwidth]{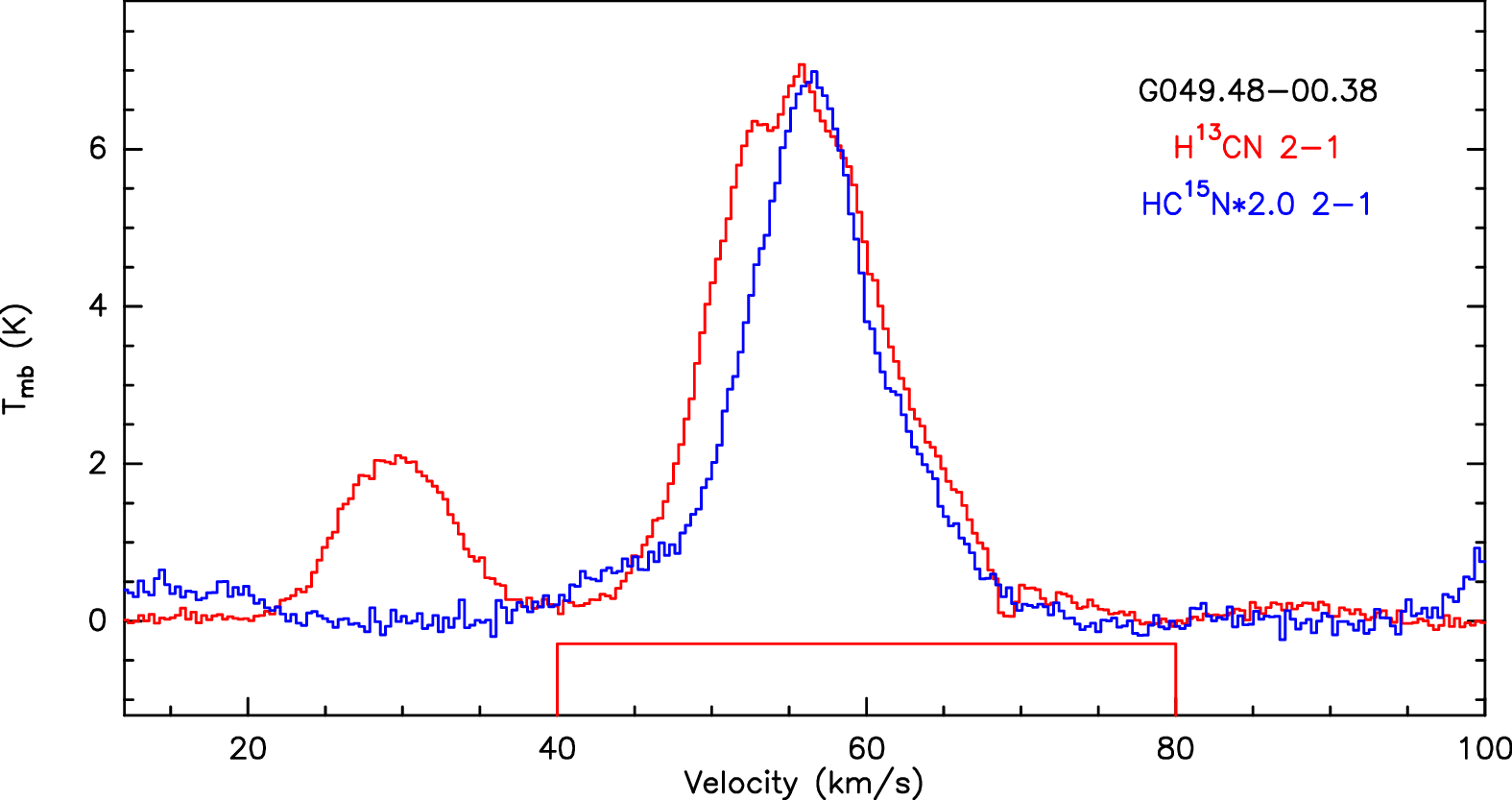}\\
\includegraphics[width=0.24\textwidth]{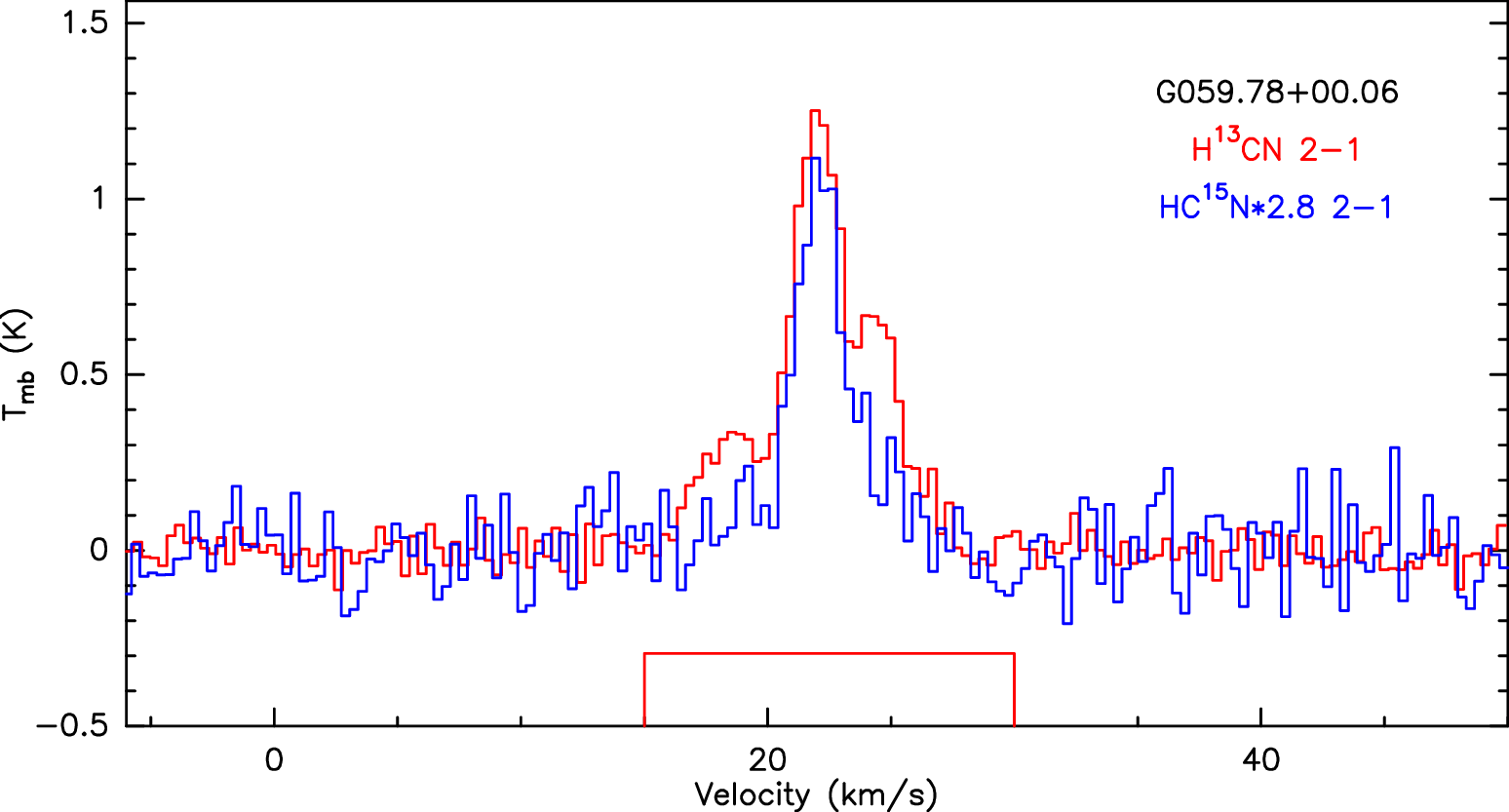}\includegraphics[width=0.24\textwidth]{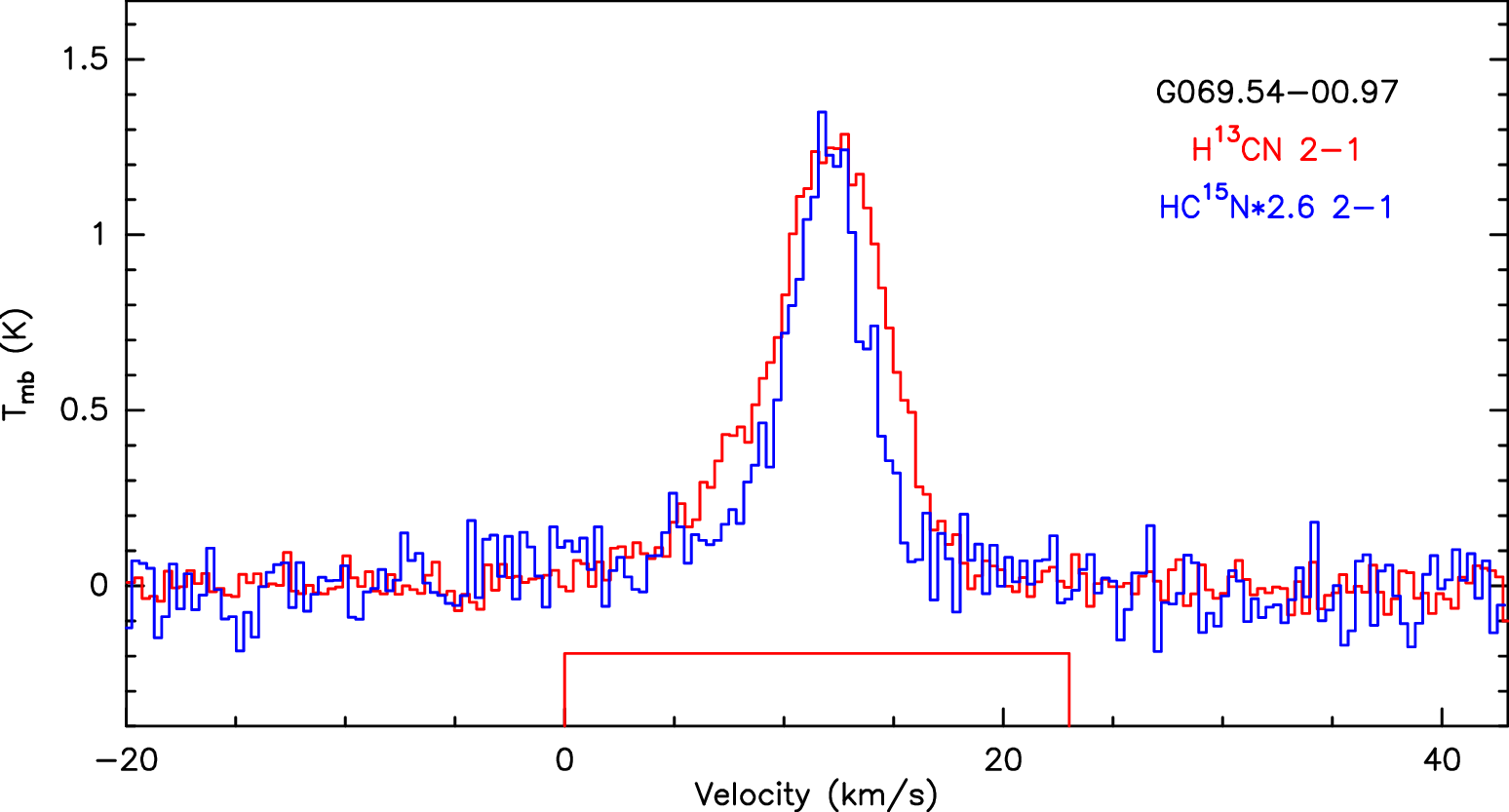}
\includegraphics[width=0.24\textwidth]{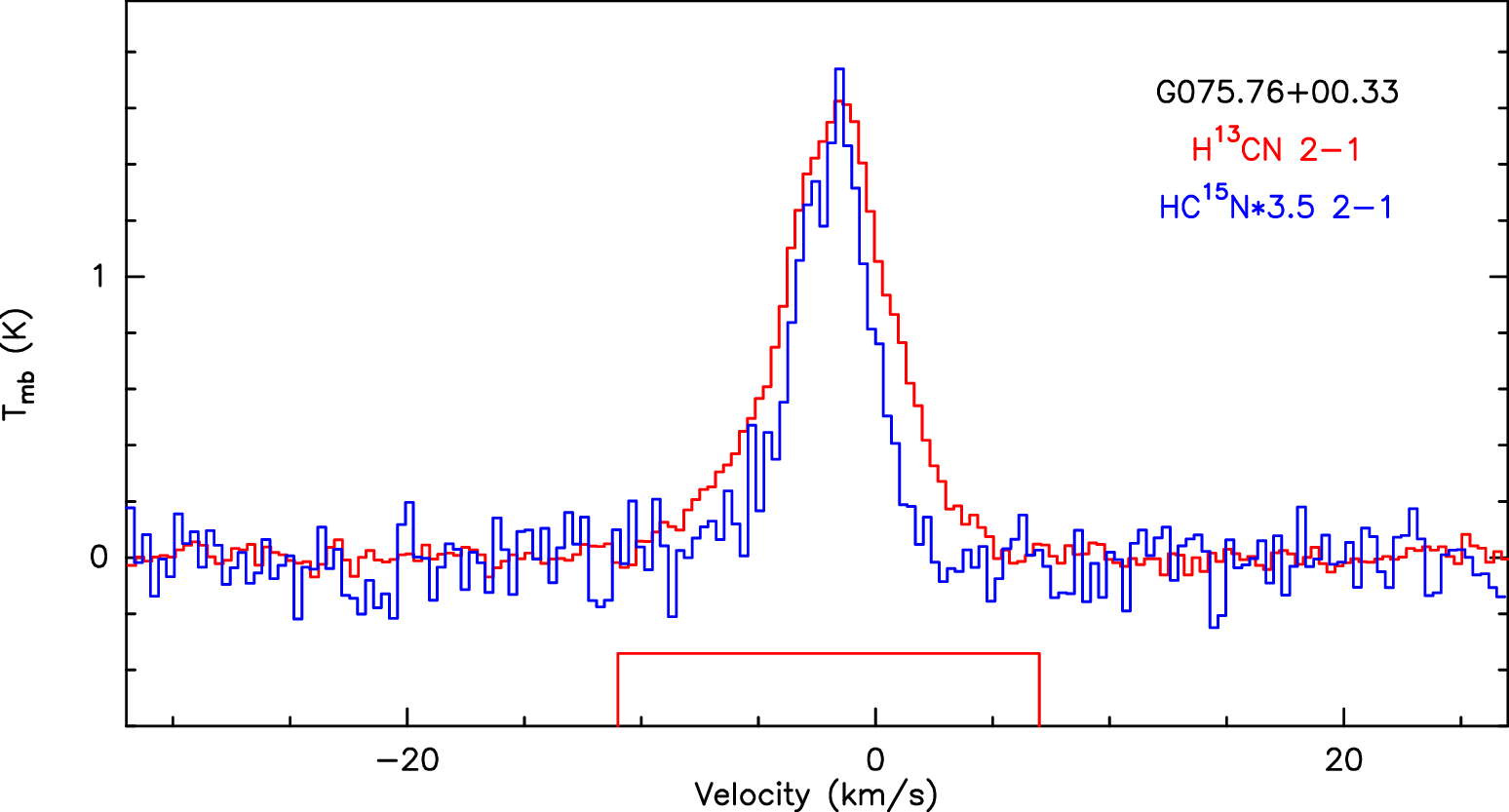}\includegraphics[width=0.24\textwidth]{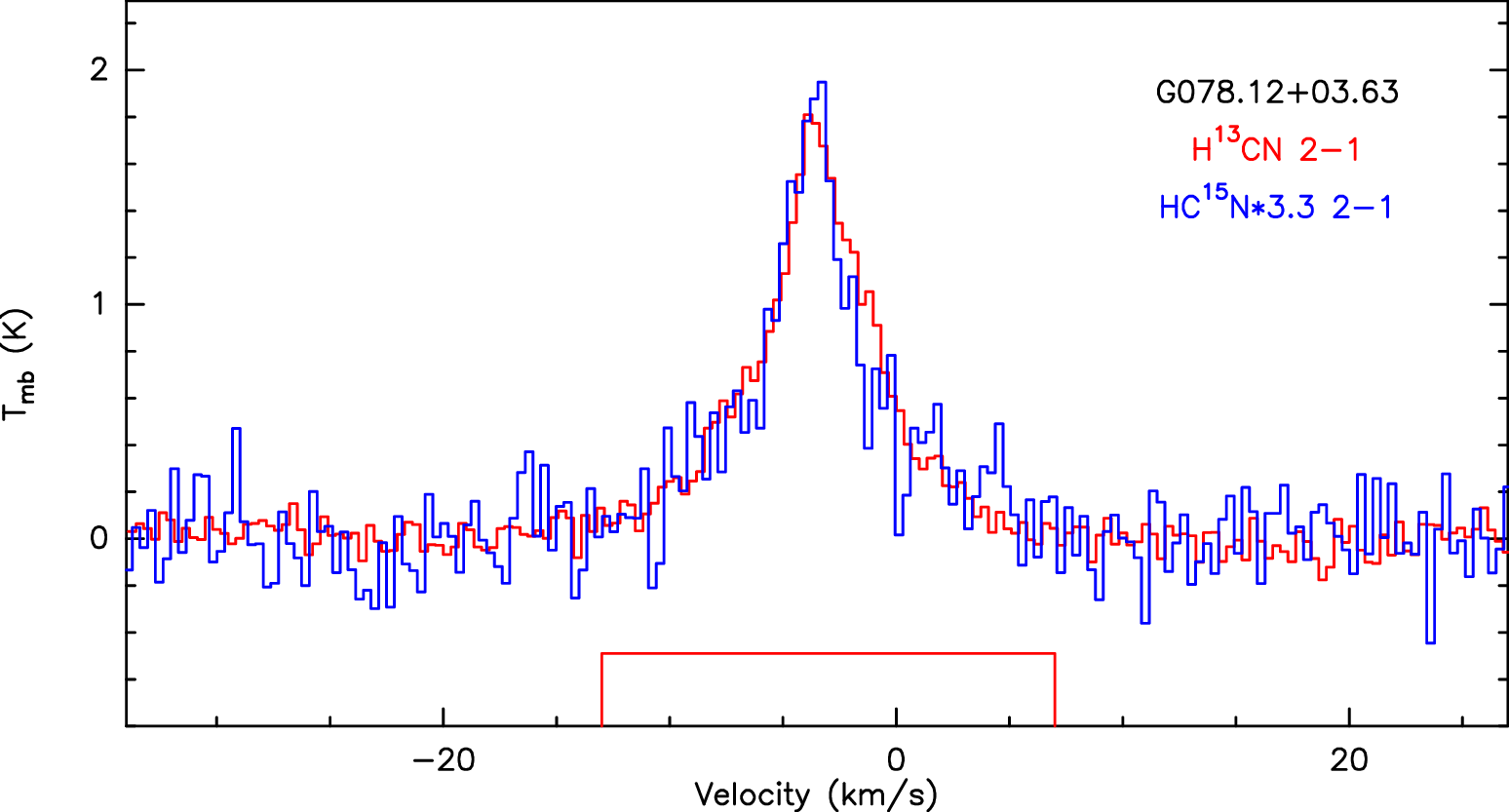}\\
\includegraphics[width=0.24\textwidth]{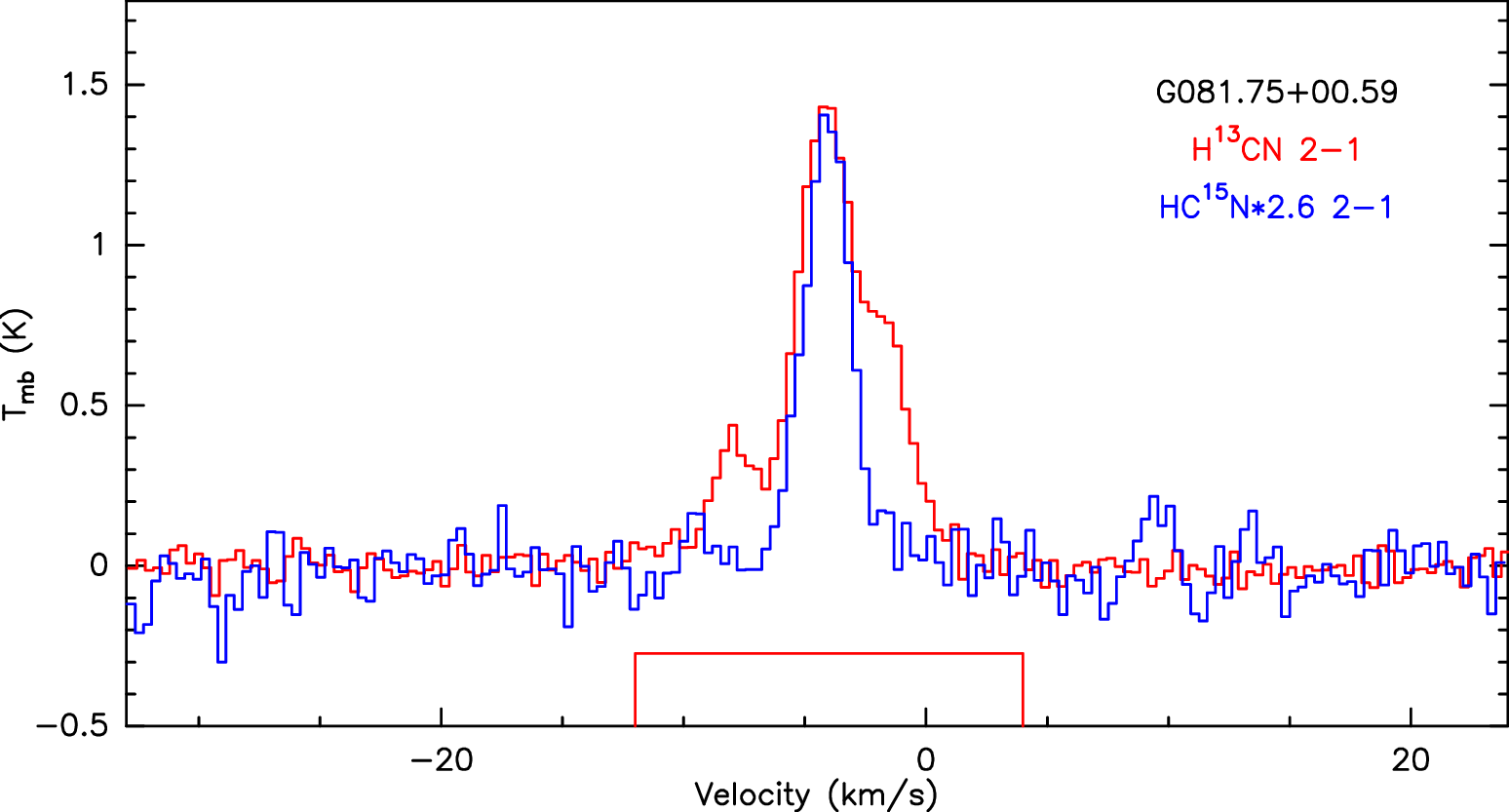}\includegraphics[width=0.24\textwidth]{Spectral_Lines/37.eps}
\includegraphics[width=0.24\textwidth]{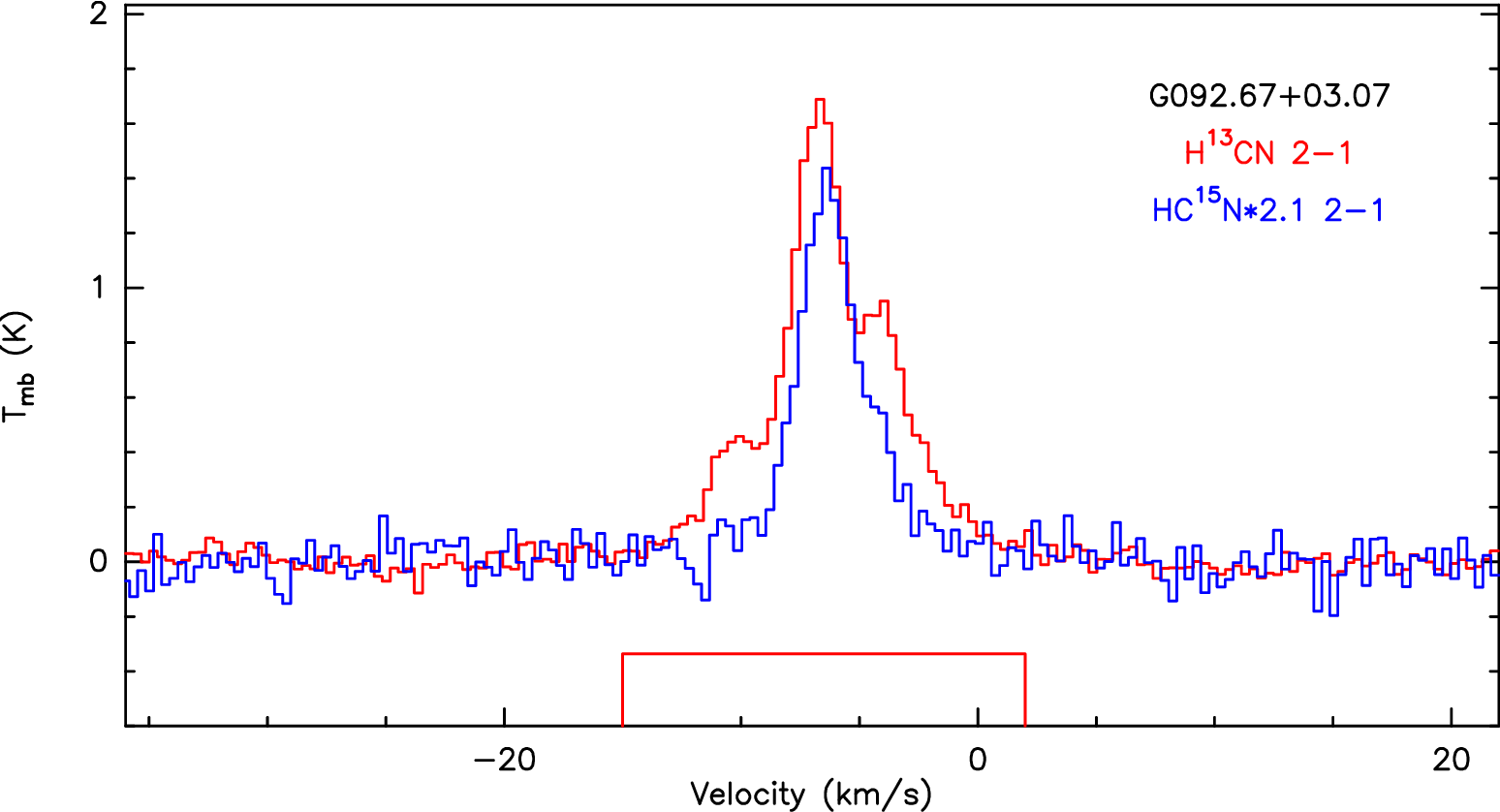}\includegraphics[width=0.24\textwidth]{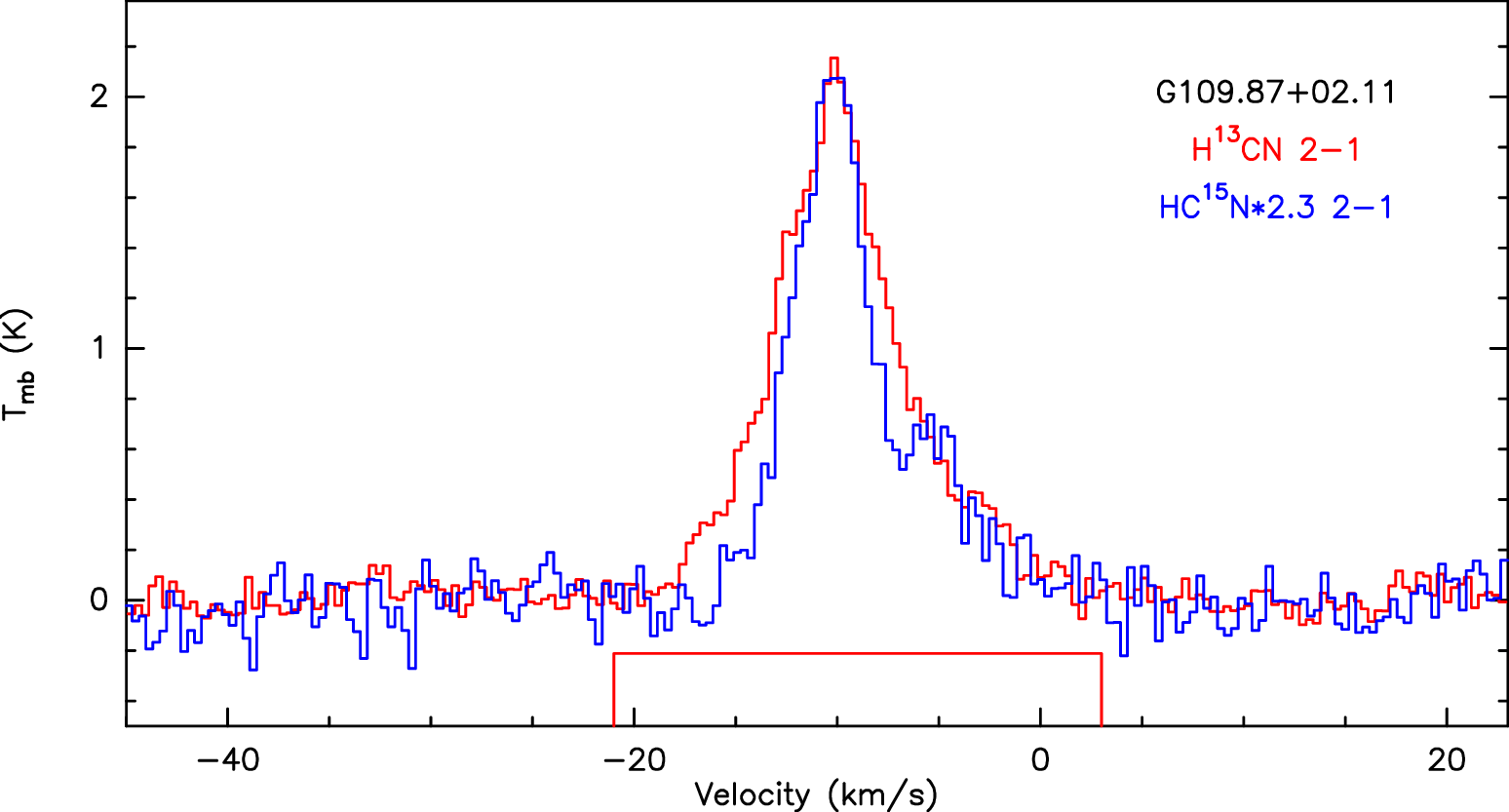}\\
\caption{Observational results of spectra for H$^{13}$CN 2-1 and HC$^{15}$N 2-1. They are plotted on the same velocity scale for each spectrum, with the red boxes below each spectrum indicating the velocity integration range for H$^{13}$CN J=2-1.}
\label{fig:spectrum1}
\end{figure}

\begin{figure*}
\addtocounter{figure}{-1}
\centering
\includegraphics[width=0.24\textwidth]{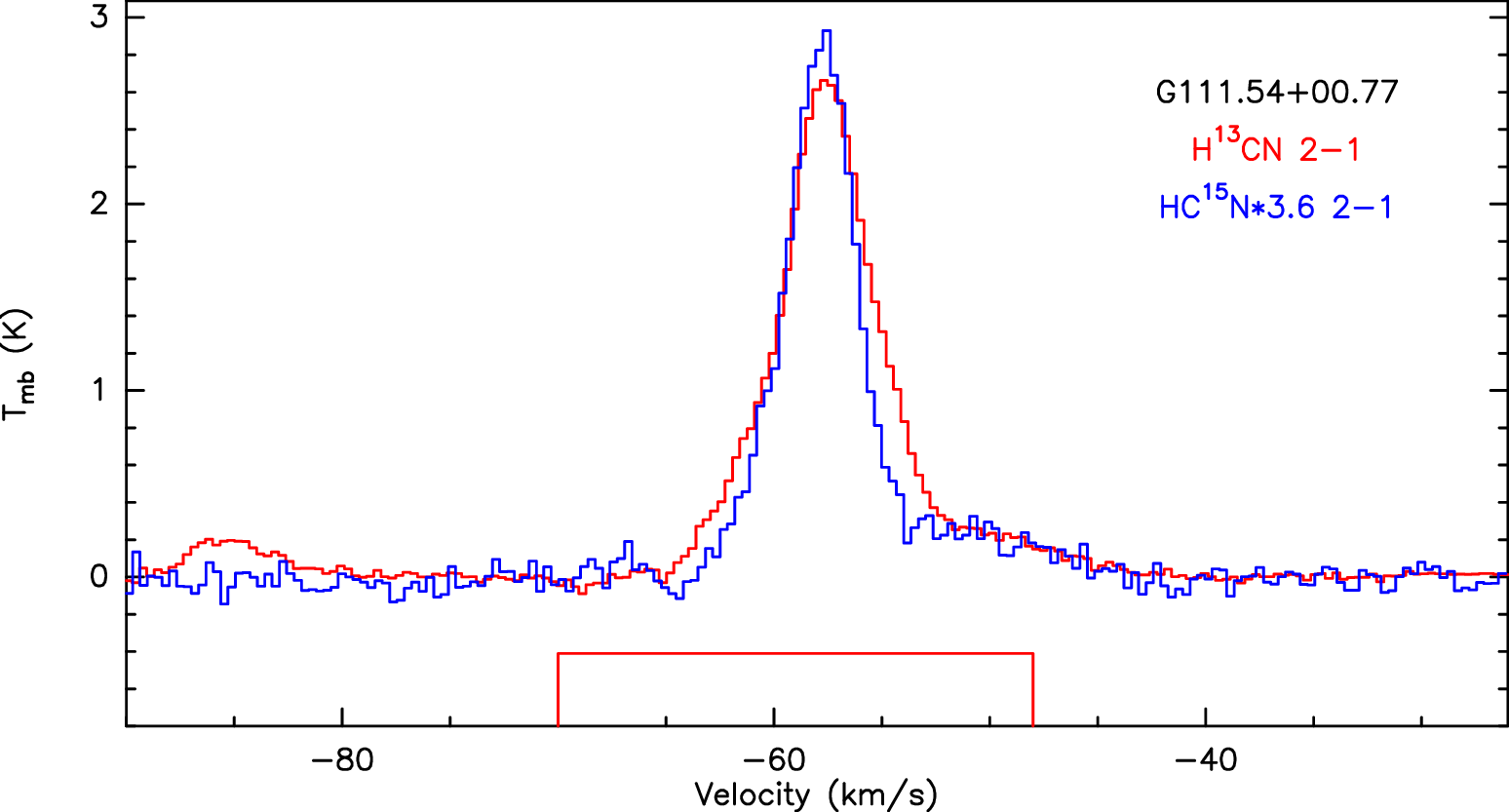}\includegraphics[width=0.24\textwidth]{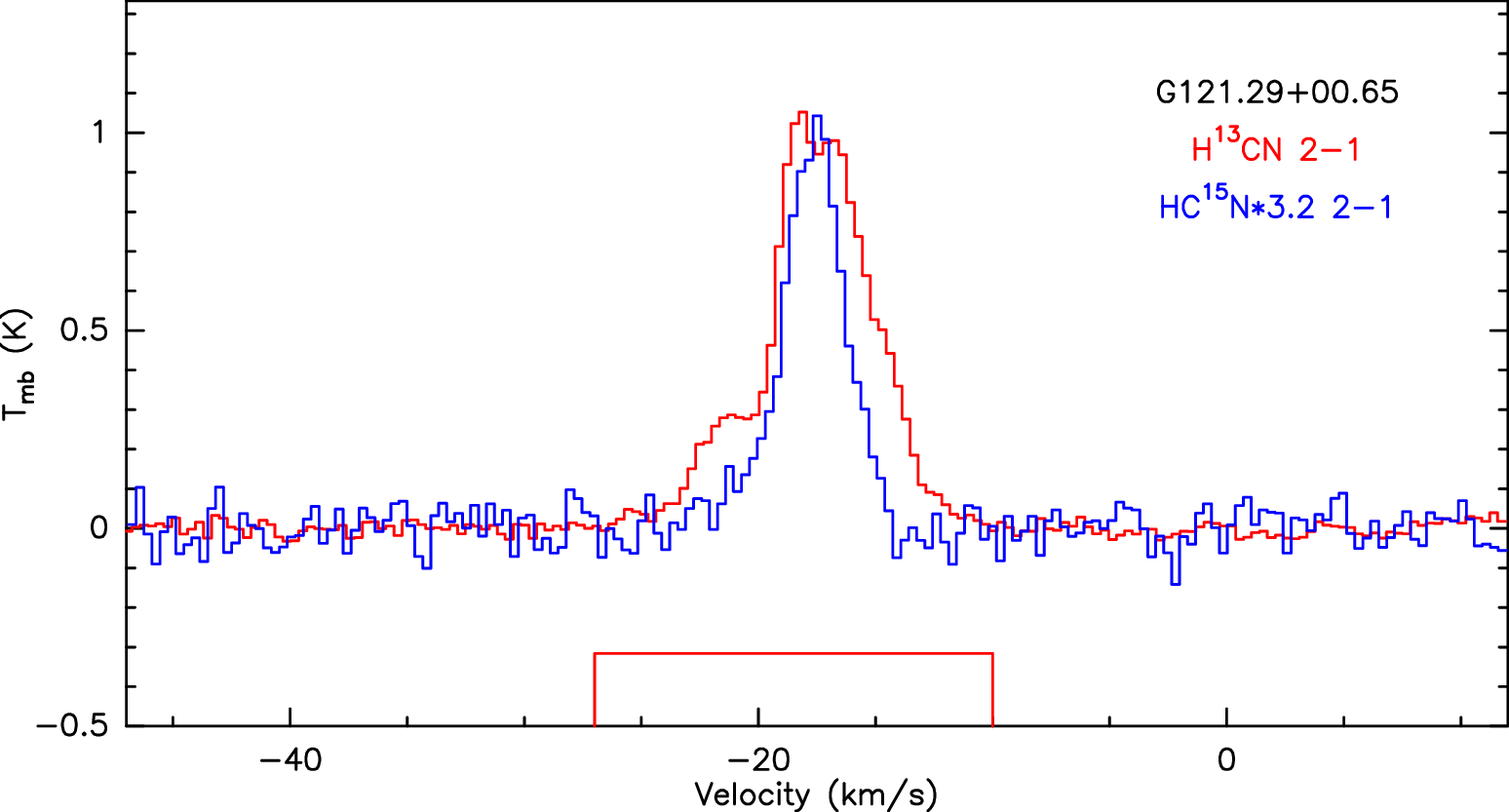}
\includegraphics[width=0.24\textwidth]{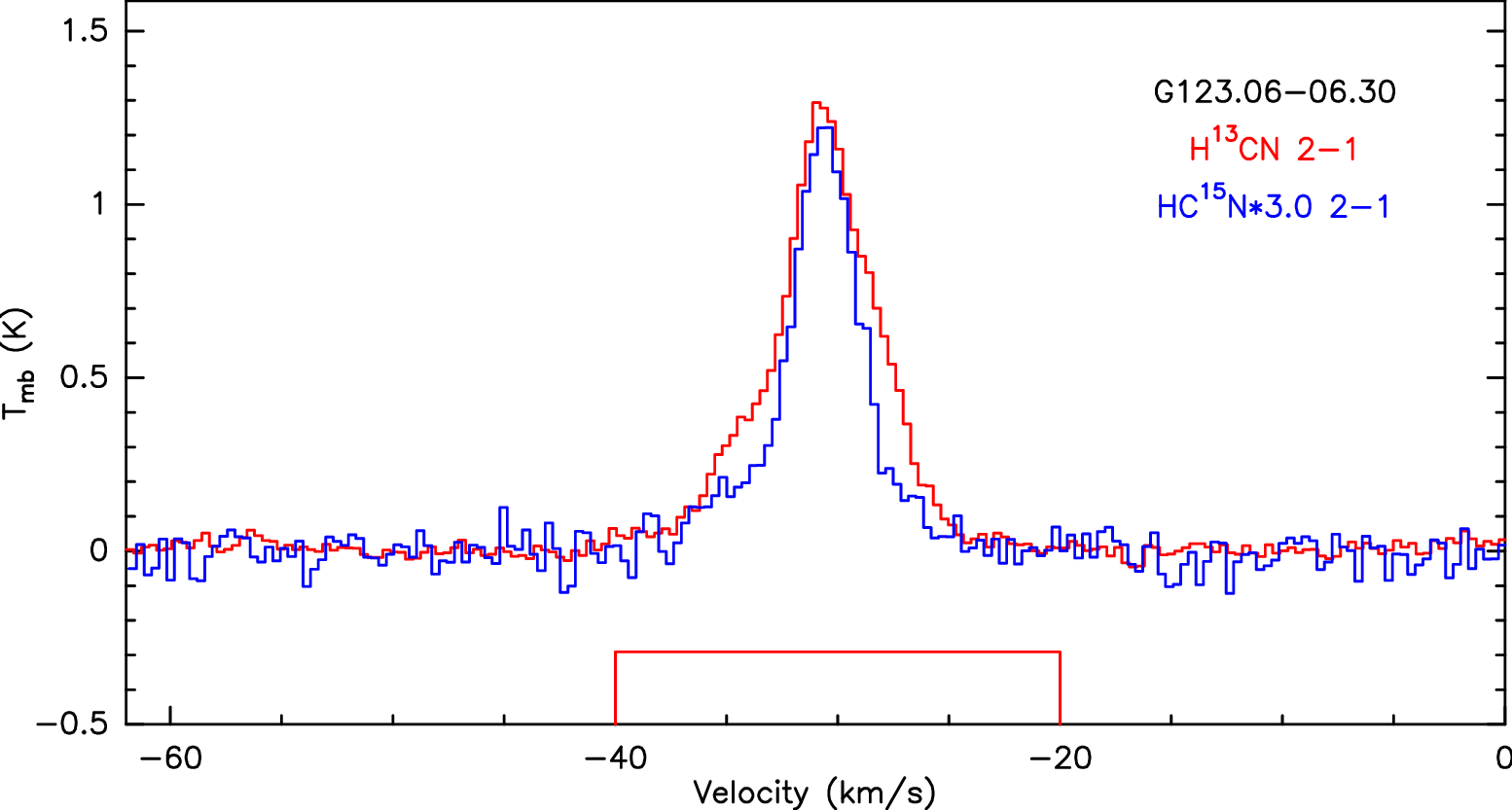}\includegraphics[width=0.24\textwidth]{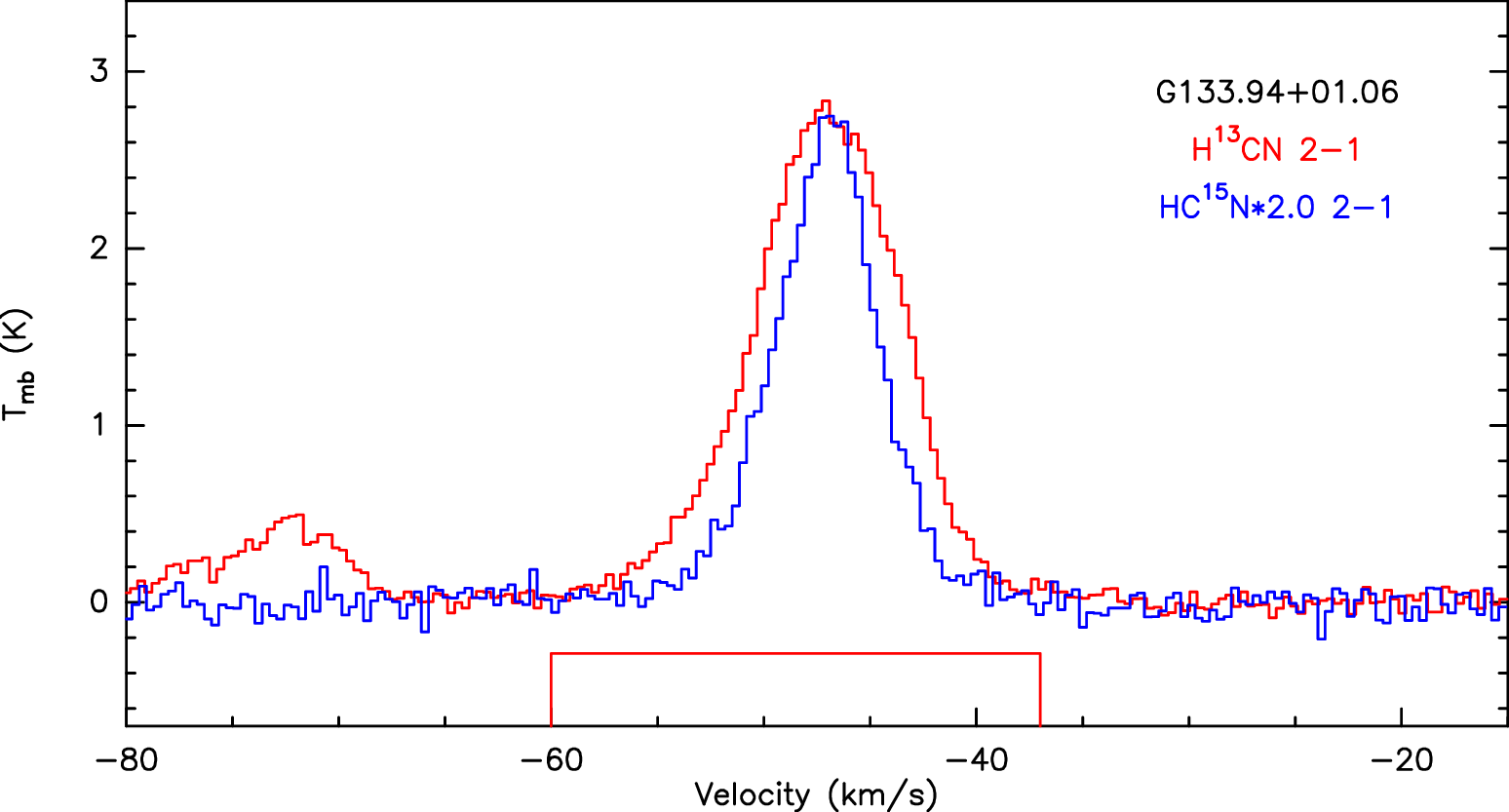}\\
\includegraphics[width=0.24\textwidth]{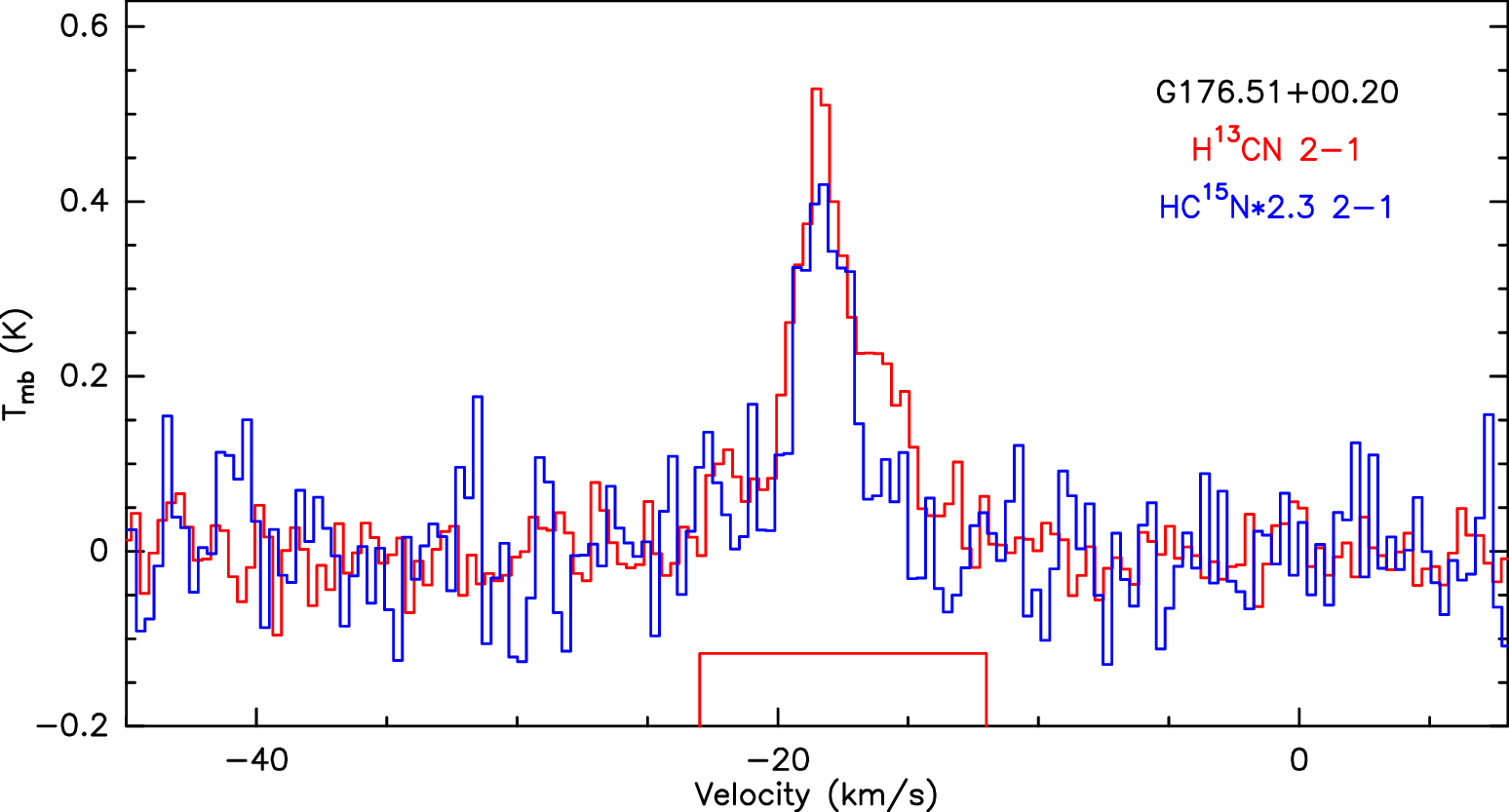}\includegraphics[width=0.24\textwidth]{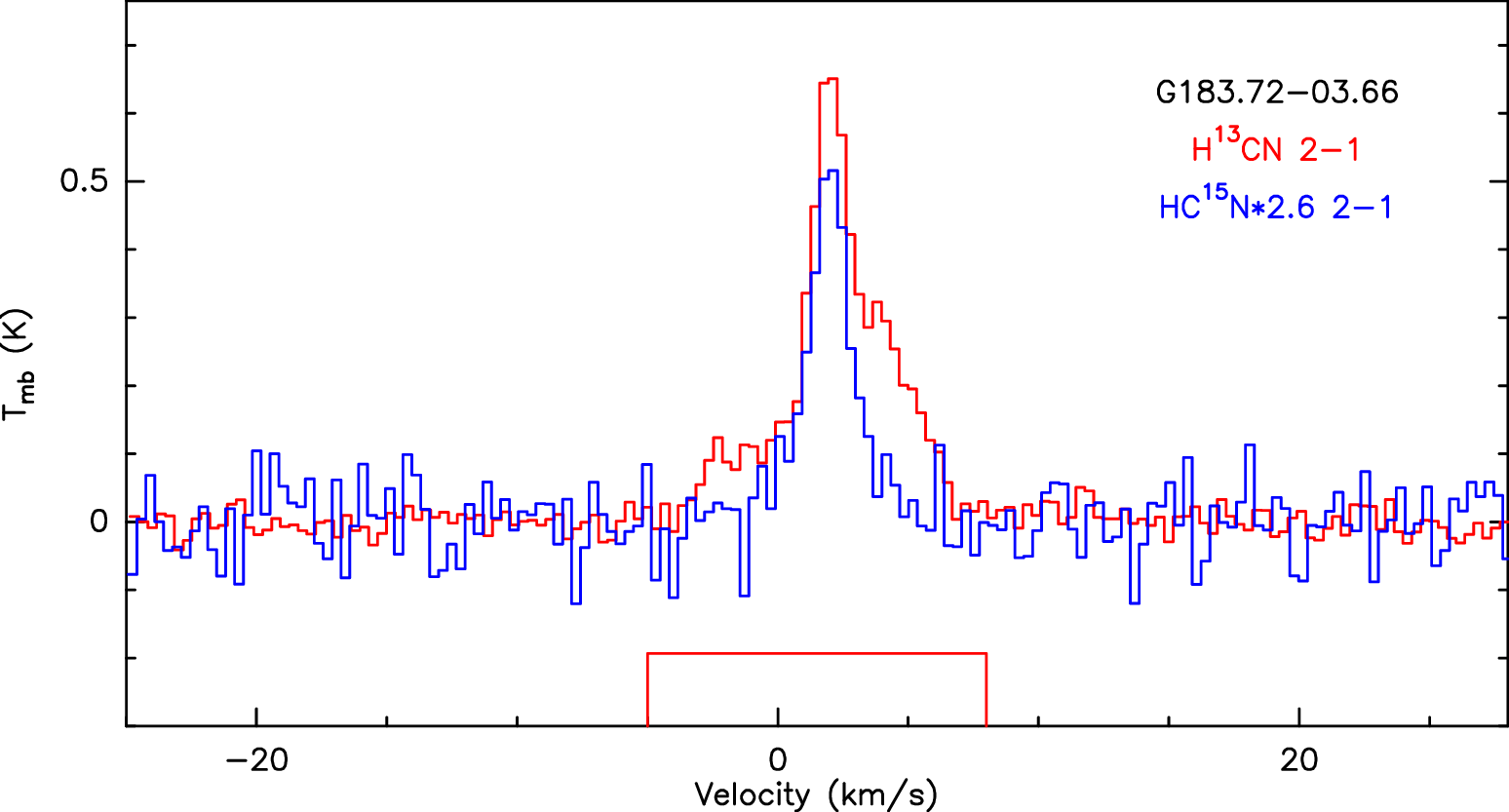}
\includegraphics[width=0.24\textwidth]{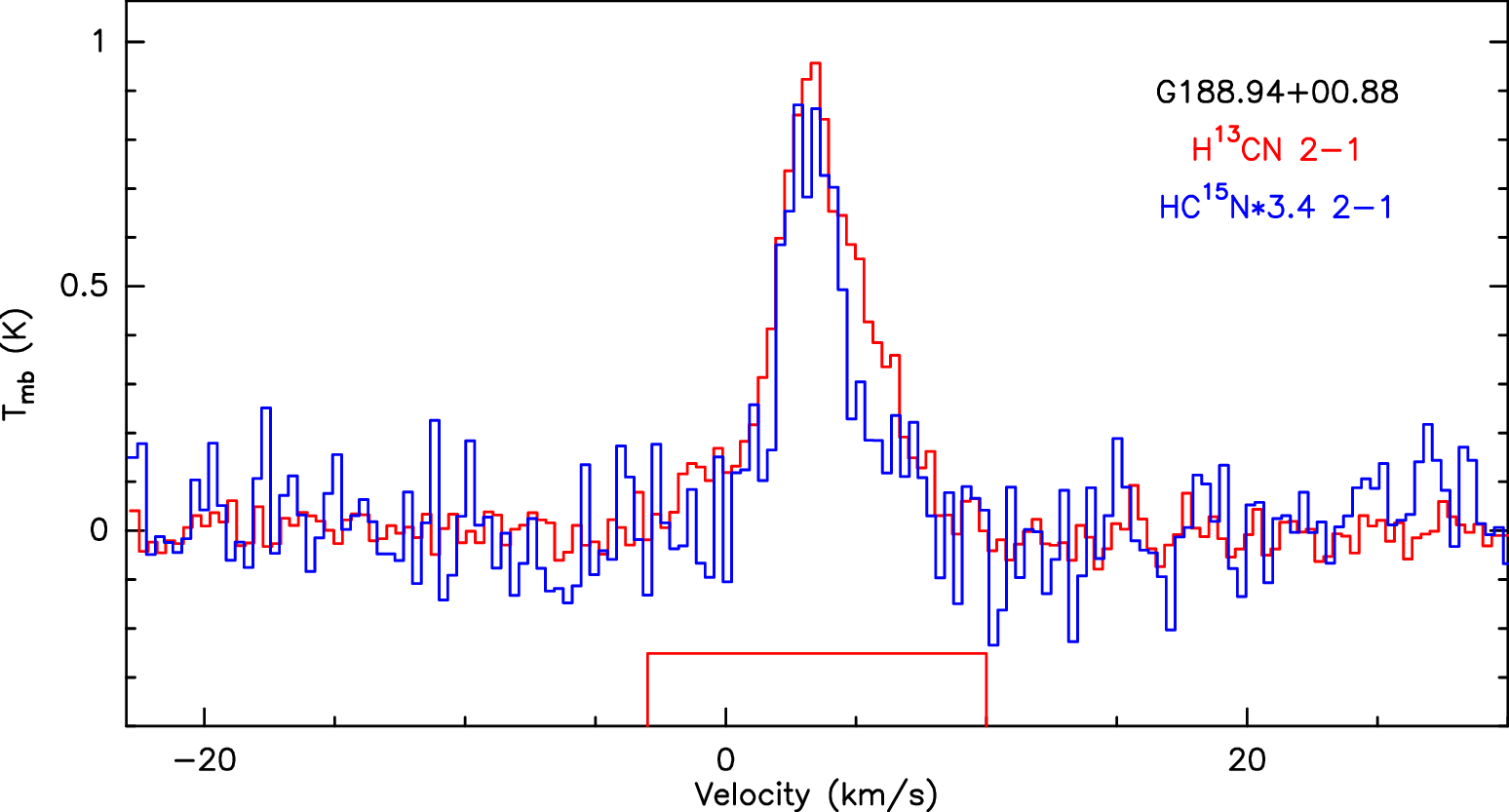}\includegraphics[width=0.24\textwidth]{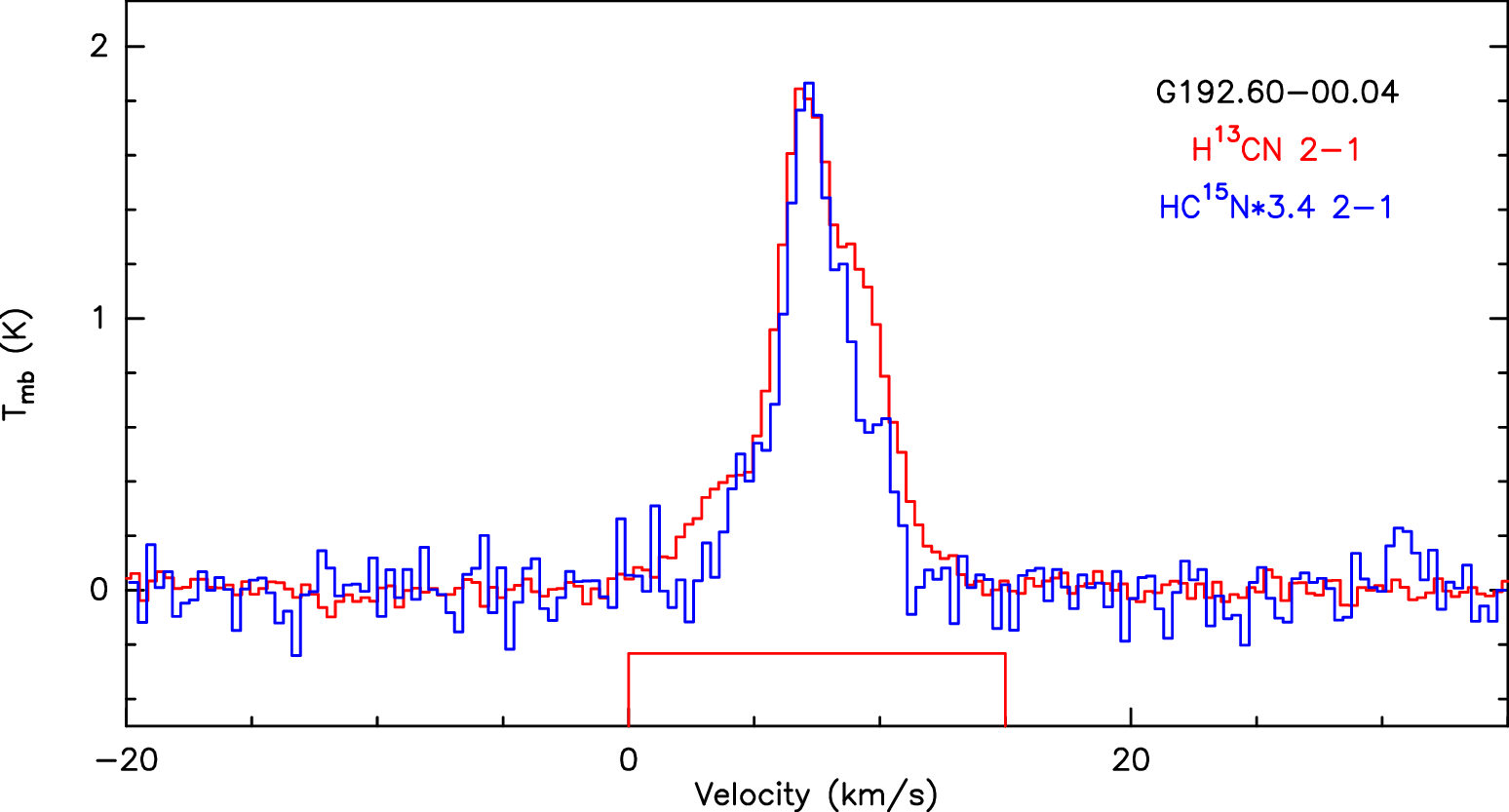}\\
\includegraphics[width=0.24\textwidth]{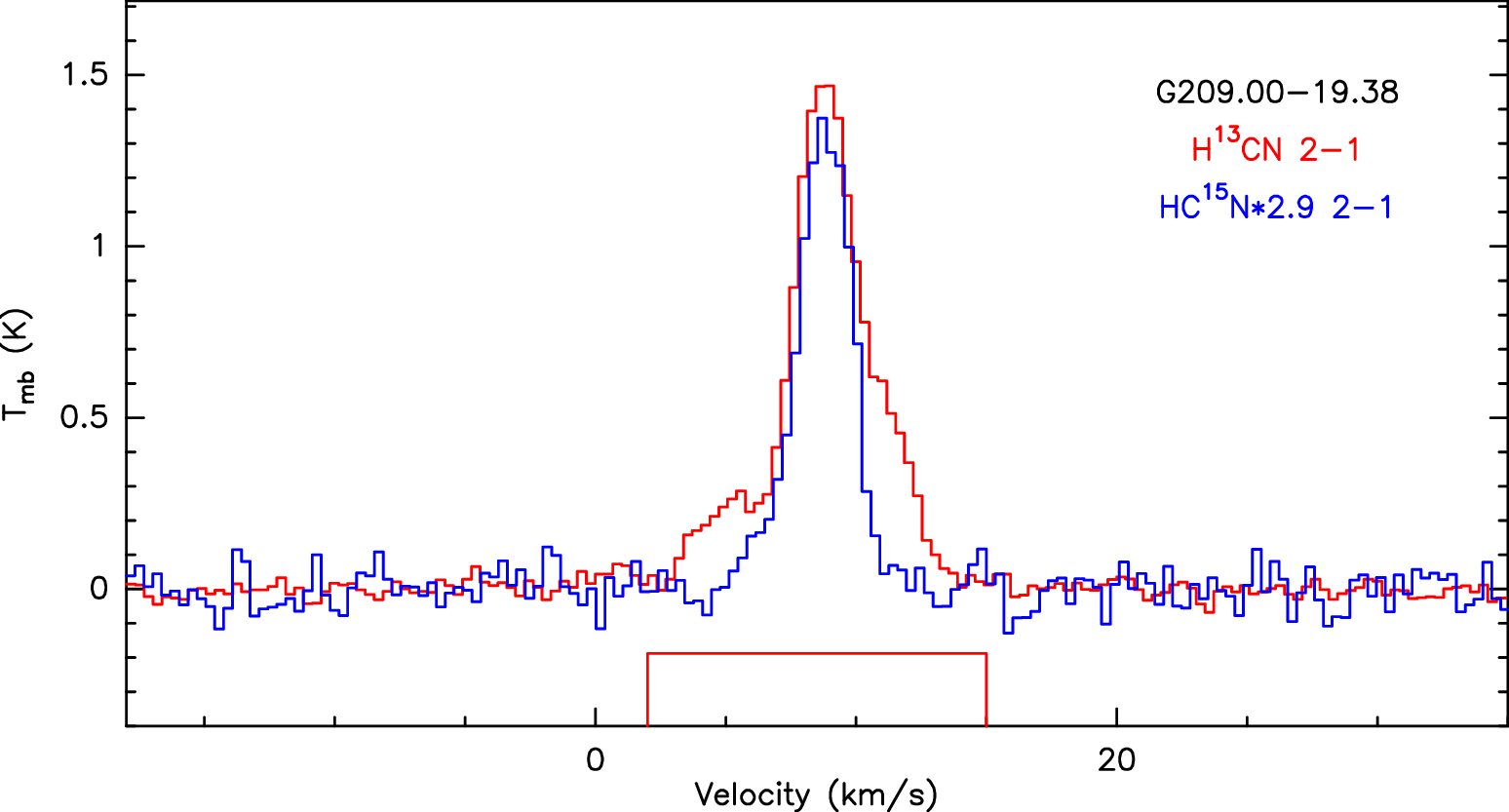}\includegraphics[width=0.24\textwidth]{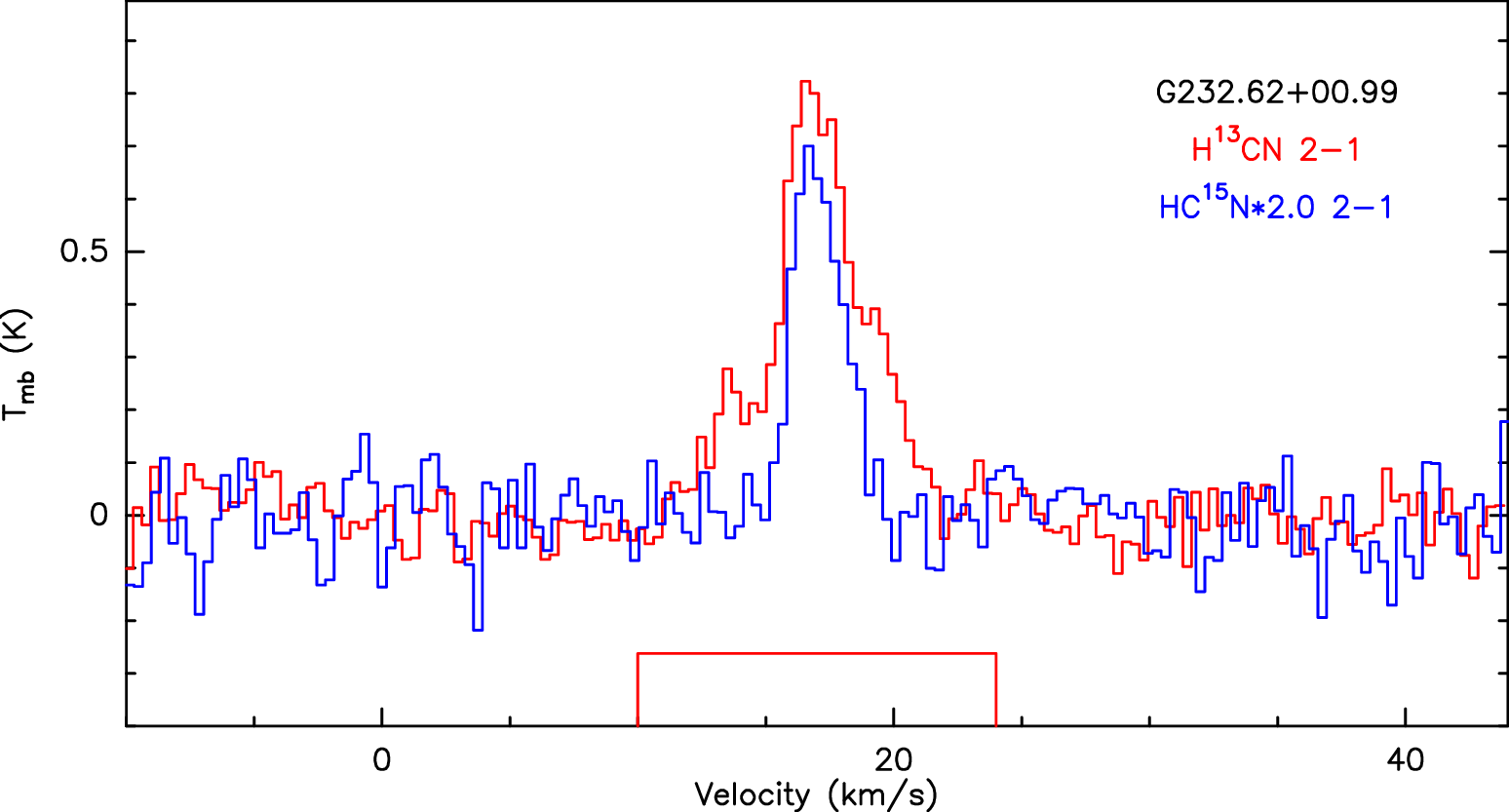}
\includegraphics[width=0.24\textwidth]{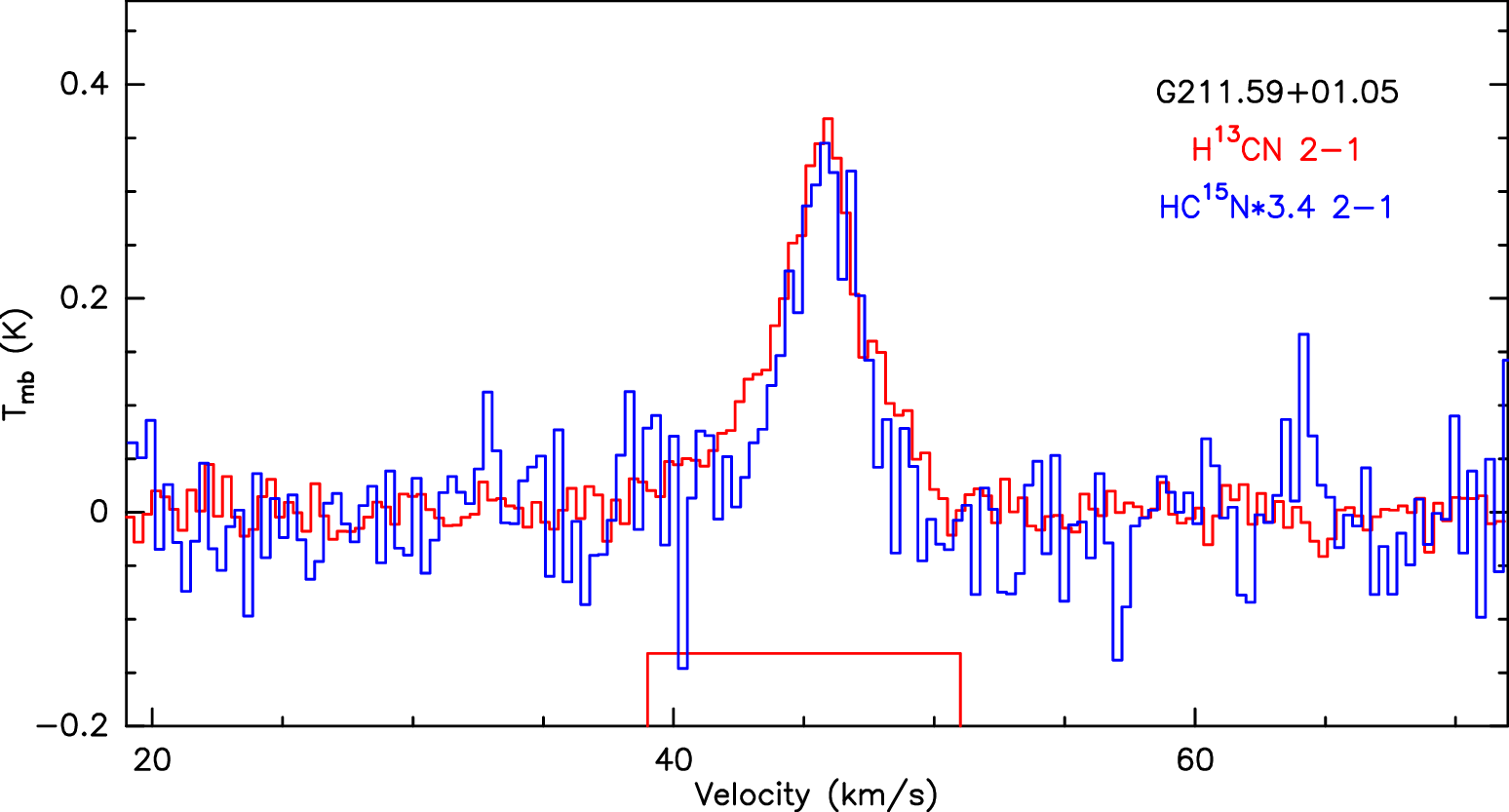}\\
\caption{Continued.}
\end{figure*}

\begin{figure*}
\centering
\includegraphics[width=0.24\textwidth]{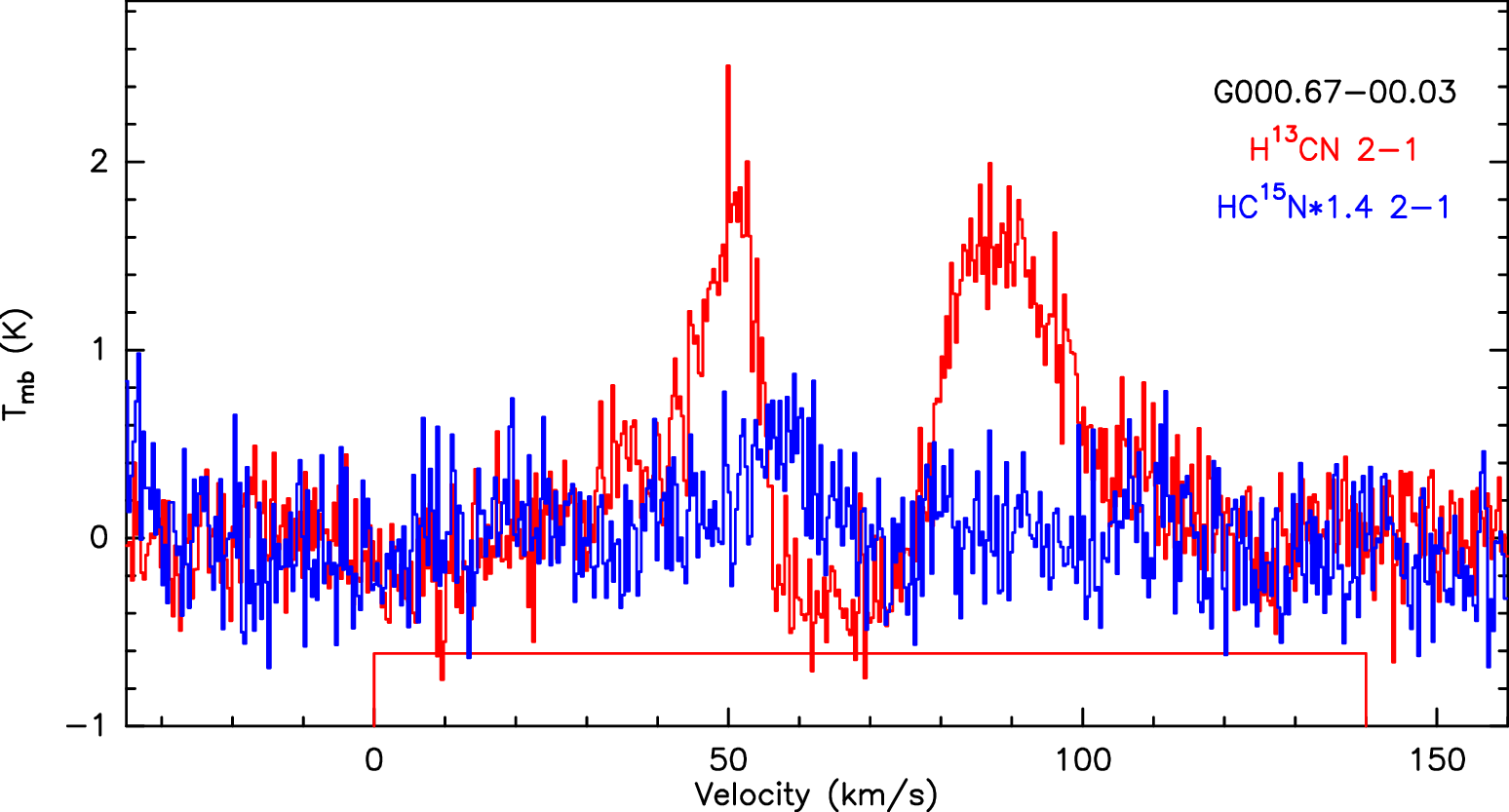}\includegraphics[width=0.24\textwidth]{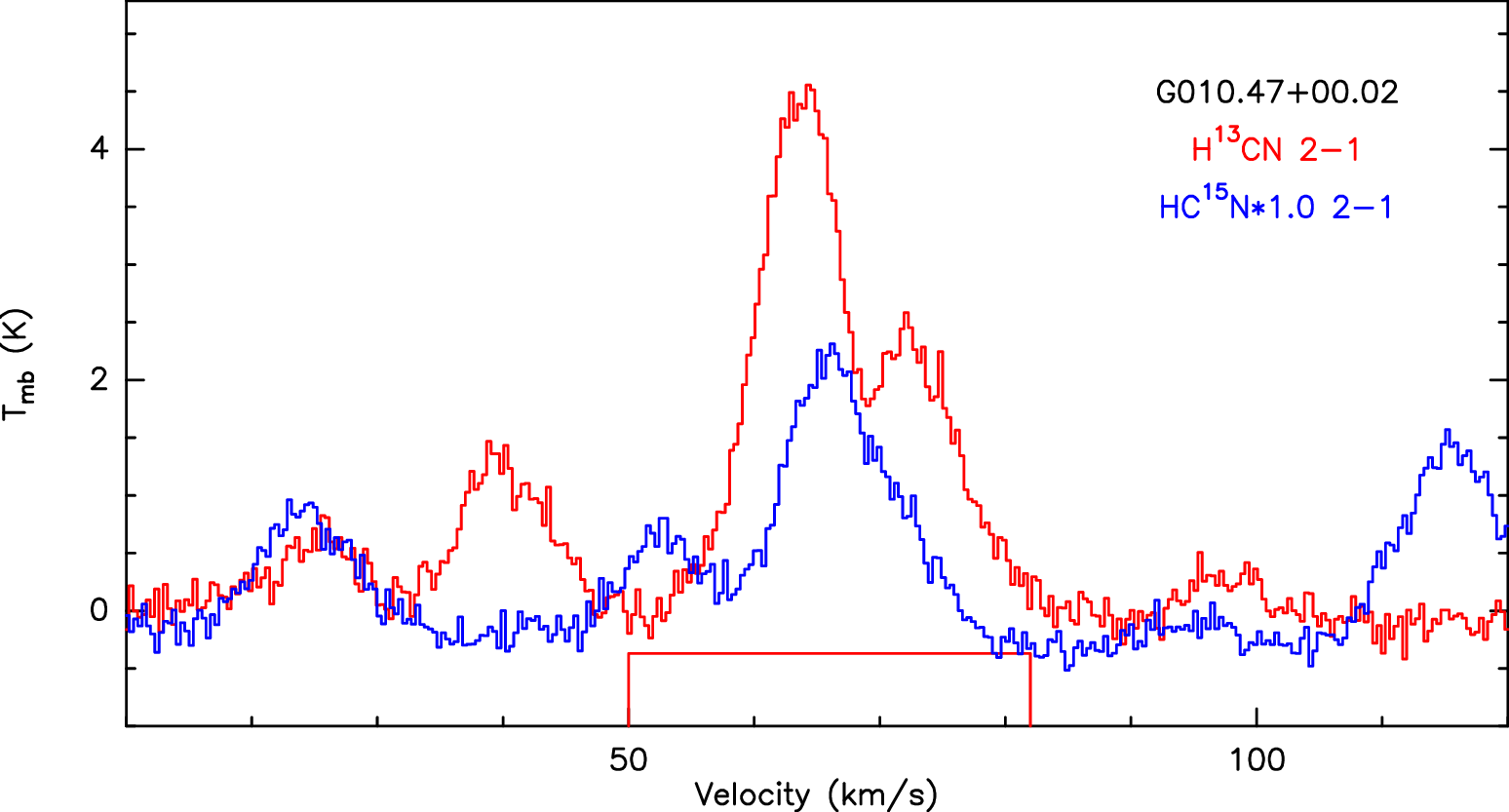}
\includegraphics[width=0.24\textwidth]{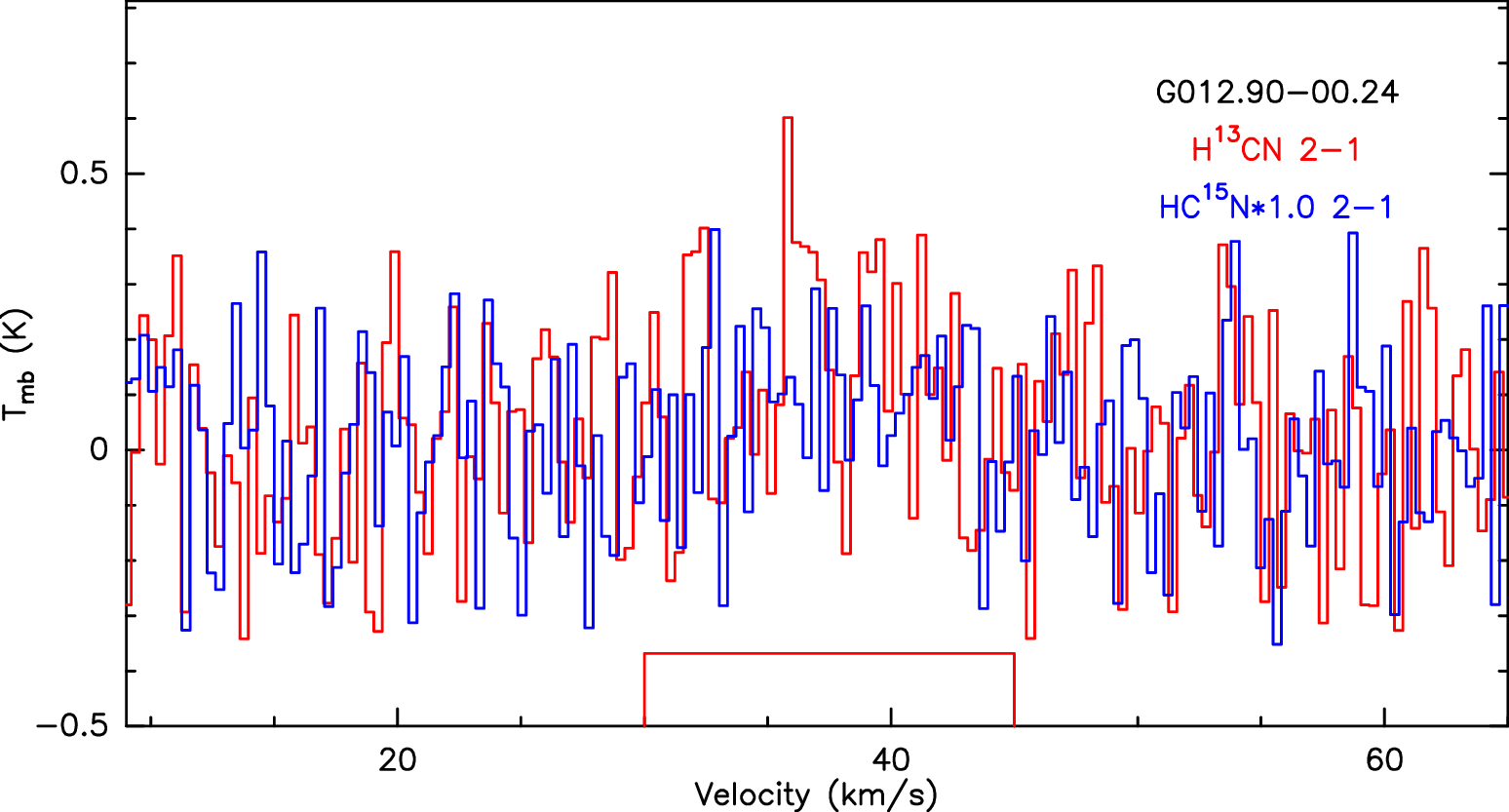}\includegraphics[width=0.24\textwidth]{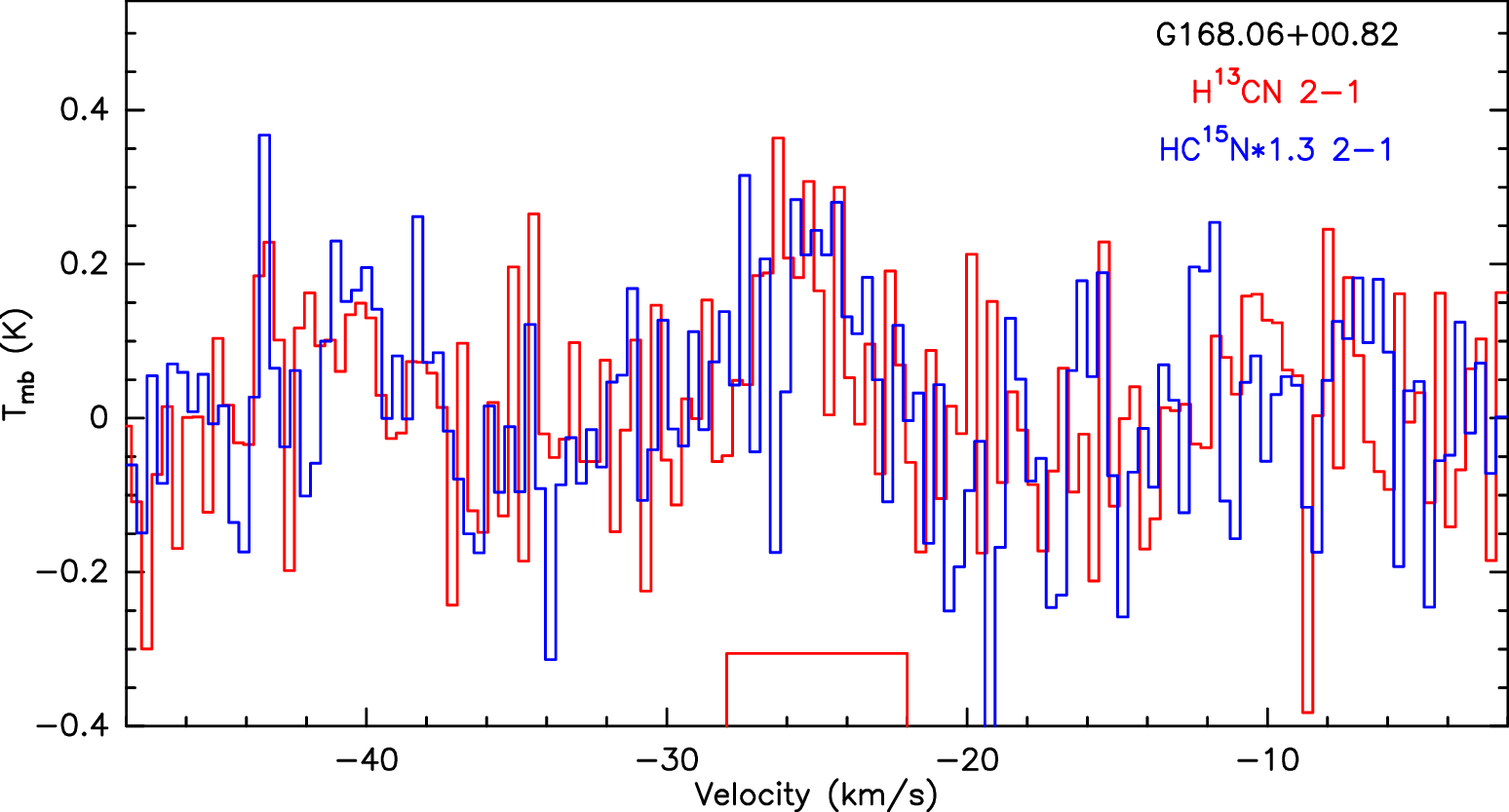}\\
\caption{Four sources (IRAS 05137+3919, G012.90-00.24, Sgr B2, and G010.47+00.02) excluded from the ratio calculations due to non-detections or contamination in the H$^{13}$CN J=2-1 and/or HC$^{15}$N J=2-1 lines.}
\label{fig:spectrum2}
\end{figure*}

\end{appendix}

\end{document}